\documentclass[aps, prd, reprint, superscriptaddress, nofootinbib,
]{revtex4-2}

\usepackage{times}
\usepackage{graphicx}
\usepackage[usenames,dvipsnames]{xcolor}
\usepackage{xspace}
\usepackage{amsmath,amsfonts,amssymb,amsthm,bm}
\usepackage{mathtools}
\usepackage[utf8]{inputenc}
\usepackage[normalem]{ulem}
\usepackage{makecell}
\usepackage{multirow}
\usepackage{float}
\usepackage{adjustbox}
\usepackage[caption=false]{subfig}
\usepackage{rotating}
\usepackage{dcolumn}

\xdefinecolor{mylinkcolor}{rgb}{0,0,0.8}
\usepackage[
	bookmarksnumbered, bookmarksopen, bookmarksopenlevel=2,
	breaklinks=true,
	colorlinks=true, filecolor=mylinkcolor, citecolor=mylinkcolor,
	linkcolor=mylinkcolor, urlcolor=mylinkcolor, menucolor=mylinkcolor,
]{hyperref}

\usepackage{setspace}
\usepackage[hang]{footmisc} % 'hang' separates the number of the footnote from the paragraph
\allowdisplaybreaks

\def\mr{\mathrm}

\def\di{\mr d}

\def\mF{\mathcal{F}}
\def\e{\mathrm{e}}
\def\s{\hspace{0.5pt}}
\def\solarmass{M_\odot}

\graphicspath{ {fig/} }

\newcommand{\AEI}{\affiliation{Max Planck Institute for Gravitational Physics (Albert Einstein Institute), Am M\"uhlenberg 1, Potsdam 14476, Germany}}
\newcommand{\Maryland}{\affiliation{Department of Physics, University of Maryland, College Park, MD 20742, USA}}
\newcommand{\PI}{\affiliation{Perimeter Institute for Theoretical Physics, 31 Caroline Street North, Waterloo, ON N2L 2Y5, Canada}}
\newcommand{\Ni}{\affiliation{Nikhef, Science Park 105, 1098 XG, Amsterdam, The Netherlands}}
\newcommand{\Cornell}{\affiliation{Cornell Center for Astrophysics and Planetary Science, Cornell University, Ithaca, New York 14853, USA}}
\newcommand{\Caltech}{\affiliation{Theoretical Astrophysics 350-17, California Institute of Technology, Pasadena, California 91125, USA}}
\newcommand{\CENTRA}{\affiliation{CENTRA, Departamento de F\'isica, Instituto Superior T\'ecnico – IST, Universidade de Lisboa – UL, Avenida Rovisco Pais 1, 1049-001 Lisboa, Portugal}}

\begin{document}
%\count\footins = 1000

\title{Accurate waveforms for eccentric, aligned-spin binary black holes: \\
The multipolar effective-one-body model \texttt{SEOBNRv5EHM}}

\author{Aldo Gamboa}
\email{aldo.gamboa@aei.mpg.de}
\AEI

\author{Alessandra Buonanno}
\AEI
\Maryland

\author{Raffi Enficiaud}
\AEI

\author{Mohammed Khalil}
\PI

\author{Antoni Ramos-Buades}
\Ni

\author{Lorenzo Pompili}
\AEI

\author{H\'ector Estell\'es}
\AEI

\author{Michael Boyle}
\Cornell

\author{Lawrence E. Kidder}
\Cornell

\author{Harald P. Pfeiffer}
\AEI

\author{Hannes R. R\"uter}
\AEI
\CENTRA

\author{Mark A. Scheel}
\Caltech

\begin{abstract}
The measurement of orbital eccentricity in gravitational-wave (GW) signals will provide unique insights into the astrophysical origin of binary systems, while ignoring eccentricity in waveform models could introduce significant biases in parameter estimation and tests of general relativity.
Upcoming LIGO-Virgo-KAGRA observing runs are expected to detect a subpopulation of eccentric signals, making it vital to develop accurate waveform models for eccentric orbits.
Here, employing recent analytical results through the third post-Newtonian order, we develop \texttt{SEOBNRv5EHM}: a new time-domain, effective-one-body, multipolar waveform model for eccentric binary black holes with spins aligned (or antialigned) with the orbital angular momentum.
Besides the dominant $(2,2)$ mode, the model includes the $(2,1)$, $(3,3)$, $(3,2)$, $(4,4)$, and $(4,3)$ modes.
We validate the model's accuracy by computing its unfaithfulness against 99 (28 public and 71 private) eccentric numerical-relativity (NR) simulations, produced by the Simulating eXtreme Spacetimes Collaboration.
Importantly, for NR waveforms with initial GW eccentricities below 0.5, the maximum $(2,2)$-mode unfaithfulness across the total mass range $20-200\, \solarmass$ is consistently below or close to $1\%$, with a median value of $ \sim 0.02 \% $, reflecting an accuracy improvement of approximately an order of magnitude compared to the previous-generation \texttt{SEOBNRv4EHM} and the state-of-the-art \texttt{TEOBResumS-Dal\'i} eccentric model.
In the quasi-circular-orbit limit, \texttt{SEOBNRv5EHM} is in excellent agreement with the highly accurate \texttt{SEOBNRv5HM} model.
The accuracy, robustness, and speed of \texttt{SEOBNRv5EHM} make it suitable for data analysis and astrophysical studies.
We demonstrate this by performing a set of recovery studies of synthetic NR-signal injections, and parameter-estimation analyses of the events GW150914 and GW190521, which we find to have no eccentricity signatures.
\end{abstract}

\date{\today}

\maketitle

%\tableofcontents

%%%%%%%%%%%%%%%%%%%%%%%%%%%
%%%%%%%%%%%%%%%%%%%%%%%%%%%
%%%%%%%%%%%%%%%%%%%%%%%%%%%

\section{Introduction}

Since the first detection of gravitational waves (GWs) passing through the Earth in September 2015 \cite{LIGOScientific:2016aoc}, about a hundred signals have been observed by the LIGO-Virgo-KAGRA (LVK) Collaboration \cite{LIGOScientific:2018mvr,LIGOScientific:2019lzm,LIGOScientific:2020ibl,LIGOScientific:2021usb,LIGOScientific:2021djp} and independent searches \cite{Venumadhav:2019lyq,Nitz:2021uxj,Nitz:2021zwj,Olsen:2022pin,Wadekar:2023gea,Mehta:2023zlk} using public data.
All these events have been associated with coalescences of binary compact objects, such as stellar-origin black holes (BHs) and neutron stars.

For current~\cite{KAGRA:2013rdx,LIGOInstrumentWhitePaper,VirgoInstrumentWhitePaper} and future~\cite{LIGOScientific:2016wof,Punturo:2010zz,TianQin:2015yph,Reitze:2019dyk,Reitze:2019iox,amaroseoane2017laser,LISA:2022kgy} GW detectors, the two main astrophysical formation channels for compact binaries are isolated-binary evolution and dynamic formation in star clusters~\cite{Bethe:1998bn,Belczynski:2001uc,Belczynski:2014iua,mennekens2014massive,Belczynski:2016obo,Eldridge:2016ymr,Marchant:2016wow,Stevenson:2017tfq,Giacobbo:2018etu,Kruckow:2018slo,Kruckow:2018slo,Mandel:2018hfr,Zevin:2020gbd,Mapelli:2020vfa,Karathanasis:2022rtr,Bouffanais:2021wcr,LIGOScientific:2021psn}.
The isolated-binary evolution channel predicts binaries with negligible eccentricity by the time they enter the detectors' frequency band;
this is because those binaries have enough time to circularize, as they lose energy and angular momentum due to the emission of GWs~\cite{Peters:1964zz,Hinder:2007qu}.
However, dense stellar environments, such as globular clusters or galactic nuclei, are expected to host systems whose gravitational signal has a detectable imprint of eccentricity \cite{heggie1975binary,sigurdsson1993primordial,Kulkarni:1993fr,Miller:2001ez,Gultekin:2004pm,Sadowski:2007dz,OLeary:2008myb,downing2010compact,downing2011compact,Antonini:2012ad,Tsang:2013mca,Rodriguez:2015oxa,Askar:2016jwt,Rodriguez:2016kxx,Rodriguez:2016avt,Stone:2016wzz,Stone:2016ryd,Petrovich:2017otm,Gondan:2017wzd,Takatsy:2018euo,Rodriguez:2018rmd,Fragione:2018vty,Rasskazov:2019gjw,Fragione:2019vgr,Chattopadhyay:2023pil,Sedda:2023qlx,1910AN....183..345V,Sigurdsson:1993tui,Sigurdsson:1994ju,kozai1962secular,lidov1962evolution,giacaglia1964notes,Wen:2002km,Kimpson:2016dgk,VanLandingham:2016ccd,Hoang:2017fvh}.
For example, this is the case for compact binaries that are dynamically formed close to merger \cite{PortegiesZwart:1999nm,Miller:2002pg,OLeary:2005vqo,Gultekin:2005fd,Samsing:2013kua,Antonini:2015zsa,Silsbee:2016djf,Antonini:2016gqe,Antonini:2017ash,Samsing:2017rat,Samsing:2017xmd,Rodriguez:2017pec,Zevin:2018kzq,Arca-Sedda:2018qgq,Fragione:2020nib,Gondan:2020svr,Vigna-Gomez:2020fvw,Tagawa:2020jnc,Britt:2021dtg}.
Therefore, the future detection of eccentric GWs will be key for the characterization of binary formation channels \cite{Sesana:2010qb,Breivik:2016ddj,Nishizawa:2016eza,Samsing:2018isx,Cardoso:2020iji,Zevin:2021rtf,Fumagalli:2023hde,Samsing:2024syt,Zeeshan:2024ovp}.

Several studies have been carried out to infer the presence of eccentricity in observed GW events.
Analyses by Refs.~\cite{Romero-Shaw:2020aaj,Romero-Shaw:2019itr,Romero-Shaw:2021ual, Romero-Shaw:2022xko} showed evidence for eccentricity in at least three GW signals of the first three observing runs of the LVK Collaboration \cite{LIGOScientific:2021usb,LIGOScientific:2021djp,LIGOScientific:2020ibl,LIGOScientific:2018mvr},
namely, the events GW190620, GW191109, and GW200208\_22.
Furthermore, Ref.~\cite{Gupte:2024jfe} analyzed 57 GW events, also from the first three LVK observing runs, and determined that the probability of one out of the 57 events being eccentric, aligned-spin binary BHs (BBHs) is greater than $\sim 99.5\%$, depending on the glitch mitigation method.
In particular, the events GW190701, GW200129, and GW200208\_22 support the hypothesis of being eccentric.
Additionally, some studies claimed that the event GW190521 also has evidence for eccentricity~\cite{Romero-Shaw:2020thy,Gayathri:2020coq}, or that it is the result of a hyperbolic capture of two nonspinning BHs~\cite{Gamba:2021gap}.
However, the analyses in Refs.~\cite{Gupte:2024jfe,Iglesias:2022xfc,Bonino:2022hkj,Ramos-Buades:2023yhy} have shown no evidence of eccentricity. 
This tension in the nature of GW190521 is probably caused by its short duration~\cite{LIGOScientific:2020iuh}, which complicates the identification of eccentricity, as it could be confused with other effects like spin precession \cite{CalderonBustillo:2020xms,Romero-Shaw:2022fbf}. 
Overall, accurate waveform models including both eccentricity and spin-precession are required to measure with high statistical significance the observation of an eccentric GW signal.

Quasicircular (QC) waveform models are commonly employed in GW analyses, such as matched-filtering LVK searches and parameter estimation studies.
However, QC waveform templates have been shown to be inefficient in detecting eccentric signals \cite{Brown:2009ng,Favata:2021vhw,Lenon:2021zac,DePorzio:2024cet}, such as when the eccentricity is larger than $
\sim 0.2$ at $ 10\, $Hz \cite{Gadre:2024ndy}.
This has motivated the incorporation of eccentricity into template banks~\cite{LIGOScientific:2019dag,LIGOScientific:2023lpe,Nitz:2019spj,Yun:2020aow,Lenon:2020oza,Wu:2020zwr,Wang:2021qsu,Nitz:2021mzz,Pal:2023dyg,Phukon:2024amh}, at least for a limited part of the binary parameter space, though no definitive eccentric signal has been detected so far.
Furthermore, even small eccentricities (e.g., $e \sim 0.1$ at $ 20 \, $Hz \cite{Divyajyoti:2023rht}) might lead to systematic biases in the extraction of source properties in inference studies \cite{Favata:2013rwa,Ramos-Buades:2019uvh,Cho:2022cdy,Guo:2022ehk,GilChoi:2022waq,Divyajyoti:2023rht,Gupte:2024jfe,Das:2024zib} and tests of general relativity \cite{Saini:2022igm,Saini:2023rto,Narayan:2023vhm,Gupta:2024gun,Shaikh:2024wyn,Bhat:2022amc,Bhat:2024hyb}.
Therefore, the effects of eccentricity in GWs cannot be ignored.

The inclusion of eccentricity in GW analyses requires the construction of accurate, robust, and fast eccentric waveform models.
Numerical-relativity (NR) simulations produce the most accurate waveforms with eccentricity effects \cite{Huerta:2019oxn,Ramos-Buades:2019uvh,Ramos-Buades:2022lgf,Bonino:2024xrv}.
However, producing these simulations is time consuming.
Furthermore, the parameter-space phenomenology of eccentric binaries is richer compared to the QC case since two eccentricity parameters need to be specified at a certain reference frequency \cite{Islam:2024rhm, Islam:2024bza, Clarke:2022fma, Carullo:2023kvj,Wang:2023wol,Knapp:2024yww,Nee:2025zdy}.
Therefore, it is crucial to complement NR results with semianalytical waveform models, which are 
built by solving Einstein's equations approximately. 

Over the past years, various eccentric waveform models have been constructed.
Based on post-Newtonian (PN) results \cite{Junker:1992kle,Gopakumar:1997bs,Gopakumar:2001dy,Memmesheimer:2004cv,Damour:2004bz,Konigsdorffer:2006zt,Arun:2007rg,Arun:2007sg,Arun:2009mc,Mishra:2015bqa,Boetzel:2019nfw,Ebersold:2019kdc,Henry:2023tka,Boetzel:2017zza, Paul:2022xfy}, References~\cite{Yunes:2009yz,Cornish:2010cd,ShapiroKey:2010cnz,Huerta:2014eca,Loutrel:2017fgu,Tanay:2016zog,Tanay:2019knc,Tiwari:2020hsu,Tiwari:2019jtz,Moore:2018kvz,Moore:2019xkm,Klein:2018ybm, Klein:2021jtd, Sridhar:2024zms} developed inspiral-only eccentric models, including spin-precession effects.
Complete inspiral-merger-ringdown (IMR) models were developed in Refs.~\cite{Huerta:2016rwp,Huerta:2017kez,Hinder:2017sxy,Islam:2024tcs,Ramos-Buades:2019uvh,Chattaraj:2022tay,Manna:2024ycx,Paul:2024ujx}, which constructed hybrid eccentric waveform models that employ PN results for the inspiral combined with information from QC NR simulations for the merger-ringdown phase.
Additionally, a NR surrogate model for eccentric, equal-mass, nonspinning binaries has been developed by interpolating existing NR simulations \cite{Islam:2021mha}.
Also, IMR eccentric models constructed upon QC models have been developed thanks to phenomenological relations observed in eccentric NR simulations \cite{Setyawati:2021gom,Wang:2023ueg,Islam:2024zqo, Islam:2024bza}.

A different approach to construct IMR eccentric waveform models is based on the effective-one-body (EOB) formalism~\cite{Buonanno:1998gg,Buonanno:2000ef,Damour:2000we,Buonanno:2005xu}.
In this approach, several analytical approximation methods (PN, post-Minkowskian, and gravitational self-force approximations) are combined with NR results to produce highly accurate waveforms.
The development of EOB models has focused mainly on QC binaries, resulting in two state-of-the-art families: \texttt{SEOBNR}~\cite{Bohe:2016gbl,Cotesta:2018fcv,Ossokine:2020kjp,Cotesta:2020qhw,Mihaylov:2021bpf,Pompiliv5,Khalilv5,RamosBuadesv5,VandeMeentv5, Mihaylovv5} and \texttt{TEOBResumS}~\cite{Nagar:2018zoe,Nagar:2019wds,Nagar:2020pcj,Gamba:2021ydi,Riemenschneider:2021ppj}.
However, the possibility of observing eccentric signals in the near future has motivated the construction of EOB models for eccentric orbits.
Historically, the inclusion of eccentricity effects in the EOB formalism started with Ref.~\cite{Bini:2012ji}, which derived 2PN analytical results for the dissipative EOB dynamics of eccentric binaries.
Then, Ref.~\cite{Hinderer:2017jcs} built the first EOB waveform model valid for eccentric, nonspinning binaries, accurate up to 1.5PN order.

The EOB eccentric, aligned-spin waveform models are now reaching a state of maturity 
for the two families \texttt{SEOBNR} and \texttt{TEOBResumS}.
In the latter, several prescriptions have been tested to improve the accuracy of eccentric waveforms \cite{Chiaramello:2020ehz,Albanesi:2021rby,Albanesi:2022ywx,Albanesi:2022xge,Albanesi:2023bgi,Nagar:2020xsk,Nagar:2021gss,Nagar:2021xnh,Placidi:2021rkh,Nagar:2023zxh,Placidi:2023ofj,Nagar:2024dzj,Nagar:2024oyk,Grilli:2024lfh}.
The resulting eccentric, aligned-spin, time-domain,  multipolar waveform model is known as \texttt{TEOBResumS-Dal\'i}.\footnote{
Note that this is a generic name that has been used for various improved eccentric versions over the last years.}
This model is valid for aligned-spins BBHs in generic orbits, and it follows a hybrid approach to incorporate analytical information about eccentric orbits.
In particular, it employs generic-orbit Newtonian prefactors for the EOB radiation-reaction (RR) force and gravitational waveform modes, as well as the 2PN results from Ref.~\cite{Bini:2012ji} for the radial component of the RR force.

In the case of the \texttt{SEOBNR} waveform models, Refs.~\cite{Cao:2017ndf,Liu:2019jpg,Liu:2021pkr,Liu:2023dgl} constructed the models \texttt{SEOBNRE} and \texttt{SEOBNREHM} by including eccentricity corrections to the waveform modes of the QC models \texttt{SEOBNRv1} \cite{Taracchini:2012ig} and \texttt{SEOBNRv4HM} \cite{Bohe:2016gbl,Cotesta:2018fcv}.
Independently, Ref.~\cite{Khalil:2021txt} derived 2PN analytical results for the EOB RR force and waveform modes valid for eccentric, aligned-spin BBHs.
Later on, Ref.~\cite{Ramos-Buades:2021adz} used these results for the waveform modes to construct the \texttt{SEOBNRv4EHM} model, which is also a time-domain, multipolar waveform model valid for aligned-spins BBHs in generic orbits.
More recently, building on the approach of Refs.~\cite{Chiaramello:2020ehz,Albanesi:2021rby,Albanesi:2022ywx,Albanesi:2022xge,Albanesi:2023bgi}, Ref.~\cite{Faggioli:2024ugn} explored the impact of different choices for the RR force in the test-mass limit.

Here, we present \texttt{SEOBNRv5EHM},\footnote{
\texttt{SEOBNRv5EHM} is publicly available through the Python package \texttt{pySEOBNR} \cite{Mihaylovv5}.
A tutorial notebook is available here: \url{https://waveforms.docs.ligo.org/software/pyseobnr}.
%{(https://git.ligo.org/waveforms/software/pyseobnr)}.
Stable versions of \texttt{pySEOBNR} are published through the Python Package Index (PyPI), and can be installed via ~\texttt{pip install pyseobnr.}}
a new waveform model for eccentric, aligned-spin BBHs that, for the first time, includes 3PN analytical information in the eccentricity corrections to the EOB RR force and waveform modes.
The foundations of the model are presented in the companion paper~\cite{Gamboa:2024imd}.
In \texttt{SEOBNRv5EHM},  the waveform modes $( \ell, |m| ) = \{(2, 2),\, (3, 3),\, (2, 1),\, (4, 4),\, (3, 2),\, (4, 3)\}$ are built in the time domain, and are applicable to bound orbits.
They are constructed upon the state-of-the-art QC, aligned-spin model \texttt{SEOBNRv5HM} \cite{Pompiliv5} and employ a QC prescription for the merger-ringdown phase, as is the case for all the existing IMR aligned-spin, eccentric waveform models.
In addition, \texttt{SEOBNRv5EHM} is the first IMR eccentric waveform model to complete a review process within the LVK Collaboration.

As discussed below, the \texttt{SEOBNRv5EHM} model can achieve unprecedented accuracy against NR waveforms.
In the zero eccentricity limit, \texttt{SEOBNRv5EHM} is in excellent agreement with the QC model \texttt{SEOBNRv5HM}.
For eccentric waveforms, \texttt{SEOBNRv5EHM} has an overall improvement of about one order of magnitude with respect to the previous-generation \texttt{SEOBNRv4EHM} model developed in 2021, and also with respect to the state-of-the-art \texttt{TEOBResumS-Dal\'i} model.\footnote{
We employ the \texttt{TEOBResumS-Dal\'i} waveform model developed in Ref.~\cite{Nagar:2024dzj}. 
More specifically, we have used the code on the \texttt{Dali} branch from the public bitbucket repository
\url{https://bitbucket.org/teobresums/teobresums/src/Dali},
with git hash ab5f8b84f23723b0daa7d122c49940a83d471e2f.
Such a version differs from the original one in Ref.~\cite{Nagar:2024dzj} due to an improved treatment of the initial conditions, which was implemented during the review of the model in the LVK Collaboration, and the absence of the $(3,2)$ and $(4,3)$ modes, which are not robust in the parameter space under investigation here.
\label{fn:dali}
}
We remark that, even if \texttt{SEOBNRv5EHM} is mostly targeting medium eccentricities, say up to 0.5 at $20$ Hz, it retains an overall better accuracy with respect to other models even for eccentricities higher than 0.5, thanks to the 3PN eccentricity corrections to the EOB modes and RR force.

Computational speed and robustness are also key characteristics of \texttt{SEOBNRv5EHM}.
Typical applications of waveform models require millions of waveform evaluations over large regions of parameter space.
In this work, we demonstrate the applicability of \texttt{SEOBNRv5EHM} by performing several inference studies with the Bayesian inference Python package \texttt{Bilby} \cite{Ashton:2018jfp,Romero-Shaw:2020owr} and its highly parallelizable version \texttt{parallel} \texttt{Bilby} \cite{Smith:2019ucc}, including injections of synthetic NR signals and analyses of real GW events.

This paper is organized as follows.
In Sec.~\ref{sec:model_dynamics}, we summarize the EOB BBH eccentric, aligned-spin dynamics employed in \texttt{SEOBNRv5EHM}.
Then, in Sec.~\ref{sec:model_waveform}, we show how the full IMR eccentric waveforms are constructed within \texttt{SEOBNRv5EHM}.
Afterward, in Sec.~\ref{sec:model_validation}, we validate the model by showing comparisons against QC and eccentric NR simulations.
Additionally, we provide information about the robustness and computational speed of  \texttt{SEOBNRv5EHM} across the parameter space of interest for current GW detectors.
In Sec.~\ref{sec:model_applications}, we demonstrate the applicability of \texttt{SEOBNRv5EHM} by performing NR injection studies and analyses of real GW events.
We conclude in Sec.~\ref{sec:conclusions} and give suggestions for future studies.
Finally, in Appendix~\ref{sec:public_ecc_NR_waveforms}, we show comparisons against a set of publicly-available eccentric NR waveforms (for ease of comparison with previous studies in the literature), in Appendix~\ref{sec:mms_2PN} we show the accuracy improvement due to the 3PN eccentricity corrections to the EOB RR force and modes in \texttt{SEOBNRv5EHM}, and in Appendix~\ref{sec:tables} we present a summary of the eccentric NR waveforms employed in this work.

%%%%%%%%%%%%%%%%%%%%%%%%%%%
%%%%%%%%%%%%%%%%%%%%%%%%%%%

\section*{Notation}

Throughout this work, we use geometrical units in which Newton's constant and the speed of light are equal to one ($G = c =1$).

We denote the individual component masses as $m_1$ and $m_2$ and the mass ratio as $q \equiv m_1/m_2 \geq 1$. 
The total mass, reduced mass, and symmetric mass ratio are defined through the relations $M \equiv m_1 + m_2$, $\mu \equiv m_1 m_2 / M$, and $\nu \equiv \mu / M $, respectively.

We restrict our analysis to individual component spins $\bm{S}_1$ and $\bm{S}_2$ that are 
aligned (or antialigned) with the binary's orbital-angular-momentum unit vector $ \bm{\hat{L}} $, and hence perpendicular to the orbital plane.
(Henceforth, for short, we refer to them as aligned-spin binaries.)
These systems are characterized by their dimensionless spin components given by
\begin{equation}
 \label{eq:comp_spins}
\chi_i \equiv \frac{\bm{S}_i \cdot \bm{\hat{L}}}{m_i^2} = \pm \frac{|\bm{S}_i|}{m_i^2} \s,  \qquad i \in \{1,2\}, 
\end{equation}
which take values in the interval $(-1,1)$, with positive spins being in the direction of the orbital angular momentum.

%Additionally, we employ the dimensionless effective spin parameter defined by
%%
%\begin{equation}
%    \label{eq:chi_eff}
%    \chi_{\text{eff}} \equiv \frac{m_1 \chi_1 + m_2 \chi_2}{M}.
%\end{equation}
%%

%%%%%%%%%%%%%%%%%%%%%%%%%%%
%%%%%%%%%%%%%%%%%%%%%%%%%%%
%%%%%%%%%%%%%%%%%%%%%%%%%%%

%\section{Binary black hole dynamics in {\texorpdfstring{\normalsize \texttt{SEOBNR\lowercase{v}5EHM}}{\texttt{SEOBNR\lowercase{v}5EHM}}}}
\section{Binary black hole dynamics in SEOBNR\lowercase{v}5EHM}
\label{sec:model_dynamics}

In the EOB formalism \cite{Buonanno:1998gg,Buonanno:2000ef,Damour:2000we,Buonanno:2005xu}, the dynamics of two gravitationally interacting bodies is mapped onto the dynamics of an effective test body in a deformed Kerr background, with the deformation parametrized by the symmetric mass ratio. The one-body dynamics is characterized by an effective Hamiltonian $H_\text{eff}$ that is related to the two-body Hamiltonian $H_\text{EOB}$ by the energy map,
\begin{equation}
\label{eq:EOBHam}
H_\text{EOB} = M \sqrt{1 + 2 \nu \left( \frac{H_\text{eff}}{\mu} - 1 \right)}.
\end{equation}

The Hamiltonian $ H_\text{eff} $ employed in the QC waveform model \texttt{SEOBNRv5HM}~\cite{Pompiliv5} reduces exactly to the aligned-spin Hamiltonian of a test mass in a Kerr background when $\nu \to 0$; it resums the full 4PN analytical information, and includes two parameters at higher PN orders that were calibrated to QC NR simulations (see Ref.~\cite{Khalilv5} for the explicit expression of the Hamiltonian).
Since the \texttt{SEOBNRv5HM} Hamiltonian is valid for generic orbits and aligned spins, the extension of \texttt{SEOBNRv5HM} to an eccentric, aligned-spin waveform model only requires the modification of the RR force acting on the binary and the generalization of the waveform modes.
We achieve this with an appropriate parametrization of the orbit, as we see next.

In general, to completely define an elliptic orbit, two parameters must be specified at a certain point of the binary evolution.
A common choice for these two parameters is the \emph{orbital eccentricity} and \emph{semilatus rectum},\footnote{
The same parameters can be used for hyperbolic orbits, but a more common choice is the initial values of energy and angular momentum of the binary at a large separation (see, e.g., Refs.~\cite{Nagar:2020xsk,Ramos-Buades:2021adz}).} with a \emph{radial phase} angle (or \emph{radial anomaly}) to track the trajectory along the orbit.
Nevertheless, eccentricity is not uniquely defined in general relativity and some definitions are gauge dependent.
Different choices exist for the eccentricity and radial anomaly parameters, and there are efforts to relate the different conventions~\cite{Shaikh:2023ypz,Vijaykumar:2024piy, Boschini:2024scu,Bonino:2024xrv}.

In \texttt{SEOBNRv5EHM}, the elliptic orbits are defined using a \textit{Keplerian parametrization} of the orbit \cite{darwin1959gravity}, in which the radial motion is parametrized as
\begin{equation}
\label{eq:Kep_param}
r = \frac{M}{u_p (1 + e \cos \zeta)},
\end{equation}
where $ r $ is the relative separation of the binary (in EOB coordinates), $u_p$ is the \emph{inverse semilatus rectum} (scaled by the total mass $ M $), $ e $ is the \emph{Keplerian eccentricity}, and $\zeta$ is the \emph{relativistic anomaly}.
The initial frequency at which these parameters are specified in \texttt{SEOBNRv5EHM} corresponds to the dimensionless \emph{orbit-averaged orbital frequency} $ \langle M \Omega \rangle $,\footnote{
In \texttt{SEOBNRv5EHM}, the dependence on the inverse semilatus rectum $ u_p $ is replaced by a dependence on the dimensionless orbit-averaged orbital frequency $ \langle M \Omega \rangle $ via a PN transformation (see Eq.~(B7) in Ref.~\cite{Gamboa:2024imd}).} where $\Omega \equiv \dot{\phi}$ is the instantaneous orbital frequency.

With this parametrization of the orbit, the eccentricity $ e $ measures the deformation of the orbit with respect to a circular configuration, and $ \zeta $ describes the relative position of the binary in the orbit at a certain value of $ \langle M \Omega \rangle $. In particular, $ \zeta = 0 $ corresponds to the closest approach configuration (known as \emph{periastron}), and $ \zeta = \pi $ corresponds to the farthest configuration (known as \emph{apastron}).
With the choice of $ \langle M \Omega \rangle $ as the starting frequency, the behavior of the system is such that the increase (decrease) of eccentricity at a fixed frequency corresponds to a decrease (increase) in the duration of the inspiral (and of the corresponding waveform)~\cite{Ramos-Buades:2023yhy}.

Thus, in \texttt{SEOBNRv5EHM} the \emph{dynamics} of eccentric, aligned-spin BBH systems is determined by 7 input parameters:
the BH masses $ m_1 $ and $ m_2 $ (or, equivalently, mass ratio $ q $ and total mass $ M $), the dimensionless spin components $ \chi_1 $ and $ \chi_2 $, and the eccentricity $ e $ and relativistic anomaly $ \zeta $ specified at the dimensionless orbit-averaged orbital frequency $ \langle M \Omega \rangle $.
This information is summarized in Table \ref{tab:notation}.

In the following subsections, we specify how the BBH dynamics is computed within the \texttt{SEOBNRv5EHM} model.
In particular, we present the equations of motion (EOM) valid for eccentric, aligned-spin BBHs, and we outline the prescription for the initial conditions.

\setlength{\extrarowheight}{8pt}
\begin{table}[h!]
    \centering
    \begin{tabular}{ c  c  }
 \hline
 \hline
   \vspace{0.1cm}
 Parameter & Description \\
 \hline
 \centering
  \vspace{0.1cm}
 $m_{1,\s 2}$ \, $ (m_1 \geq m_2) $ &  mass components \\
 $\chi_{1,\s 2}$ \, $ (|\chi_{1,\s2}| < 1 )$ &  \makecell[cc]{dimensionless spin\\ components }  \\
 $e$ & Keplerian eccentricity       \\
  \vspace{0.1cm}
 $\zeta$ &  relativistic anomaly  \\
  \vspace{0.1cm}
 $ \langle M \Omega \rangle $ &  \makecell[cc]{dimensionless orbit-averaged\\ orbital frequency}  \\
 %\hline
 \hline
 \hline
    \end{tabular}
 \caption{
Input parameters of the \texttt{SEOBNRv5EHM} model that characterize the BBH eccentric, aligned-spin dynamics.
}
%\vspace{-15pt}
\label{tab:notation}
\end{table}

%%%%%%%%%%%%%%%%%%%%%%%%%%%
%%%%%%%%%%%%%%%%%%%%%%%%%%%

\subsection{Eccentric, aligned-spin effective-one-body dynamics}

In \texttt{SEOBNRv5EHM}, the binary dynamics is computed by numerically solving Hamilton's EOM supplemented with a RR force (with radial and azimuthal components $ \mathcal F_r $ and $ \mathcal F_\phi $, respectively) and coupled to three additional equations for the Keplerian parameters $ e $, $ \zeta $, and $ x \equiv \langle M \Omega \rangle ^{2/3} $,
\begin{subequations}
\label{eq:EOMprStr}
\begin{align}
\dot{r} &= \xi(r) \frac{\partial H_\text{EOB}}{\partial p_{r_*}}(r,p_{r_*},p_\phi),
\label{eq:dot_r}
\\
\dot{\phi} &= \frac{\partial H_\text{EOB}}{\partial p_\phi}(r,p_{r_*},p_\phi), \\
\dot{p}_{r_*} &= - \xi(r) \frac{\partial H_\text{EOB}}{\partial r}(r,p_{r_*},p_\phi) + \xi(r) \mF_r, 
\label{eq:dot_prstar}
\\
\dot{p}_\phi &= \mF_\phi,
\\
\dot{e} &= -\frac{\nu e x^4}{M} \left[\frac{\left(121 e^2+304\right)}{15 \left(1-e^2\right)^{5/2}} + \text{3PN expansion}\right], \label{eq:e}\\
\dot{\zeta} &= \frac{x^{3/2}}{M}\left[\frac{ (1+e \cos\zeta)^2}{\left(1-e^2\right)^{3/2}} + \text{3PN expansion}\right], \label{eq:zeta}\\
x &= (M \Omega)^{2/3} \left[\frac{\left(1-e^2\right) }{(1+e \cos\zeta)^{4/3}} + \text{3PN expansion}\right], \label{eq:xOmega}
\end{align}
\end{subequations}
where the dot denotes differentiation with respect to time $ t $, the variables $ (r, \s \phi, \s p_{r_*}\!, \s p_\phi) $ are phase-space polar coordinates defined in the binary's center of mass, and we employ the tortoise radial momentum $p_{r_*}$ since it improves the stability of the EOM near the event horizon~\cite{Damour:2007xr,Pan:2009wj}.
This momentum is conjugate to the tortoise coordinate $r_*$ defined by $\di r / \di r_* \equiv \xi(r)$, and it is related to the canonical radial momentum $ p_r $ by the relation $ p_{r_*} \!=  p_r \s \xi(r) $, where $\xi(r)$ is given by Eq.~(44) of Ref.~\cite{Khalilv5}.
The equations for $ e $, $ \zeta $, and $ x $, contain the \emph{full} 3PN information for the eccentric dynamics.
We refer the reader to the companion paper \cite{Gamboa:2024imd} for the derivation and details about these equations.

The RR force accounts for the energy and angular-momentum losses by the system due to the emission of GWs.
In \texttt{SEOBNRv5EHM}, we write the components of the RR force as a sum over waveform modes~\cite{Buonanno:2000ef,Buonanno:2005xu} multiplied by suitable eccentricity corrections, such that
\begin{subequations}
\label{eq:eccRRforce}
\begin{align}
\mF_\phi &= \mF_\phi^\text{modes}\, \mF_\phi^\text{ecc}(x,e,\zeta), \\
\mF_r &= \frac{p_r}{p_\phi} \ \mF_\phi^\text{modes} \,\mF_r^\text{ecc}(x,e,\zeta),\\
\mF_\phi^\text{modes} &= -\frac{M^2 \Omega}{8 \pi} \sum_{\ell=2}^8 \sum_{m=1}^{\ell} m^2\left|d_L \s h_{\ell m}^{\text{F}}\right|^2 \!,
\end{align}
\end{subequations}
where $ \mF_\phi^\text{ecc} $ and $ \mF_r^\text{ecc} $ correspond to the 3PN nonspinning eccentricity corrections to the RR force \cite{Gamboa:2024imd}, $d_L$ is the luminosity distance between the binary and the observer, and $h_{\ell m}^\text{F}$ are the factorized \emph{eccentric} waveform modes, given by
\begin{equation}
\label{eq:eob_modes}
h_{\ell m}^\text{F}=h_{\ell m}^\text{F,\,qc} (x) \,h_{\ell m}^\text{ecc}(x,e,\zeta).
\end{equation}
Here, $ h_{\ell m}^\text{F,\s\s qc} $ represents the factorization of QC modes employed in \texttt{SEOBNRv5HM} \cite{Pompiliv5}, and $ h_{\ell m}^\text{ecc} $ are the nonspinning eccentricity corrections accurate up to the 3PN order \cite{Gamboa:2024imd} [see Sec.~\ref{sec:eob_fact_modes} for a description of the factors entering Eq.~\eqref{eq:eob_modes}]. 
The complete 3PN expressions for the eccentricity corrections to the RR force and modes ($ \mF_\phi^\text{ecc} $, $ \mF_r^\text{ecc} $, and $ h_{\ell m}^\text{ecc} $) are provided in the companion paper~\cite{Gamboa:2024imd}.

In the QC limit ($e \to 0$), the eccentricity corrections to the RR force and modes reduce to 1.
Furthermore, from Eqs.~\eqref{eq:e} and \eqref{eq:xOmega} we get
\begin{equation}
\left.\dot{e}\right|_{e\to0} = 0,  \qquad
\left.x\right|_{e\to0} = (M \Omega)^{2/3},
\end{equation}
and no dependence in the EOM on $\zeta$.
This way, we recover the evolution equations of \texttt{SEOBNRv5HM}, and inherit the accurate 
calibration of this model to NR QC simulations~\cite{Pompiliv5}.

%%%%%%%%%%%%%%%%%%%%%%%%%%%
%%%%%%%%%%%%%%%%%%%%%%%%%%%

\subsection{Initial conditions}
\label{sec:ICs}

To initialize an eccentric, aligned-spin BBH system, we require a prescription to compute the initial values $ (r_0,\, \phi_0,\, p_{r_* 0},\, p_{\phi 0}) $ that physically correspond to the given input parameters of the model $ (m_{1,\s2},\, \chi_{1,\s2},\, e,\, \zeta,\, \langle M \Omega \rangle) $.
By convention, we set $ \phi_0 = 0 $.
The other initial values are computed using the postadiabatic (PA) approximation~\cite{Buonanno:2000ef,Damour:2002vi,Buonanno:2005xu,Damour:2012ky}, where we assume that the system evolves through a sequence of conservative orbits, which slowly inspiral inwards due to the emission of GWs.
The details are given in Ref.~\cite{Gamboa:2024imd}, and here we just outline the main equations employed in such a prescription.
We note that our method to determine the initial conditions is a generalization of the procedure employed in \texttt{SEOBNRv4EHM} \cite{Khalil:2021txt, Ramos-Buades:2021adz, Ramos-Buades:2023yhy}.

In the PA approximation, the initial values of the dynamical variables are split into conservative 
and dissipative parts
\begin{subequations}
\label{eq:PA_approx}
\begin{align}
r_0
&= r_{ 0, \s \text{cons}} + r_{ 0, \s \text{diss}} ,
\label{eq:} \\
p _{\phi 0}
&= p_{ \phi 0, \s \text{cons}} + p_{\phi 0, \s \text{diss}} ,
\label{eq:} \\
p _{r_* 0}
&= p_{ r_* 0, \s \text{cons}} + p_{r_* 0, \s \text{diss}} ,
\label{eq:}
\end{align}
\end{subequations}
where $ \left( r_{ 0, \s \text{cons}} ,  \, p_{ \phi 0, \s \text{cons}}, \, p_{ r_* 0, \s \text{cons}} \right) $ satisfy the conservative EOM, and $ \left( r_{ 0, \s \text{diss}} ,  \, p_{ \phi 0, \s \text{diss}}, \, p_{ r_* 0, \s \text{diss}} \right) $ are the first-order PA contributions to the dynamics, which start at 2.5PN order.

The values of $ r_{0, \s \text{cons}} $ and $  p_{\phi 0, \s \text{cons}} $ are determined by numerically finding the roots of the equations
\begin{subequations}
\label{eq:root_cons}
\begin{align}
\Omega(\langle\Omega\rangle, e, \zeta)  &= 
  \frac{\partial  H_ \text{EOB}[r, p_{r_* } (\langle\Omega\rangle, e, \zeta) , p_\phi] }{\partial p_\phi}  ,
\label{eq:root_cons_b}
\\
\dot p_{r_*} (\langle\Omega\rangle, e, \zeta)  
&= 
- \xi (r) \, \frac{\partial  H_ \text{EOB}[r, p_{r_* } (\langle\Omega\rangle, e, \zeta) , p_\phi] }{\partial r} ,
\label{eq:root_cons_a} 
\end{align}
\end{subequations}
where $ \Omega(\langle\Omega\rangle, e, \zeta) $, $ p_{r_* } (\langle\Omega\rangle, e, \zeta) $, and $ \dot p_{r_*} (\langle\Omega\rangle, e, \zeta) $ are 3PN formulas without dissipative contributions, derived in 
Ref.~\cite{Gamboa:2024imd}. 

Subsequently, the values of $ r_{0, \s \text{diss}} $ and $ p_{\phi 0, \s \text{diss}} $ are directly determined with PN formulas given in Ref.~\cite{Gamboa:2024imd}, while the initial value $ p_{r_* 0} $ (including its conservative and dissipative parts) is obtained by numerically finding the root of
\begin{equation} 
\label{eq:dissipative_equation}
\left(  \dot r _{ \text{cons}} + \dot r _{ \text{diss}} \right)|_0 
= \xi(r) \, \frac{\partial H _{ \text{EOB}}(r_0, p_{r_*} , p_{\phi 0}) }{\partial p_{r_*}},
\end{equation}
where $  \dot r _{ \text{cons}} $ and $ \dot r_{ \text{diss}} $ are formulas given in Ref.~\cite{Gamboa:2024imd} that combine PN expansions with $ H _{ \text{EOB}} $ and the RR force.

This new prescription is convenient since it is constructed in such a way as to recover the QC initial conditions of \texttt{SEOBNRv5HM} \cite{Pompiliv5, Buonanno:2005xu} when $ e = 0 $.
Another advantage of this new prescription is the possibility of extending the parameter space of the model.
For some eccentric systems, certain combinations of the input parameters $ (\langle M \Omega \rangle_0, e_0, \zeta_0) $ are associated with a very small binary separation;
that is the case for high eccentricity and/or high frequency systems initialized close to periastron.
This is a problematic situation as many approximations become less accurate due to the high velocities involved.
To avoid this issue, in \texttt{SEOBNRv5EHM} we implement a set of three \emph{secular evolution equations} for $ (x = \langle M \Omega \rangle_0^{2/3}\!, e_0, \zeta_0) $ that are evolved backward in time whenever the input parameters are such that the starting separation of the binary satisfies $ r_0 < 10 \, M $.
More specifically, we evolve backward Eqs.~\eqref{eq:e}, \eqref{eq:zeta}, and
\begin{equation}
\label{eq:dotx}
\dot{x} =  \frac{2  \nu x^5}{3 M} \left[\frac{96 +292 e^2 + 37 e^4}{5 \left(1-e^2\right)^{7/2}} + \text{3PN expansion} \right],
\end{equation}
until the starting separation of the binary satisfies $ r_0 = 10 \, M $.
The full 3PN expressions of these evolution equations, as well as the PN equation employed to estimate the starting separation, are provided in the companion paper~\cite{Gamboa:2024imd}.
This way, we get new initial values for $ (\langle M\Omega \rangle, e, \zeta) $, which are then employed with the PA prescription for EOB initial conditions given above.
Thus, we obtain a system whose evolution eventually passes through the original input values $ (\langle  M\Omega \rangle_0, e_0, \zeta_0) $ to a good approximation depending on how high the values for eccentricity and frequency are.
For completeness, in Appendix \ref{sec:secular} we assess the impact of using or not using this set of secular evolution equations in determining the initial conditions for eccentric binaries.

We leave for future work the assessment of different prescriptions for the initial conditions of eccentric binaries.
Here, we focus on the parametrization in terms of $ (\langle M \Omega \rangle, e, \zeta) $ as this 
reduces directly to the one employed in QC models, and facilitates the integration of the model into the codes employed by the GW community.

%%%%%%%%%%%%%%%%%%%%%%%%%%%
%%%%%%%%%%%%%%%%%%%%%%%%%%%
%%%%%%%%%%%%%%%%%%%%%%%%%%%

%\section{The {\texorpdfstring{\normalsize \texttt{SEOBNR\lowercase{v}5EHM}}{\texttt{SEOBNR\lowercase{v}5EHM}}} waveform model}

\section{The SEOBNR\lowercase{v}5EHM waveform model}
\label{sec:model_waveform}

Gravitational waves have two degrees of freedom encoded in the \emph{polarizations} $ h_+ $ and $ h_\times $, which can be decomposed as
\begin{equation}
h_+-i \s h_\times =  \sum^{\infty}_{\ell =2} \sum_{m=-\ell }^{\ell }{} \!_{-2} Y_{\ell,\s m}(\iota,\s \varphi)\, h_{\ell m}(\bm{\Pi}; \s t),
\label{eq:gw_polarizations_decomp}
\end{equation}
where $ _{-2}Y_{\ell,\s m}$ are the $-2$ spin-weighted spherical harmonics that depend on the inclination $ \iota $ and azimuthal $ \varphi $ angles of the line of sight measured in the source frame, and $h_{\ell m}(\bm{\Pi};\s t)$ are the \emph{gravitational-waveform modes}, which depend on the binary's intrinsic parameters $ \bm{\Pi} $.
Therefore, given a prescription for the modes $ h_{\ell m} $ and the intrinsic parameters of the source, we can completely characterize the associated waveform.

In \texttt{SEOBNRv5EHM}, we model the full IMR eccentric, aligned-spin waveform modes for $( \ell, |m| ) = \{(2, 2),\, (3, 3),\, (2, 1),\, (4, 4),\, (3, 2),\, (4, 3)\}$.
As in previous EOB waveform models, we split them into an inspiral-plunge and a merger-ringdown part, according to
\begin{equation}
\label{eq:h_match}
h_{\ell m}(\bm{\Pi};t)= \begin{cases}h_{\ell m}^{\text {insp-plunge }}(\bm{\Pi};t), & t<t_{\text {match }} \\ h_{\ell m}^{\text {merger-RD }}(\bm{\Pi};t), & t>t_{\text {match }} \end{cases},
\end{equation}
where the matching time $t_{\text {match}}$ is related to the peak of the $(2,2)$ mode for QC binaries (see Sec.~\ref{sec:attachment}).\footnote{
Note that we do not employ the notation $ t _{ \text{match}} ^{\ell m} $ for the matching time as in Ref.~\cite{Pompiliv5} since we do not include the $(5,5)$ mode in \texttt{SEOBNRv5EHM}, and this is the only mode whose matching time is calculated differently.
%Hence, there is no need to write a dependence on $\ell$ or $m$, as the $(5,5)$ mode is the only one whose attachment time is calculated in a slightly different way. 
\label{fn:t_attach}
}

Furthermore, we follow the convention (see Appendix C of Ref.~\cite{Albanesi:2024xus}) in which the merger time ($ t = 0 $) is defined as the time of the \emph{last peak} of the frame-invariant amplitude $ A _{ \text{inv}} $, given a threshold value of $ 10\% $ with respect to the largest peak of $ A_{ \text{inv}} $, where 
\begin{equation}
%\label{eq:}
A_{ \text{inv}} = \sqrt{\sum_{\ell, \,|m| \leq \ell} |h_{\ell m}|^2} \, .
\end{equation}
This choice reduces to the one employed in \texttt{SEOBNRv5HM} \cite{Pompiliv5}, while including the possibility that the last peak of $ A_{ \text{inv}}  $ is not the largest one, which can occur for some high-eccentricity binaries, as noted in Refs.~\cite{Gold:2012tk,Carullo:2023kvj,Albanesi:2024xus}.
The threshold value of $ 10\%$ is chosen to avoid selecting amplitude peaks caused by mode-mixing during the ringdown phase.
Overall, the waveform behavior during the coalescence of highly eccentric binaries requires further investigation once more NR simulations are available.

In the following subsections, we specify how the inspiral-plunge part of the modes $ h_{\ell m}^{\text {insp-plunge}} $ is computed using the BBH eccentric dynamics and improved with information from QC NR simulations.
Then, we present the phenomenological ansatz that is employed for the merger-ringdown part of the signal $ h_{\ell m}^{\text {merger-RD}} \!$, which assumes that the binary has sufficiently circularized by the end of the inspiral.

%%%%%%%%%%%%%%%%%%%%%%%%%%%
%%%%%%%%%%%%%%%%%%%%%%%%%%%

\subsection{Inspiral-plunge modes}
\label{sec:modes_inspiral}

Following previous EOB models, the inspiral-plunge part of the modes is computed as
\begin{equation}
%\label{eq:}
h_{\ell m}^{\text {insp-plunge}}  =  N_{\ell m} \, h_{\ell m}^{\text {F}},
\end{equation}
where $ N_{\ell m} $ are the so-called nonquasicircular (NQC) corrections that guarantee that the modes’ amplitude and frequency
agree with values coming from QC NR simulations (known as NR input values), and $  h_{\ell m}^{\text {F}} $ are the \emph{factorized} EOB waveform modes with eccentricity contributions.
We now discuss how these elements are constructed.

%%%%%%%%%%%%%%%%%%%%%%%%%%%

\subsubsection{EOB factorized eccentric modes}
\label{sec:eob_fact_modes}

In \texttt{SEOBNRv5EHM}, we employ the following factorization of the waveform modes:
\begin{equation}
\label{eq:fact_QCecc} 
h_{\ell m}^\text{F}  = h_{\ell m}^\text{F, qc}(x) \, h_{\ell m}^\text{ecc} ( x, e, \zeta).
\end{equation}
The first term $ h_{\ell m}^\text{F,\,qc} $ represents the factorization employed in the QC model \texttt{SEOBNRv5HM} \cite{Pompiliv5},
\begin{equation}
\label{eq:qcFactModes}
h_{\ell m}^\text{F, qc} = h_{\ell m}^\text{N, qc}  \, \hat{S}_\text{eff} \, T_{\ell m}^\text{qc} \, f_{\ell m}^\text{qc} \, \e^{i \delta_{\ell m}^\text{qc}},
\end{equation}
where $h_{\ell m}^\text{N,\,qc}$ is the leading PN order for QC orbits, which is given for every $ \left( \ell, m \right) $ by~\cite{Damour:2008gu,Pan:2010hz}
\begin{equation}
\label{eq:hnewt}
h_{\ell m}^\text{N, qc}=\frac{\nu M}{d_L} \, n_{\ell m} \, c_{\ell+\epsilon_{\ell m}}(\nu) \, v_\Omega^{\ell+\epsilon_{\ell m}} \,Y_{\ell-\epsilon_{\ell m},-m}\!\left(\frac{\pi}{2}, \s \phi\right),
\end{equation}
where $Y_{\ell, \s m}$ is the scalar spherical harmonic, $\epsilon_{\ell m}$ is the parity of the mode (such that $\epsilon_{\ell m} = 0$ if $\ell+m$ is even, and $\epsilon_{\ell m} = 1$ if $\ell+m$ is odd), the factors $n_{\ell m}$ and $c_{k}(\nu)$ are given by Eqs.~(5)--(7) of Ref.~\cite{Damour:2008gu}, while $ v_\Omega $ depends on the orbit-averaged angular velocity, and it is given by\footnote{
In \texttt{SEOBNRv5HM}, Eq.~\eqref{eq:hnewt} is written in terms of $v_\phi \equiv M \Omega \,r_\Omega$, with $r_\Omega \equiv \nolinebreak \left(M \s \partial H_{\rm EOB}/\partial p_\phi \right)^{-2/3}  |_{p_r=0}$, instead of $ v_\Omega $.
Historically, the purpose of $v_\phi$ was to improve the modes' accuracy when there was no calibration to NR simulations.
However, the generalization of $v_\phi$ to eccentric orbits becomes complicated due to the definition of $ r_\Omega $.
Hence, in \texttt{SEOBNRv5EHM} we directly employ the factor $ v_\Omega $ instead of $ v_\phi $.
The impact of this choice is assessed in Sec.~\ref{sec:comparison_nr_qc} (see Fig.~\ref{fig:qc_mismatch_spaghetti_22}), where we show that it has a minimal effect on the agreement with QC NR waveforms.
\label{fn:vphi}
}
\begin{equation}
v_\Omega \equiv \langle M \Omega \rangle^{1/3} .
\end{equation}

The second factor in Eq.~\eqref{eq:qcFactModes} is the dimensionless effective source term $\hat{S}_\text{eff}$, which is given by
\begin{equation} \label{eq:Seff}
\hat{S}_\text{eff}= \left\{
        \begin{array}{ll}
            H_\text{eff} / \mu, & \quad \ell + m \text{ even} \\
          \langle M \Omega \rangle^{1/3} \, p_\phi/(M \mu), & \quad \ell + m \text{ odd}
        \end{array}
    \right. .
\end{equation}
The remaining factors $ T_{\ell m}^\text{qc} $, $ f_{\ell m}^\text{qc} $ and $\delta_{\ell m}^\text{qc} $ are specified in Ref.~\cite{Pompiliv5}, and they depend directly on the orbit-averaged angular frequency $ \langle M \Omega \rangle $ and on the EOB Hamiltonian $ H _{ \text{EOB}} $.

The second term in Eq.~\eqref{eq:fact_QCecc}, $ h_{\ell m}^\text{ecc} $, contains nonspinning eccentricity corrections up to the 3PN order and depends on the Keplerian parameters $ ( x = \langle M \Omega \rangle^{2/3}\!,\, e, \,\zeta) $.
It is constructed such that the PN expansion of Eq.~\eqref{eq:fact_QCecc} gives the original PN-expanded EOB modes and satisfies $h_{\ell m}^\text{ecc}  \to 1$ in the $ e \to 0 $ limit.
The complete expressions of these eccentricity corrections are given in the companion paper  \cite{Gamboa:2024imd}.

%%%%%%%%%%%%%%%%%%%%%%%%%%%

\subsubsection{Nonquasicircular corrections}
\label{sec:nqcs}

In the QC model \texttt{SEOBNRv5HM}, the functional form of the NQC corrections is given by
\begin{equation}
\label{eq:nqcs}
	\begin{aligned}
	N_{\ell m} &=\left[1+\frac{p_{r_*}^2}{(r \Omega)^2}\left(a_1^{h_{\ell m}}+\frac{a_2^{h_{\ell m}}}{r}+\frac{a_3^{h_{\ell m}}}{r^{3 / 2}}\right)\right] \\
	&\quad \times \exp \left[i\left(b_1^{h_{\ell m}} \frac{p_{r_*}}{r \Omega}+b_2^{h_{\ell m}} \frac{p_{r_*}^3}{r \Omega}\right)\right],
	\end{aligned}
\end{equation}
where $(a_1^{h_{\ell m}}\!$, $a_2^{h_{\ell m}}\!$, $a_3^{h_{\ell m}}\!$, $b_1^{h_{\ell m}}\!$, $b_2^{h_{\ell m}})$ are constants determined by requiring that the amplitude, its first and second derivatives, as well as the frequency of the modes and its first derivative, agree for every $ (\ell, m) $ mode with the NR input values at the matching time $ t _{ \text{match}} $ \cite{Taracchini:2013rva,Bohe:2016gbl,Cotesta:2018fcv,Pompiliv5}.
The objective of the NQC corrections is to improve the behavior of the modes near merger.
Hence, for QC binaries, the combination of the dynamical variables $ \left( r, p_{r_*}, \Omega \right) $ in Eq.~\eqref{eq:nqcs} is such that $ N_{\ell m} \to 1$ in the early inspiral since $ |p_{r_*}| \ll  r \Omega $.

However, in an eccentric binary system, the functional form in Eq.~\eqref{eq:nqcs} is not applicable since the values of $ p_{r_*}$ and $ r \Omega $ oscillate due to the orbital eccentricity.
Therefore, one needs to give an alternative prescription for computing the NQC corrections to avoid the introduction of unphysical features in the waveform modes due to the eccentricity oscillations.
For example, in \texttt{SEOBNRv4EHM}, the NQC corrections were computed using the orbit-averaged dynamical variables $ \left( \bar r, \,\bar p_{r_*}, \,\bar \Omega \right) $ with the same functional form as in Eq.~\eqref{eq:nqcs} \cite{Ramos-Buades:2021adz}.
Nevertheless, the orbit-averaging procedure introduces its own complications:
it becomes difficult to compute the orbit average of very low eccentricity systems, it is not possible to compute the orbit average of systems with very few cycles, and the orbit-averaging procedure could make the model nonrobust under small perturbations of the input parameters. 

For these reasons, in \texttt{SEOBNRv5EHM}, we compute the NQC corrections by employing a \emph{background QC dynamics}. More specifically, after obtaining the eccentric dynamics, we calculate the inspiral evolution of a BBH QC system with the same values of masses and spins as in the eccentric system.
Then, we employ the dynamical variables $ \left(  r _{ \text{QC}}, \, p_{r_*, \text{QC}},  \,\Omega_{ \text{QC}} \right) $ of this QC system to compute the NQC corrections~\eqref{eq:nqcs} using the QC NR input values from the \texttt{SEOBNRv5HM} model \cite{Pompiliv5}.
We note that the background QC dynamics is such that it has a duration larger or equal to the duration of the eccentric dynamics.
The starting frequency of such QC system is estimated using the first terms from the PN relation for $x(\tau)$ given by Eq.~(6) of Ref.~\cite{Blanchet:2023bwj}, where the time difference $ \tau $ is replaced by the duration of the eccentric dynamics.
In addition, to reduce the computational cost, the background QC dynamics is calculated through the use of a postadiabatic approximation \cite{Damour:2012ky,Nagar:2018gnk,Rettegno:2019tzh} for the inspiral, as in \texttt{SEOBNRv5HM}.

The eccentric and background QC dynamics are aligned at a given value of the separation for both systems.
We assume that, by the end of the inspiral, the eccentric dynamics has circularized sufficiently, so that the final behavior is the same for both the eccentric and QC systems.
In other words, we assume that there exists a time $ t _{ *} $ such that, for all the subsequent times $ t \geq t _{ *}  $, we have $ r \approx r _{ \text{QC}} $, where $ r $ is the relative separation in the eccentric dynamics and $ r _{ \text{QC}} $ is the relative separation in the background QC dynamics.
Given this assumption, we align both dynamics at the times $ t _{ \text{ref,\,ecc}} > t_*$ and $ t _{ \text{ref,\,qc}} $ at which
\begin{subequations}
\label{eq:t_r_ref_background_qc}
\begin{align}
r  (t _{ \text{ref,\,ecc}}) &= r _{ \text{QC}} (t _{ \text{ref,\,qc}}) = r _{ \text{ref}},
\\
 r _{ \text{ref}} &\equiv \max \left( r_\text{final}, \, r _{ \text{final,qc}}\right),
 \label{eq:r_ref}
 \end{align}
\end{subequations}
where $  r ^{ \text{final}} $ and $ r _{ \text{QC}} ^{ \text{final}} $ are the final values of the separation in the eccentric and QC dynamics, respectively.
Note that these final values by themselves are not robust under perturbations of the input parameters due to the stopping conditions of the EOM not being robust under perturbations.
Hence, one may think that the reference point \eqref{eq:t_r_ref_background_qc} will be a significant source of nonrobustness in the model.
However, this is not the case for systems that sufficiently circularize because these systems satisfy $ r \approx r _{ \text{QC}} $ at the end of the inspiral.
These are precisely the systems targeted by \texttt{SEOBNRv5EHM} (for reference, in Sec.~\ref{sec:robustness}, we study the robustness of the model across the eccentric binary parameter space).
Naturally, future models valid for higher eccentricities at merger will probably require a different prescription for the NQC corrections.

%%%%%%%%%%%%%%%%%%%%%%%%%%%
%%%%%%%%%%%%%%%%%%%%%%%%%%%

\subsection{Matching time}
\label{sec:attachment}

In \texttt{SEOBNRv5HM}, the time at which the inspiral-plunge modes are matched to the merger-ringdown modes according to Eq.~\eqref{eq:h_match} is computed as (see footnote \ref{fn:t_attach}) \cite{Pompiliv5}
\begin{equation}
\label{eq:t_attach_QC}
t _{ \text{match,\,qc}}
= t_\text{ISCO} + \Delta t_\text{ISCO} ^{ 22},
\end{equation}
where $\Delta t_\text{ISCO}^{ 22}$ is a calibration parameter determined with QC NR simulations, and $ t_\text{ISCO}$ is the time at which the relative separation of the binary is equal to the Kerr geodesic innermost stable circular orbit (ISCO) radius \cite{Bardeen:1972fi},
\begin{equation}
\label{eq:r_is_rISCO}
r(t_\text{ISCO}) = r_\text{ISCO}.
\end{equation}
Here, $r_{\text{ISCO}}$ is calculated with the values of final mass and spin $ \chi $ of the remnant BH, which are computed with fitting formulas obtained from QC NR simulations \cite{Jimenez-Forteza:2016oae, Hofmann:2016yih}.

The reason for selecting the ISCO radius as a reference point in Eq.~\eqref{eq:t_attach_QC} is because it is independent of features in the late dynamics, like the stopping conditions of the EOM, or the existence of a peak in the orbital frequency. In this way, the prescription for $ t _{ \text{match,\,qc}} $ becomes robust.
Furthermore, this reference time is uniquely defined since the relative separation decreases monotonically for a QC system. 

For eccentric binaries, two difficulties arise with Eq.~\eqref{eq:t_attach_QC}.
First, the value of $\Delta t_\text{ISCO}^{ 22}$ is calibrated using QC NR simulations, so formally we would need to calibrate it against eccentric NR simulations to obtain the correct time at which the $(2, 2)$-mode eccentric amplitude has its final peak.
The second difficulty appears because the relative separation does not monotonically decrease for eccentric orbits.
Thus, in general, we could have more than one solution to Eq.~\eqref{eq:r_is_rISCO}.
This poses a problem for the model's robustness because slight variations of the input parameters could cause multiple roots to appear or disappear, which is translated into sudden non-negligible changes in the attachment time.

To address the first difficulty, we use the calibration parameter $ \Delta t_\text{ISCO}^{ 22} $ determined from QC NR simulations as a first approximation.
Naturally, this choice is a source of systematic errors for high-eccentricity and high-spin systems, since these cases are expected to have the most significant differences when compared against QC binaries \cite{Nee:2025zdy}.
As more eccentric NR simulations become available, we will improve the model by properly calibrating the value of the last peak of the $(2, 2)$-mode amplitude.

To overcome the second difficulty, we employ the background QC dynamics used in the computation of the NQC corrections (see Sec.~\ref{sec:nqcs}).
Specifically, we use the reference times $ t _{ \text{ref,\,qc}} $ and $ t _{ \text{ref,\,ecc}} $ defined in Eq.~\eqref{eq:t_r_ref_background_qc} to infer the matching time of the eccentric modes $ t _{ \text{match,\,ecc}} $, according to
\begin{subequations}
\label{eq:t_attach_ecc}
\begin{align}
\Delta t _{ \text{match,\,qc}} &\equiv t _{ \text{match,\,qc}} - t _{ \text{ref,\,qc}},
\\
t _{ \text{match,\,ecc}} &= t _{ \text{ref,\,ecc}} + \Delta t _{ \text{match,\,qc}}.
\end{align}
\end{subequations}

Hence, Eq.~\eqref{eq:t_attach_ecc} determines the time at which the inspiral-plunge modes are matched to the merger-ringdown modes.
This prescription for the matching time inherits the robustness of the computation of the reference times $ t _{ \text{ref,\,qc}} $ and $ t _{ \text{ref,\,ecc}} $, and by construction satisfies Eq.~\eqref{eq:t_attach_QC} in the QC limit.
Naturally, this prescription will be less accurate and less robust for systems with high eccentricities near merger.

%%%%%%%%%%%%%%%%%%%%%%%%%%%
%%%%%%%%%%%%%%%%%%%%%%%%%%%

\subsection{Merger-ringdown modes}

Having assumed that the eccentric binary has circularized sufficiently by the end of the inspiral, in \texttt{SEOBNRv5EHM}, we employ a QC prescription for the merger-ringdown part of the waveform. 
This assumption is shared among all the existing eccentric waveform models, and it is a source of error for high eccentricity systems. In particular, we use the same phenomenological ansatz for the merger-ringdown modes as in the \texttt{SEOBNRv5HM} model \cite{Pompiliv5}.
This ansatz is informed by QC NR simulations and QC test-mass limit (TML) waveforms.

Physically, the employed ansatz for the merger-ringdown represents a Kerr BH that vibrates 
down to a state of equilibrium emitting quasinormal modes (QNMs) (or damped oscillations).
Therefore, in this waveform model, the merger-ringdown $(2, 2)$, $(3, 3)$, $(2, 1)$, and $(4, 4)$ modes have a monotonic amplitude and frequency evolution.
In contrast, the $(3, 2)$ and $(4, 3)$ modes have postmerger oscillations (a phenomenon known as \emph{mode mixing}).
These oscillations are mostly related to the mismatch between the \textit{spherical} harmonic basis and the \textit{spheroidal} harmonics employed in computing QNMs in BH perturbation theory~\cite{Buonanno:2006ui, Kelly:2012nd}.

The merger-ringdown ansatz is~\cite{Damour:2014yha,Bohe:2016gbl, Cotesta:2018fcv, Pompiliv5}
\begin{equation}
h_{\ell m}^{\text {merger-RD }}(\bm{\Pi}; t)=\nu \tilde{A}_{\ell m}(t) \e^{i \tilde{\phi}_{\ell m}(t)} \e^{-i \sigma_{\ell m 0}\left(t-t_{\text {match}}\right)},
\label{eq:ansatz_hlm}
\end{equation}
where $\sigma_{\ell m 0}$ is the complex frequency of the least-damped QNM of the remnant BH.
This frequency is obtained for each ($ 
\ell $, $ m $) mode as a function of the BH's final mass and spin using the \texttt{qnm} Python package of Ref.~\cite{Stein:2019mop}.
The functions $\tilde{A}_{\ell m}$ and $\tilde{\phi}_{\ell m}$ are determined by \cite{Bohe:2016gbl, Cotesta:2018fcv, Pompiliv5}
\begin{subequations}
\begin{align}
\tilde{A}_{\ell m}(t) &=c_{1, c}^{\ell m} \tanh \left[c_{1, f}^{\ell m}\left(t-t_{\text {match}}\right)+c_{2, f}^{\ell m}\right]+c_{2, c}^{\ell m},
\label{eq:ansatz_a}
\\
\tilde{\phi}_{\ell m}(t) &=
\phi_{\text {match }}^{\ell m}-d_{1, c}^{\ell m} \log \left[\frac{1+d_{2, f}^{\ell m} \e^{-d_{1, f}^{\ell  m}\left(t-t_{\text {match}}\right)}}{1+d_{2, f}^{\ell m}}\right],
\label{eq:ansatz_phi}
\end{align}
\end{subequations}
where $ c_{i, c}^{\ell m} $ ($ i \in \{ 1, 2\} $) and $ d_{1, c}^{\ell m} $ are constants that ensure that the modes are continuously differentiable at $ t = t _{ \text{match}} $, while $ c_{i, f}^{\ell m} $ and $ d_{i, f}^{\ell m} $ ($ i \in \{ 1, 2\} $) are coefficients obtained from polynomial fits in $ \nu $ and $ \chi $ informed by QC NR and QC TML waveforms, and $\phi_{\text {match }}^{\ell m}$ is the phase of the inspiral-plunge $(\ell, m)$ mode at $t=t_{\text {match}}$ (see Ref.~\cite{Pompiliv5} for more details).

Finally, we note that the merger-ringdown
modes \eqref{eq:ansatz_hlm} are completely independent of the NQC corrections \eqref{eq:nqcs} and the inspiral-plunge modes.
This will be advantageous for future improvements of the model with respect to the merger-ringdown of eccentric binaries.

%%%%%%%%%%%%%%%%%%%%%%%%%%%
%%%%%%%%%%%%%%%%%%%%%%%%%%%
%%%%%%%%%%%%%%%%%%%%%%%%%%%

\section{Model validation}
\label{sec:model_validation}

In this section, we validate the eccentric, aligned-spin \texttt{SEOBNRv5EHM} waveform model by studying its accuracy, speed, and robustness across different regions of the parameter space of eccentric binaries.
First, following Refs.~\cite{Cotesta:2018fcv,Pompiliv5}, we introduce a set of \textit{faithfulness functions}, which are employed as metrics that quantify the agreement between two waveforms as measured by a GW detector.
With these metrics, we assess the accuracy of \texttt{SEOBNRv5EHM} by comparing its waveforms against a set of QC and eccentric NR waveforms at our disposal.
Accordingly, we show comparisons between \texttt{SEOBNRv5EHM} and the QC waveform model \texttt{SEOBNRv5HM}~\cite{Pompiliv5}, and between two other eccentric waveform models: \texttt{SEOBNRv4EHM} \cite{Khalil:2021txt,Ramos-Buades:2021adz} and \texttt{TEOBResumS-Dal\'i} \cite{Nagar:2024dzj}.
Afterward, we perform waveform evaluation benchmarks and compare the results for \texttt{SEOBNRv5EHM} against other approximants within the \texttt{SEOBNRv5} family.
Finally, we quantify the robustness of \texttt{SEOBNRv5EHM} by determining the confident region of parameter space where the model successfully generates physically sensible eccentric waveforms and where it is stable under variations of the input parameters.

%%%%%%%%%%%%%%%%%%%%%%%%%%%
%%%%%%%%%%%%%%%%%%%%%%%%%%%

\subsection{Faithfulness functions}
\label{sec:faithfulness}

The gravitational signal from an eccentric, aligned-spin BBH system observed by a GW detector depends on 6 \emph{intrinsic} parameters (introduced in Sec.~\ref{sec:model_dynamics}) and 7 \emph{extrinsic} parameters.
The intrinsic parameters are related to physical properties of the binary:
the BH masses $m_1$ and $m_2$ (or, equivalently, mass ratio $ q $ and total mass $ M $), the dimensionless spin components $\chi_1$ and $\chi_2$, the eccentricity $e$ of the orbit, and the radial phase parameter $\zeta$.
The extrinsic parameters characterize the source with respect to a GW detector:
the angular position of the line of sight measured in the source frame ($\iota$, $\varphi$), the sky location of the source in the detector frame $(\Theta,\Phi)$, the polarization angle $\Psi$, the luminosity distance of the source $d_L$, and the time of coalescence $t _{ \text{c}}$.
In this subsection, $ e $ and $ \zeta $ will denote generic eccentric parameters, not necessarily equal to the Keplerian parameters that we employ in \texttt{SEOBNRv5EHM}.

In general, we can express the observed GW signal as
\begin{align}
\label{eq:h_detector}
h(t) =&  \; F_+(\Theta,\Phi,\Psi) \, h_+(\iota, \varphi,d_L,\bm{\Pi},t_{ \text{c}};t)
\nonumber
\\
&+ F_\times(\Theta,\Phi,\Psi) \, h_\times(\iota, \varphi,d_L,\bm{\Pi},t_{ \text{c}};t),
\end{align}
where $F_+$ and $F_\times$ are the antenna-pattern functions of the detector \cite{Sathyaprakash:1991mt,Finn:1992xs}, $ \,h_+ $ and $ h_\times $ are the GW polarizations defined in Eq.~\eqref{eq:gw_polarizations_decomp}, and $\bm{\Pi}=\{m_{1, \s 2},\,\chi_{1, \s2},\, e,\, \zeta\}$.

The waveform in Eq.~\eqref{eq:h_detector} can be rewritten in terms of an effective polarization angle $\kappa (\Theta,\Phi,\Psi)$ as
\begin{equation}
h(t) =  \mathcal{A}(\Theta,\Phi)\left ( h_+ \cos \kappa + h_+ \sin \kappa\right ),
\label{eq:eq14.1}
\end{equation}
where the definition of $\mathcal{A}(\Theta,\Phi)$ is displayed in Refs.~\cite{Cotesta:2018fcv,Ossokine:2020kjp}, and we have removed the dependences of $h_+$, $h_\times$, and $\kappa$ to ease the notation.

\subsubsection{Waveform inner product}
\label{sec:inner_product}

Given two waveforms $h_1$ and $h_2$, we introduce their inner product within the frequency band $ [f_{\text{min}}, f_{\text{max}}]$ as \cite{Sathyaprakash:1991mt,Finn:1992xs}
\begin{equation}
\langle h_1 | h_2 \rangle \equiv 4 \,\Re \int^{f_{\text{max}}}_{f_{\text{min}}} \frac{\tilde{h}_1(f) \,\tilde{h}^*_2(f)}{S_\text{n}(f)} \, df,
\label{eq:inner_product}
\end{equation}
where the tilde $ \, \tilde{} \, $ denotes the Fourier transform, the star $ ^{*} $ denotes the complex conjugate, and $S_\text{n}$ is the one-sided power-spectral density (PSD) of the detector's noise.
Additionally, we define the normalized waveform $h$ as
\begin{equation}
\label{eq:normalization}
\hat h \equiv \frac{ h}{\sqrt{  \langle h | h \rangle }}.
\end{equation}

In this work, we employ the \texttt{A+} PSD of the LIGO detectors, which is expected to be the one at design sensitivity of the fifth observing run (O5)~\cite{sensitivity_curves}, and we set $f_{\text{min}}=10\,$Hz and $f_{\text{max}}=2048\,$Hz.
For NR waveforms whose \emph{orbit-averaged} minimum frequency content is above $ 10\, $Hz, we set $f_{\text{min}} = 1.35 \, f_{\text{start}}$, where $f_{\text{start}}$ is the starting \emph{orbit-averaged} frequency of the NR $(2,2)$ mode.
This choice for $ f _{ \text{min}} $ coincides with the one employed in Refs.~\cite{Pompiliv5,RamosBuadesv5} for QC waveforms.
This prescription ensures that only meaningful frequency content is considered for eccentric waveforms that naturally have an oscillatory (nonmonotonic) frequency.
In contrast, if one employs the starting \emph{instantaneous} frequency of an eccentric NR waveform, this could lead to a configuration in which the waveforms are being compared over a frequency band that is incomplete, or that is strongly contaminated by Fourier transform artifacts.\footnote{
We note that the problem of selecting an appropriate lower bound to the inner product \eqref{eq:inner_product} has been discussed in Ref.~\cite{Shaikh:2023ypz}.
There, it is proposed to use the \emph{pericenter} frequencies to determine $f_{\text{min}}$ (in their notation, $f_{\text{low}}$ corresponds to our $f_{\text{min}}$).
This choice ensures that all the physical frequency content of the eccentric waveform is above $f_{\text{min}}$. However, depending on the eccentricity value, with this choice one would discard a significant portion of the early inspiral frequency content, as seen in their Fig.~3.
In this work, we employ the starting orbit-averaged frequency of the NR $(2,2)$ mode to determine $f_{\text{min}}$ since we are interested in including as much as possible the information from the early inspiral in the comparisons between NR and template waveforms.
Further work is required to completely understand the impact of these choices for $f_{\text{min}}$.
}
Similarly, in calculating the inner product \eqref{eq:inner_product}, we ensure that the eccentric waveforms $ h_1 $ and $ h_2 $ have the same time to merger.
Otherwise, spurious frequencies could contaminate the comparison between the two waveforms.
For all our results, we employ a sampling rate such that all waveforms have a time spacing of $ 0.1 \,M$ with $ M = 10 \, \solarmass $.
Additionally, we taper the time-domain waveforms using a Planck window function \cite{McKechan:2010kp} before transforming them into the frequency domain, as in Ref.~\cite{Pompiliv5}.

%%%%%%%%%%%%%%%%%%%%%%%%%%%

\subsubsection{QC systems}
\label{sec:faithfulness_qc}

For QC, aligned-spin waveforms containing higher-order $(\ell, m)$ modes (HMs), the agreement between a \emph{signal} $h_\text{s}$ and a \emph{template} $h_\text{t}$ with respect to a GW detector can be quantified by computing the \emph{faithfulness
function} \cite{Cotesta:2018fcv,Ossokine:2020kjp},
\begin{equation}
\label{eq:faithfulness_qc_hom}
\mathcal{F} ^{ \text{qc}}(M,\iota_{\text{s}},\varphi_{\text{s}},\kappa_{\text{s}}) =  \!  \max_{t_\text{c\s t},\s \varphi_{\text{t}},\s \kappa_{\text{t}}} \! \left[\left .  \left \langle \hat{h}_\text{s} \Big | \hat{h}_\text{t} \right \rangle \right \vert_{\substack{
\vspace{-0.3cm} \\
\iota_{\text{s}} = \iota_{\text{t}} \\
\boldsymbol{\Pi} ^{ \text{qc}}_{\text{s}, 0} = \boldsymbol{\Pi} ^{ \text{qc}}_{\text{t}, 0}
%\boldsymbol{\Pi} ^{ \text{qc}}_\text{s}(t_{\text{s}} = t_{\text{s}\s 0}) = \boldsymbol{\Pi}^{ \text{qc}}_{\text{t}}(t_\text{t} = t_{\text{t}\s 0})
%\boldsymbol{\Pi}_\text{s} = \boldsymbol{\Pi}_{\text{t}}
}}  \right ]\!,
\end{equation}
where the subindices ``s'' and ``t'' denote that the corresponding quantity is associated with the signal or the template, respectively.
In Eq.~\eqref{eq:faithfulness_qc_hom}, the inclination angles are set to be the same, $ \iota _{ \text{s}} = \iota _{ \text{t}} $, while the coalescence time, azimuthal angle, and effective polarization angle of the template $(t_\text{c\s t}, \varphi_{\text{t}}, \kappa_\text{t})$ are adjusted to maximize the faithfulness of the template with respect to the signal;
this is a common choice for comparing waveforms with HMs \cite{Cotesta:2018fcv,Ossokine:2020kjp,Garcia-Quiros:2020qpx}.
The intrinsic parameters characterizing the signal and the template are assumed to be the same at the start of the waveforms, hence $ \boldsymbol{\Pi} ^{ \text{qc}}_{\text{s}, 0} \equiv \boldsymbol{\Pi}^{ \text{qc}}_\text{s}(t_{\text{s}} = t_{\text{s}\s 0}) = \boldsymbol{\Pi}^{ \text{qc}}_{\text{t}}(t_\text{t} = t_{\text{t}\s 0}) \equiv \boldsymbol{\Pi} ^{ \text{qc}}_{\text{t}, 0}$, where $\bm{\Pi}^{ \text{qc}}=\{m_{1, \s 2},\, \chi_{1, \s2} \}$ \cite{Ossokine:2020kjp, RamosBuadesv5}.
The optimizations over the template's coalescence time $t_\text{c\s t}$ and azimuthal angle $\varphi_{\text{t}}$ are performed numerically, while the optimization over the template's effective polarization angle $ \kappa_\text{t} $ is done analytically as described in Refs.~\cite{Harry:2017weg, Capano:2013raa}.
Note that the total mass dependence in Eq.~\eqref{eq:faithfulness_qc_hom} appears since it determines the physical frequency of the waveform:
by increasing (decreasing) the total mass, less (more) content of the waveform will be inside the detector's frequency band.
Additionally, there is no dependence on the parameters $ (\Theta,\Phi, d_L)$ or the function $ \mathcal A (\Theta,\Phi)$ because the waveforms in Eq.~\eqref{eq:faithfulness_qc_hom} are normalized.

When the waveforms only contain the $(2, \pm 2)$ modes, the faithfulness function in Eq.~\eqref{eq:faithfulness_qc_hom} can be simplified.
This simplification is possible due to the angular dependence of the $_{-2}Y_{2 \pm2}$ harmonics, and the fact that $\iota$, $\varphi$, and $\kappa$ are degenerate
\cite{Capano:2013raa}.
The resulting faithfulness function for QC waveforms with only $(2, \pm 2)$ mode content can be expressed as 
\begin{equation}
\label{eq:faithfulness_qc_22}
%\mathcal{F} ^{ \text{qc}} _{ \text{22}} (M) = \!  \max_{t_\text{c\s t}, \varphi_{\text{t}}} \left[\left . \frac{ \langle h_\text{s}|h_\text{t} \rangle}{\sqrt{  \langle h_\text{s} | h_\text{s} \rangle  \langle h_\text{t} | h_\text{t} \rangle}}\right \vert_{
\mathcal{F} ^{ \text{qc}} _{22} (M) =   \max_{t_\text{c\s t},\, \varphi_{\text{t}}} \left[\left . \left \langle \hat{h}_\text{s} \Big | \hat{h}_\text{t} \right \rangle \right \vert_{
\substack{
\vspace{-0.3cm} \\
\boldsymbol{\Pi} ^{ \text{qc}}_{\text{s}, 0} = \boldsymbol{\Pi} ^{ \text{qc}}_{\text{t}, 0}
%\boldsymbol{\Pi}^{ \text{qc}}_\text{s}(t_{\text{s}} = t_{\text{s}\s 0}) = \boldsymbol{\Pi}^{ \text{qc}}_{\text{t}}(t_\text{t} = t_{\text{t}\s 0})
%\boldsymbol{\Pi}_\text{s} = \boldsymbol{\Pi}_{\text{t}}
}
} \,\right ],
\end{equation}
where we numerically optimize over the coalescence time of the template, and we analytically optimize over the azimuthal angle of the template \cite{Harry:2017weg, Capano:2013raa}.

%%%%%%%%%%%%%%%%%%%%%%%%%%%

\subsubsection{Eccentric systems}
\label{sec:faithfulness_ecc}

For eccentric waveforms, the previous faithfulness functions need to be generalized to account for the gauge-dependent nature of eccentricity in GR.
In general, the signal $ h _{ \text{s}} $ and the template $ h _{ \text{t}} $ will be characterized by different eccentricity parameters.
This raises a problem with the condition $ \boldsymbol{\Pi}_\text{s}(t_{\text{s}} = t_{\text{s}\s 0}) = \boldsymbol{\Pi}_{\text{t}}(t_\text{t} = t_{\text{t}\s 0}) $, where $\bm{\Pi}=\{m_{1, \s 2},\,\chi_{1, \s2},\, e,\, \zeta\}$ since, in general, there is no unique way of mapping the definitions of eccentricity $ e $ and radial anomaly $ \zeta $ between the two waveforms.
Therefore, when comparing two eccentric waveforms, optimizations over the initial eccentricity, radial anomaly, and starting frequency have to be performed to account for the different definitions employed \cite{Ramos-Buades:2021adz,Knee:2022hth}.
A disadvantage of this process is that these optimizations also take into account the waveforms' systematic errors, not just the different definitions of eccentric parameters.
To avoid this issue, Ref.~\cite{Bonino:2024xrv} developed a pipeline for mapping the different definitions employed in eccentric waveform models to a given eccentric waveform.
However, this pipeline cannot be robustly applied to eccentric NR waveforms with a significant amount of noise at the early stages of the waveform;
in these cases, some waveform cycles need to be removed from the comparison.
Since one of our objectives is to compare models against NR waveforms as completely as possible (to assess the accuracy of the models at different eccentricities), we decided to use an optimization method to find the best-fitting waveforms corresponding to a given eccentric NR waveform.

More precisely, we follow a procedure for the optimization over eccentric parameters similar to the one presented in Ref.~\cite{Ramos-Buades:2021adz}, which was employed for the \texttt{SEOBNRv4EHM} model.
In this way, to compute the agreement between an eccentric waveform template $ h _{ \text{t}} $ (e.g., a \texttt{SEOBNRv5EHM} waveform) and a signal $ h _{ \text{s}} $ (e.g., a NR waveform), we perform additional maximizations over the template's initial eccentricity $e_{\text{t}}$ and starting frequency $\langle M \Omega \rangle_{\text{t}}$ with fixed radial anomaly $\zeta _{ \text{t}}=\pi$.\footnote{
Another alternative would be to fix the starting frequency of the template $\langle M \Omega \rangle_{\text{t}}$ to that of the signal $\langle M \Omega \rangle_{\text{s}}$, and optimize over the template's initial eccentricity $e_{\text{t}}$ and radial anomaly $\zeta _{ \text{t}}$.
We do not follow this method to avoid the generation of systems whose reference values are specified at anomalies close to periastron ($\zeta _{ \text{t}}=0$) since these are strong-field, high-velocity configurations for highly eccentric systems.
The waveform models employed in this work provide different treatments for these challenging configurations, so generating waveforms at apastron is more convenient across all the models.
}

For eccentric waveforms whose only content is the $ (2, \pm 2) $ modes, we define the faithfulness function as
\begin{equation}
\label{eq:faithfulness_ecc_22}
\mathcal{F} ^{ \text{ecc}} _{ \text{22}} (M) = \max_{
%\substack{
%t_\text{c\s t}, \varphi_{\text{t}},
%\\
%e_{\text{t}}, \langle M \Omega \rangle_{ \text{t}}
%}
t_\text{c\s t},\s \varphi_{\text{t}},\s e_{\text{t}},\s \langle M \Omega \rangle_{ \text{t}}
}
%\! \left[\left . \frac{ \langle h_\text{s}|h_\text{t} \rangle}{\sqrt{  \langle h_\text{s} | h_\text{s} \rangle  \langle h_\text{t} | h_\text{t} \rangle}}\right \vert_{
\! \left[\left .  \left \langle \hat{h}_\text{s} \Big | \hat{h}_\text{t} \right \rangle\right \vert_{
\substack{
\vspace{-0.3cm} \\
\zeta _{ \text{t}} = \pi
\\
\boldsymbol{\Pi} ^{ \text{qc}}_{\text{s}, 0} = \boldsymbol{\Pi} ^{ \text{qc}}_{\text{t}, 0}
%\boldsymbol{\Pi}^{ \text{qc}} _\text{s}(t_{\text{s}} = t_{\text{s}\s 0}) = \boldsymbol{\Pi}^{ \text{qc}} _{\text{t}}(t_\text{t} = t_{\text{t}\s 0})
}
} \,\right ] \!.
\end{equation}
The optimizations in Eq.~\eqref{eq:faithfulness_ecc_22} are done as follows \cite{Ramos-Buades:2021adz}:
\begin{itemize}
\item[1)]
Define a total mass interval $ [M _{ \text{min}}, \s M _{ \text{max}}] $ for the analysis, and fix $ M $ to the lower bound $ M _{ \text{min}}$.
This way, more of the inspiral part of the signal $ h _{ \text{s}} $ is in the frequency band of the faithfulness calculation.
\item[2)]
Estimate a starting value of eccentricity $ e_0 $ of the signal and create an array with $N_e$ values in the interval $ [e_0 - \delta e, \s e_0 + \delta e] $ for a given $ \delta e $.
In the case of $ e_0 - \delta e < 0 $, the lower bound is set to zero.
To estimate the starting eccentricity, we employ the formula proposed by Mora and Will \cite{Mora:2002gf},
\begin{equation}
\label{eq:e_est_0}
e_{\Omega _{ \text{orb}}} = \frac{\sqrt{\Omega _{ \text{orb}} ^{\text{p}} }  - \sqrt{\Omega_{ \text{orb}}^{ \text{a}} \vphantom{\Omega_{\text{orb}}^{\text{p}}} } }{\sqrt{\Omega_{ \text{orb}}^{ \text{p}}}  + \sqrt{\Omega_{ \text{orb}}^{ \text{a}}  \vphantom{\Omega_{\text{orb}}^{\text{p}}} } },
\end{equation}
where $ \Omega _{ \text{orb}} ^{\text{p}}$  and $\Omega_{ \text{orb}}^{ \text{a}}$ are interpolants of the signal's orbital angular frequency at consecutive periastron and apastron passages, i.e., interpolants of the maxima and minima of $ \Omega _{ \text{orb}}(t)$ from the signal dynamics (namely, from the NR simulation).
Equation~\eqref{eq:e_est_0} is evaluated at the start of the signal, such that $ e_0 = e_{\Omega _{ \text{orb}}}(t=0)  $.\footnote{
The definition in Eq.~\eqref{eq:e_est_0} depends on the binary dynamics, so it would not be applicable for signals (waveform approximants) which have no associated binary dynamics.
In such cases, it would be straightforward to use any other definition of eccentricity from the waveform, as the only purpose of this step is to give an \emph{estimation} of the initial eccentricity around which a grid is created of starting eccentricities and frequencies.
}
\item[3)]
For every $ e \in [e_0 - \delta e, \s e_0 + \delta e] $, determine the starting frequency $ \langle M \Omega \rangle _{ e} $ at which $ h _{ \text{t}} $ has the same time to merger as $ h _{ \text{s}} $.
Then, for all the $ N_e $ eccentricities, create an array with $N_\Omega$ values of starting frequencies in the interval $ [\langle M \Omega \rangle _{ e} - \delta \Omega, \s \langle M \Omega \rangle _{ e}] $ for a given $ \delta \Omega $.
\item[4)]
Compute the faithfulness for each point in the resulting grid of eccentricities and starting frequencies by optimizing analytically over the template's azimuthal angle.
\item[5)]
Store the values $(e_{\text{opt}},\s \langle M \Omega \rangle_{\text{opt}})$ that provide the maximum faithfulness for $ M = M _{ \text{min}}$.
\item[6)]
Employ the values $(e_{\text{opt}},\s \langle M \Omega \rangle_{\text{opt}})$ to calculate the faithfulness for the rest of the total masses.
This is calculated by numerically optimizing over the coalescence time of the template, and analytically optimizing over the azimuthal angle of the template as in Eq.~\eqref{eq:faithfulness_qc_22}.
\end{itemize}

In Sec.~\ref{sec:comparison_nr_eccentric}, we apply this procedure to compare the eccentric waveform models \texttt{SEOBNRv4EHM}, \texttt{SEOBNRv5EHM}, and \texttt{TEOBResumS-Dal\'i} to eccentric NR waveforms.
In particular, we set: $ M _{ \text{min}} = 20 \, \solarmass $,
$ M _{ \text{max}} = 200 \, \solarmass $,
$ \delta e = 0.1 $,
$ \delta \Omega =\nolinebreak 0.1 \, \langle M \Omega \rangle_e $,
$ N_e = 100 $,
and
$ N_\Omega = 400 $.
Naturally, increasing the grid points $ N_e $ and $ N_\Omega$ would improve the faithfulness of the best-fitting template $ h _{ \text{t}} $ with respect to the given signal $ h _{ \text{s}} $, at the expense of a higher computational cost.
We verified that the chosen values of $ N_e $ and $ N_\Omega$ produce a robust faithfulness value for the NR waveforms and the waveform approximants employed in this work.

Generalizing Eq.~\eqref{eq:faithfulness_qc_hom} to be applicable for eccentric waveforms with HMs as in Eq.~\eqref{eq:faithfulness_ecc_22} would be computationally prohibitive.
Thus, we follow Ref.~\cite{Ramos-Buades:2021adz}, and we optimize over the template parameters $ (t_\text{c\s t}, \varphi_{\text{t}}, \kappa_{\text{t}}) $ employing the values $(e_{\text{opt}},\s \langle M \Omega \rangle_{\text{opt}})$ coming from the $ (2, 2) $-mode optimization done in Eq.~\eqref{eq:faithfulness_ecc_22} for $ M = M _{ \text{min}} $.
Hence, the faithfulness function that we use to compute the agreement of eccentric waveforms with HMs is given by
\begin{equation}
\label{eq:faithfulness_ecc_hom}
%\mathcal{F} ^{ \text{ecc}}(M,\iota_{\text{s}},\varphi_{\text{s}},\kappa_{\text{s}}) = \! \! \! \max_{t_\text{c\s t}, \varphi_{\text{t}}, \kappa_{\text{t}}} \! \left[\left . \frac{ \langle h_\text{s}|h_\text{t} \rangle}{\sqrt{  \langle h_\text{s} | h_\text{s} \rangle  \langle h_\text{t} | h_\text{t} \rangle}}\right \vert_{\substack{
\mathcal{F} ^{ \text{ecc}}(M,\iota_{\text{s}},\varphi_{\text{s}},\kappa_{\text{s}}) =   \max_{t_\text{c\s t},\, \varphi_{\text{t}}, \s\kappa_{\text{t}}} \! \left[\left .  \left \langle \hat{h}_\text{s} \Big | \hat{h}_\text{t} \right \rangle\right \vert_{\substack{
\vspace{-0.2cm} \\
\iota_{\text{s}} = \iota_{\text{t}},  \,\, \zeta _{ \text{t}} = \pi, \, \, e_{\text{t}} = e _{ \text{opt}} 
\\
\langle M \Omega \rangle_{\text{t}} = \langle M \Omega \rangle_{\text{opt}}
\\
\boldsymbol{\Pi} ^{ \text{qc}}_{\text{s}, 0} = \boldsymbol{\Pi} ^{ \text{qc}}_{\text{t}, 0}
}}  \right ]\!.
\end{equation}
%

%%%%%%%%%%%%%%%%%%%%%%%%%%%

\subsubsection{Averaged faithfulness and the mismatch function}
\label{sec:faithfulness_avg}

To reduce the dimensionality of faithfulness functions, we introduce the \emph{sky-and-polarization-averaged, signal-to-noise (SNR)-weighted faithfulness} \cite{Ossokine:2020kjp,Cotesta:2018fcv},
%
%\begin{widetext}
\begin{equation}
\label{eq:faithfulness_avg_snr}
%\overline{\mathcal{F}}_{\text{SNR}}(M, \iota _{ \text{s}}) = \sqrt[3]{\frac{\int_{0}^{2\pi} d\varphi_{\text{s}} \int_{0}^{2\pi} d\kappa_ {\text{s}} \ \mathcal{F}^{3}(M_{\text{s}},\iota_{\text{s}}, \varphi_{\text{s}},\kappa_{\text{s}}) \ \text{SNR}^3(\iota_{\text{s}},\varphi_{\text{s}},\kappa_{\text{s}})}{\int_{0}^{2\pi} d\varphi_{\text{s}} \int_{0}^{2\pi} d\kappa_{\text{s}} \ \text{SNR}^3(\iota_{\text{s}},\varphi_{\text{s}},\kappa_{\text{s}})}} \, ,
\overline{\mathcal{F}}_{\text{SNR}}(M, \iota _{ \text{s}}) = \sqrt[3]{\frac{\int_{0}^{2\pi} d\varphi_{\text{s}} \int_{0}^{2\pi} d\kappa_ {\text{s}} \ \mathcal{F}^{3} \ \text{SNR}^3}{\int_{0}^{2\pi} d\varphi_{\text{s}} \int_{0}^{2\pi} d\kappa_{\text{s}} \ \text{SNR}^3}} \, ,
\end{equation}
%\end{widetext}
%
where $ \mathcal{F} = \mathcal{F}(M_{\text{s}},\iota_{\text{s}}, \varphi_{\text{s}},\kappa_{\text{s}}) $ is the faithfulness function given either by Eq.~\eqref{eq:faithfulness_qc_hom} for QC waveforms, or by Eq.~\eqref{eq:faithfulness_ecc_hom} for eccentric waveforms, and the SNR is defined as
\begin{equation}
\text{SNR}(\iota_{\text{s}},\varphi_{\textrm{s}}, \kappa_{\text{s}}, \Theta_\text{s}, \Phi_\text{s}, d_L) = \sqrt{\langle h_{\text{s}} | h_{\text{s}} \rangle} \,.
\label{eq:SNR}
\end{equation}

In practice, the integrals in Eq.~\eqref{eq:faithfulness_avg_snr} are computed numerically.
Specifically, for a given inclination $ \iota_{\text{s}} $, % \in \{0, \s \pi/3, \s \pi/2\} $.
we evaluate the integrands in a uniform grid of $ 8 \kern-0.08em \times \kern-0.08em 8 $ values for $ \varphi_{\textrm{s}}, \kappa_{\text{s}} \in [0, \s 2 \pi] $.

Finally, we introduce the \emph{unfaithfulness} (or \emph{mismatch}) for any of the previous faithfulness functions as
\begin{equation}
\mathcal{M} \equiv 1-\mathcal{F} \,.
\label{eq:mismatch}
\end{equation}
The lower the value of mismatch between two waveforms, the better they agree with respect to a given GW detector.

%%%%%%%%%%%%%%%%%%%%%%%%%%%
%%%%%%%%%%%%%%%%%%%%%%%%%%%

\subsection{Comparison against quasicircular numerical-relativity waveforms}
\label{sec:comparison_nr_qc}

\begin{figure*}%[ht]
\includegraphics[width=\textwidth]
{v5HM_v5EHM_Dali_QCsims_spaghetti_22_HoM_iota_1_05}
\vspace{-20pt}
\caption{
Waveform mismatches between different aligned-spin approximants and the 441 SXS QC NR simulations used in this work, calculated over a range of total masses between $10$ and $300 \,\solarmass$.
The first column corresponds to the QC \texttt{SEOBNRv5HM} model, the second column to the eccentric \texttt{SEOBNRv5EHM} model, and the third column to the eccentric \texttt{TEOBResumS-Dal\'i} model.
The top panels show the $(2,2)$-mode mismatch $ \mathcal M_{22} $ for each NR waveform, and the bottom panels show the sky-and-polarization averaged, SNR-weighted mismatch $\overline{\mathcal{M}}_\mr{SNR}$ (which takes into consideration higher-order modes) for inclination $\iota _{ \text{s}} = \pi/3$.
The colored lines highlight cases with the worst maximum mismatch for each model (solid lines for $ \mathcal M_{22} $ and dashed lines for $\overline{\mathcal{M}}_\mr{SNR}$) with the legends indicating the parameters of the worst-case mismatch for the corresponding model. 
}
\label{fig:qc_mismatch_spaghetti_22}
\end{figure*}

We validate the eccentric, aligned-spin \texttt{SEOBNRv5EHM} model in the zero eccentricity limit by computing waveform mismatches against a set of 441 BBH QC NR waveforms produced with the pseudo-Spectral Einstein code (\texttt{SpEC}) by the Simulating eXtreme Spacetimes (SXS) Collaboration \cite{SXS:catalog,Boyle:2019kee,Chu:2015kft,Blackman:2017dfb,Hemberger:2013hsa,Scheel:2014ina,Lovelace:2014twa,Abbott:2016apu,Blackman:2015pia,Lovelace:2016uwp,Varma:2018mmi,Abbott:2016nmj,Varma:2019csw,Kumar:2015tha,Mroue:2013xna,Yoo:2022erv},\footnote{
This set of 441 QC NR waveforms was employed in the calibration and validation of the QC model \texttt{SEOBNRv5HM}.
%, and its parameter space distribution is depicted in Figure 2 of Ref.~\cite{Pompiliv5}.
The properties of all the 441 waveforms are listed in the ancillary file
\url{https://arxiv.org/src/2303.18039v1/anc/NR_simulations.json} 
%\href{https://arxiv.org/src/2303.18039v1/anc/NR_simulations.json} {\texttt{arxiv.org/src/2303.18039v1/anc/\\NR\_simulations.json}}
of Ref.~\cite{Pompiliv5}.
}
and by comparing these results against the ones from the QC model \texttt{SEOBNRv5HM}~\cite{Pompiliv5}.
The parameter space distribution of these QC NR waveforms is depicted in Fig.~2 of Ref.~\cite{Pompiliv5}.
For comparison, we also compute the mismatches associated to the eccentric, aligned-spin model \texttt{TEOBResumS-Dal\'i} \cite{Nagar:2024dzj}.
The $ (2,2) $-mode mismatches are computed with Eq.~\eqref{eq:faithfulness_qc_22} over a range of total masses $ M \in\nolinebreak {[10, 300]}\, \solarmass$, and the results for the three waveform approximants \texttt{SEOBNRv5HM}, \texttt{SEOBNRv5EHM}, and \texttt{TEOBResumS-Dal\'i} are shown in the top panels of Fig.~\ref{fig:qc_mismatch_spaghetti_22}.

From Fig.~\ref{fig:qc_mismatch_spaghetti_22}, we see an excellent agreement between the QC $ \texttt{SEOBNRv5HM} $ and eccentric $ \texttt{SEOBNRv5EHM} $ models in the zero eccentricity limit.
This is due to the parametrization employed in \texttt{SEOBNRv5EHM}, which directly uses the eccentricity $ e $ in the eccentric corrections to the EOB RR force and waveform modes.
Additionally, \texttt{SEOBNRv5EHM} manifests an overall accuracy improvement with respect to \texttt{TEOBResumS-Dal\'i} in the QC limit.
As with QC models, the most challenging cases correspond to systems with high mass ratios and high spins. % (e.g., see Figure 5 of Ref.~\cite{Pompiliv5}).
However, all these cases have mismatches below  $ 3 \times 10^{-3} $ for both \texttt{SEOBNRv5} models.
In contrast, the \texttt{TEOBResumS-Dal\'i} model has various cases with mismatches above $ 3 \times 10^{-3} $, and even reaching the $ 1\% $ level. Overall, the bulk of \texttt{TEOBResumS-Dal\'i}  $ (2,2) $-mode mismatches is above the one for the \texttt{SEOBNRv5} models.

The accuracy of the GW polarizations (including HMs) is quantified with the sky-and-polarization-averaged, SNR-weighted faithfulness given by Eqs.~\eqref{eq:faithfulness_qc_hom} and \eqref{eq:faithfulness_avg_snr}, calculated for an inclination $ \iota _{ \text{s}} = \pi/3 $, over the same range of total masses $ M \in [10, 300]\, \solarmass$.
The associated mismatches are shown in the bottom panels of  Fig.~\ref{fig:qc_mismatch_spaghetti_22}.
For \texttt{SEOBNRv5HM} and \texttt{SEOBNRv5EHM}, we include the modes $( \ell, |m| ) = \{(2, 2),\, (3, 3),\, (2, 1),\, (4, 4),\, (3, 2),\, (4, 3)\}$.
For \texttt{TEOBResumS-Dal\'i}, we include the modes $( \ell, |m| ) = \{(2, 2),\, (3, 3),\, (2, 1),\, (4, 4)\}$, which are the modes supported by this model (see footnote \ref{fn:dali}).
For the NR waveforms, we include all modes up to $ \ell = 4 $ with $ |m| > 0 $.

When including HMs, \texttt{SEOBNRv5EHM} also demonstrates identical results to \texttt{SEOBNRv5HM} and shows an overall improvement with respect to \texttt{TEOBResumS-Dal\'i}.
In particular, the latter model has a significant amount of cases with mismatches above $ 10^{-2} $ (even reaching the $ \sim 30\% $ level), whereas \texttt{SEOBNRv5EHM} is always below the $ 3\% $ level.
Apart from this, for both eccentric models the observed behavior of mismatches when including HMs is completely analogous to the one observed for QC models (e.g., see Fig. 7 of Ref.~\cite{Pompiliv5}, which was produced with Advanced LIGO PSD \cite{Barsotti:2018} and all NR modes up to $ \ell = 5 $).
In particular, the increase in mismatch with respect to the total mass indicates that the modeling of higher-order modes needs to be improved.

All of our results demonstrate a remarkable agreement between the waveform models \texttt{SEOBNRv5HM} and \texttt{SEOBNRv5EHM} in the QC limit, and an overall improvement of \texttt{SEOBNRv5EHM} with respect to the \texttt{TEOBResumS-Dal\'i} model for zero eccentricities.
This is particularly relevant in the context of inference studies because having an eccentric waveform model with an accurate QC limit is important to avoid biases in the estimation of binary parameters \cite{Bonino:2022hkj, Ramos-Buades:2023yhy}.

%%%%%%%%%%%%%%%%%%%%%%%%%%%
%%%%%%%%%%%%%%%%%%%%%%%%%%%

\subsection{Comparison against eccentric numerical-relativity waveforms}
\label{sec:comparison_nr_eccentric} 

We assess the accuracy of the eccentric, aligned-spin \texttt{SEOBNRv5EHM} model by computing waveform mismatches against a catalog of 99 BBH eccentric NR waveforms produced with the \texttt{SpEC} code from the SXS Collaboration \cite{SXS:catalog}.
Of these waveforms, 18 correspond to aligned-spin binaries, and the rest correspond to nonspinning ones. 
To perform a more structured analysis, we split this catalog into two subsets, depending on the initial value of the GW eccentricity $ e _{ \text{gw}} $ as defined in Refs.~\cite{Ramos-Buades:2022lgf,Shaikh:2023ypz}.\footnote{
Throughout this paper, we use version \texttt{1.0.2} of the \texttt{gw\_eccentricity} package from the public repository \url{https://github.com/vijayvarma392/gw_eccentricity} to calculate the GW eccentricity $ e _{ \text{gw}} $ and GW mean anomaly $ l _{ \text{gw}} $ from the waveforms.
\label{fn:e_gw}
}
Namely,
\begin{itemize}
\item [1)]
84 waveforms with initial GW eccentricities $ e _{ \text{gw}} < 0.5 $ and mass ratios up to $ q = 18 $.
This subset includes 28 publicly available waveforms \cite{Hinder:2017sxy, Boyle:2019kee} plus 56 private waveforms \cite{Ramos-Buades:2022lgf}---16 of these waveforms correspond to aligned-spin binaries, with different mass ratios.
\item [2)]
15 private waveforms with initial GW eccentricities $ 0.5 < e _{ \text{gw}} \lesssim 0.88 $ and mass ratios up to $ q = 10 $ \cite{Ramos-Buades:2022lgf}.
Two of these waveforms correspond to aligned-spin binaries, but with equal masses.
\end{itemize}

\begin{figure}%[H]
%\centering
%\begin{subfigure}{\linewidth}
\hspace*{-10pt}
\includegraphics[width=1.04\linewidth]{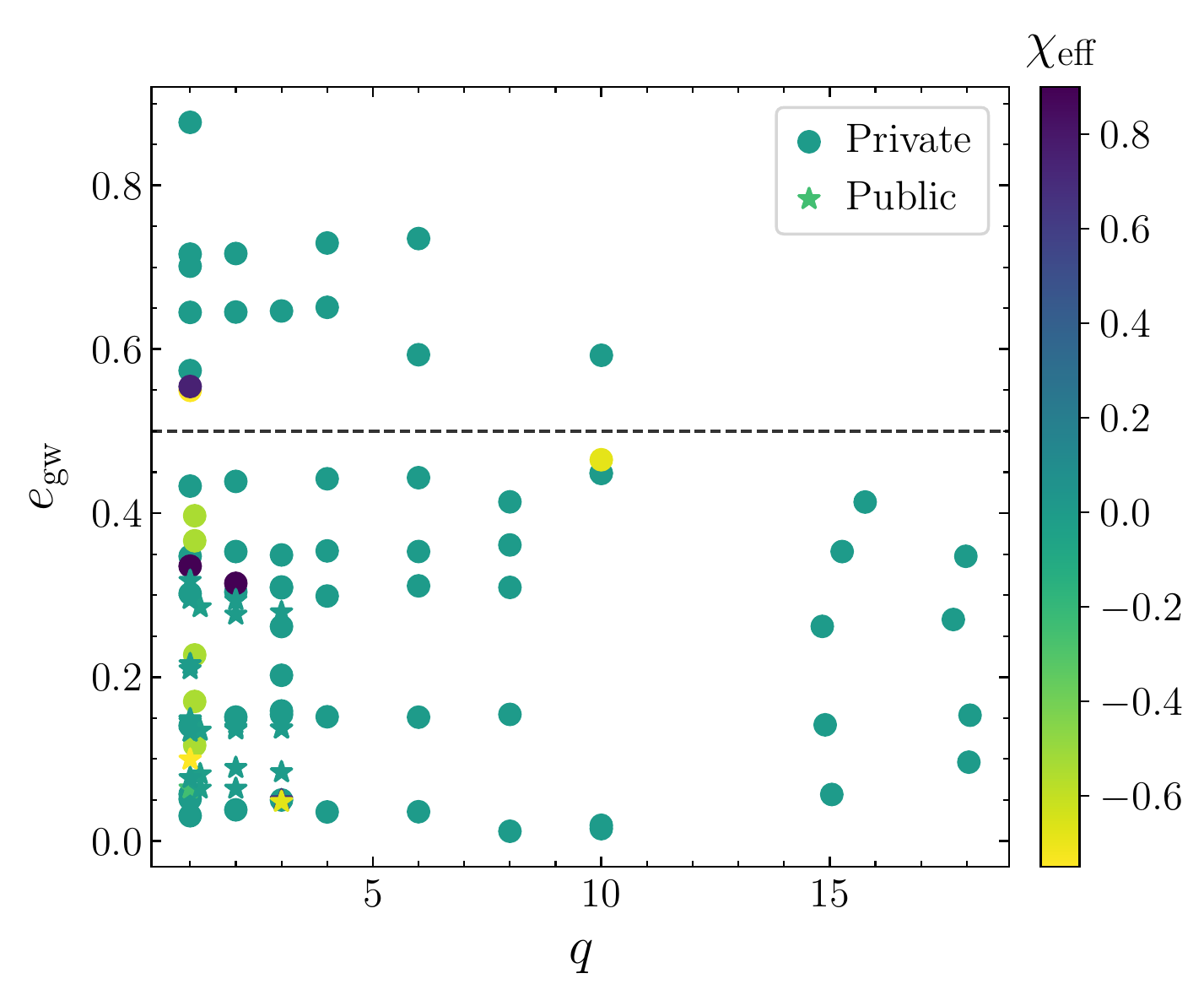}
\vspace{-20pt}
\caption{
Parameter-space distribution of the 99 eccentric SXS NR waveforms employed in this work, in terms of the mass ratio $q = m_1 / m_2$, dimensionless effective spin $\chi_{\text{eff}}=(\chi_1 m_1 + \chi_2 m_2)/M$, and initial GW eccentricity $ e _{ \text{gw}} $.
We represent the 28 publicly available NR waveforms with stars and the private NR waveforms with circles.
The dashed horizontal line indicates the value $ e _{ \text{gw}} = 0.5 $, which was employed to separate the 99 waveforms into two subsets, depending on the initial value of $ e _{ \text{gw}} $.
}
\label{fig:param_space_NR_ecc_sims}
\end{figure}

In Fig. \ref{fig:param_space_NR_ecc_sims}, we show the parameter-space distribution of these waveforms in terms of mass ratio $ q $, dimensionless effective spin parameter $ \chi _{ \text{eff}} $, and initial GW eccentricity $ e _{ \text{gw}} $.
Additionally, we provide a summary of the properties of each NR waveform in Table~\ref{tab:simulation_data} of Appendix~\ref{sec:tables}.\footnote{
The full list of NR eccentric waveforms is also provided as an ancillary file. % in \url{https://arxiv.org/src/2412.XXXXX/anc/ancillary_ecc_NR_sims.json}.
For each waveform, we list the mass ratio $ q $, dimensionless spin components $ \chi _1 $ and $ \chi _2 $, the GW eccentricity $ e _{ \text{gw}} $ at a reference $ (2,2) $-mode orbit-averaged dimensionless frequency $\langle M \omega_{22} \rangle$ (both associated to their starting values calculated with \texttt{gw\_eccentricity}), number of periastron passages $ N_p $, and time to merger $ t _{ \text{merger}} / M$.
We also provide the optimum values of initial eccentricity $ e_0 $, initial dimensionless orbit-averaged orbital frequency $ \langle M \Omega \rangle_0 $, and maximum $ (2,2) $-mode mismatch $ \max_M \mathcal M_{22} $ (calculated over a range of total masses $ M \in [ 20, 200] \, \solarmass $), corresponding to the best-fitting waveform for each eccentric waveform model employed in this work.
\label{fn:ancillary}
}

The mismatches against the 99 eccentric NR simulations are computed with the method and metrics described in Secs.~\ref{sec:faithfulness_ecc} and \ref{sec:faithfulness_avg}, for three different eccentric, aligned-spin waveform models: \texttt{SEOBNRv4EHM}, \texttt{SEOBNRv5EHM}, and \texttt{TEOBResumS-Dal\'i}.
This way, we show the improvement of \texttt{SEOBNRv5EHM} with respect to its predecessor \texttt{SEOBNRv4EHM}, and we show the comparison between the two state-of-the-art eccentric waveform models \texttt{SEOBNRv5EHM} and \texttt{TEOBResumS-Dal\'i}.
For the computation of mismatches, we apply \emph{exactly} the same code and settings for each of these models to ensure a fair comparison.
The corresponding optimum parameters that produce the best-fitting waveform for each model and for each eccentric NR simulation are provided within the ancillary file to this work.\footnote{
We note that the inferred best-fitting waveforms typically have more cycles than the corresponding NR waveforms.
This originates from the method employed to find the best-fitting input parameters, which optimizes over eccentricity and starting frequency while keeping the time to merger larger or equal to that of the NR waveform.
As commented in Sec.~\ref{sec:inner_product}, before computing the mismatch, we trim the model waveform so that it has the same time to merger as the given NR waveform.
\label{fn:extra_cycles}
}

%%%%%%%%%%%%%%%%%%%%%%%%%%%
%%%%%%%%%%%%%%%%%%%%%%%%%%%
%%%%%%%%%%%%%%%%%%%%%%%%%%%

\subsubsection{Accuracy of the eccentric $ (2,2) $ mode}
\label{sec:accuracy_22_mode}

\begin{figure*}%[H]
%\centering
%\begin{subfigure}{\linewidth}
\hspace*{-6pt}
\subfloat{\includegraphics[width=0.49\linewidth]{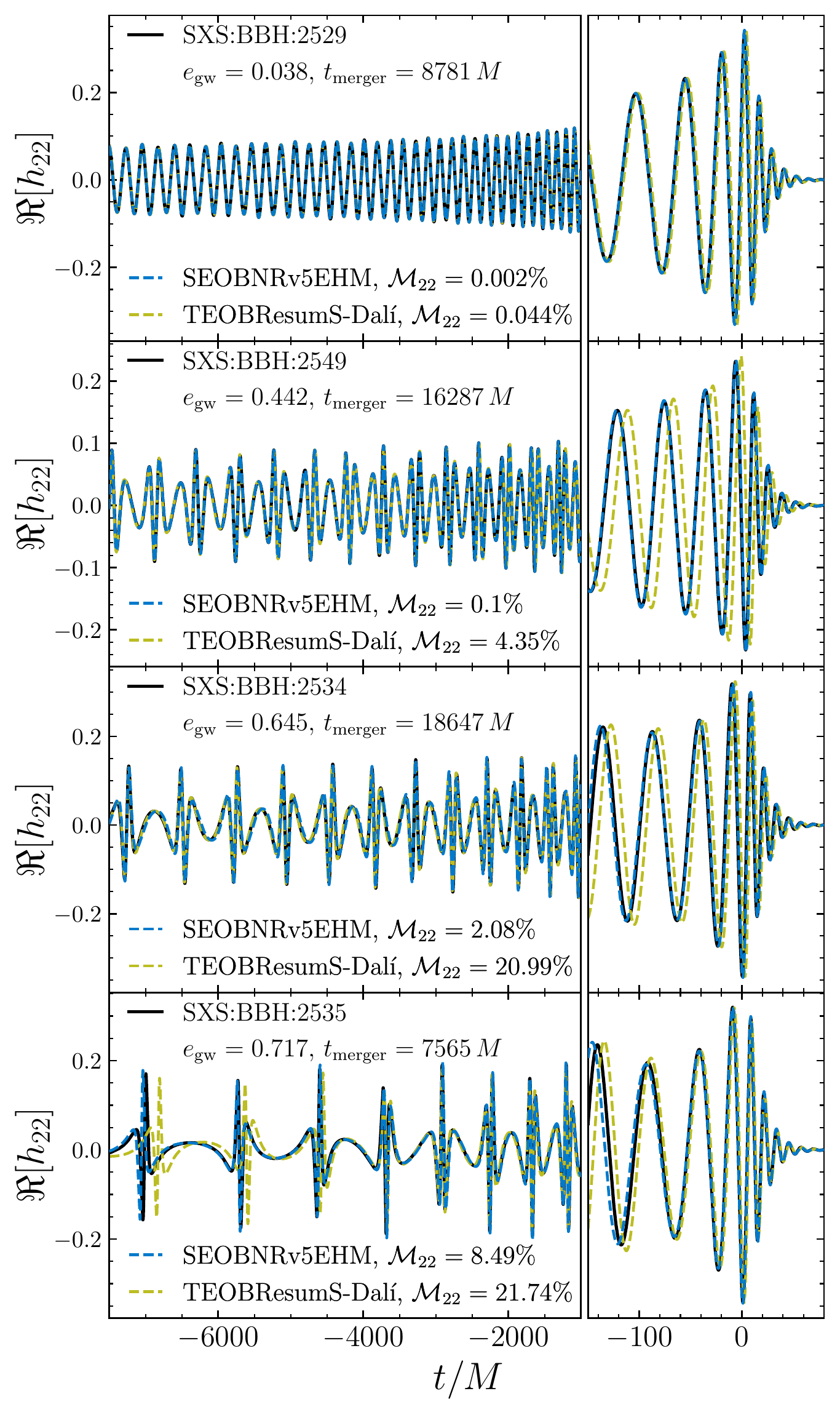}}
\hfill
\subfloat{\includegraphics[width=0.488\linewidth]{22_mode_phase_v5EHM_Dali}}
\vspace{-10pt}
\caption{
\textit{Left column:}
Real part of the $ (2,2) $ mode for different eccentric, nonspinning NR waveforms (black) with increasing eccentricity (one for each panel) along with the best-fitting waveforms for the state-of-the-art eccentric, aligned-spin models \texttt{SEOBNRv5EHM} (blue, dashed) and \texttt{TEOBResumS-Dal\'i} (olive, dashed), in geometric units.
For each NR waveform, we display the initial value of the GW eccentricity $ e _{ \text{gw}} $, the time to merger $ t _{ \text{merger}} $ (measured from the start of the waveform), and the $ (2,2) $-mode mismatch $ \mathcal M_{22} $ for the best-fitting waveforms of each model calculated for a binary with total mass $ M = 20 \, \solarmass $.
\textit{Right column:}
$ (2,2) $-mode phase difference between the eccentric NR waveforms on the left panels and the best-fitting waveforms for each model.
}
\label{fig:22_mode_waveforms}
\end{figure*}

We start by providing in Fig.~\ref{fig:22_mode_waveforms}, a proof of concept for our method to compute the best-fitting template waveform given an eccentric signal, demonstrating its effectiveness even for highly eccentric systems.
In the left column of this figure, we show the real part of the $ (2,2) $ mode for 4 nonspinning NR waveforms with increasing eccentricity (one for each panel, from top to bottom), and the corresponding best-fitting waveforms for the state-of-the-art eccentric, aligned-spin models \texttt{SEOBNRv5EHM} and \texttt{TEOBResumS-Dal\'i}, in geometric units.
For each NR waveform, we indicate the initial value of the GW eccentricity $ e _{ \text{gw}} $, the time to merger $ t _{ \text{merger}} $ (measured from the beginning of the waveform), and the $ (2,2) $-mode mismatch $ \mathcal M_{22} $ for the best-fitting waveforms of \texttt{SEOBNRv5EHM} and \texttt{TEOBResumS-Dal\'i} computed for a binary with total mass $ M = 20 \, \solarmass $.
In the left panels of the left column, we display the inspiral part of the waveforms from $ -7500 \, M $ to $ -1000 \, M $, and in the right panels we show the merger-ringdown part.
Additionally, in the right column of Fig.~\ref{fig:22_mode_waveforms}, we show the $ (2,2) $-mode phase difference between the same eccentric, nonspinning NR waveforms and the corresponding best-fitting waveforms of each model.
In this comparison, we add integer multiples of $ 2 \pi $ to the phase of the best-fitting templates to account for the extra cycles that the template waveforms have with respect to the NR waveforms, as explained in footnote~\ref{fn:extra_cycles}.
The nonspinning NR waveforms employed for these plots are:
\texttt{SXS:BBH:2529} ($ q = 2 $, $ e _{ \text{gw}} = 0.038 $, $ t _{ \text{merger}} = 8781 \,M$),
%\item
\texttt{SXS:BBH:2549} ($ q = 4 $, $ e _{ \text{gw}} = 0.442 $, $ t _{ \text{merger}} = 16287 \,M$),
%\item
\texttt{SXS:BBH:2534} ($ q = 2 $, $ e _{ \text{gw}} = 0.645 $, $  t _{ \text{merger}} = 18647 \,M$), and
%\item
\texttt{SXS:BBH:2535} ($ q = 2 $, $ e _{ \text{gw}} = 0.717$, $ t _{ \text{merger}} = 7565 \,M$).
The rest of parameters are specified in Table~\ref{tab:simulation_data} of Appendix~\ref{sec:tables} (see also footnote \ref{fn:ancillary}).

The plots in Fig.~\ref{fig:22_mode_waveforms} provide insight into the source of accuracy in the \texttt{SEOBNRv5EHM} model:
the GW phasing is precise, even for high eccentricity waveforms.
This is particularly noticeable in the second and third rows:
for these cases, the waveform length extends up to $ 16287 \, M $ and $ 18647 \, M $, respectively.
Thanks to the improved and extended analytical content in \texttt{SEOBNRv5EHM}~\cite{Gamboa:2024imd}, we obtain smaller GW dephasings compared to the \texttt{TEOBResumS-Dal\'i} model.
This is the case notwithstanding higher values of eccentricities, as testified in the last rows of Fig.~\ref{fig:22_mode_waveforms}.
We remark that the waveforms shown in these plots have the least mismatch among all the waveforms generated across the grid of eccentricities and starting frequencies for each model.

\begin{figure*}%[H]
\includegraphics[width=\linewidth]
{v4EHM_v5EHM_Dali_eccsims_alllowecc_spaghetti_22_HoM_iota_1_05}

\vspace{5pt}
\includegraphics[width=\linewidth]
{v4EHM_v5EHM_Dali_eccsims_allhighecc_spaghetti_22_HoM_iota_1_05}
\vspace{-20pt}
\caption{
\emph{Top panel:}
Mismatches of a subset of 84 eccentric NR waveforms with initial GW eccentricities $ e _{ \text{gw}} < 0.5 $ against different eccentric, aligned-spin waveform models: \texttt{SEOBNRv4EHM} (first column), \texttt{SEOBNRv5EHM} (second column), and \texttt{TEOBResumS-Dal\'i} (third column), calculated over a range of total masses $ M \in [20, 200] \,\solarmass$.
The color of each curve indicates the initial value of the GW eccentricity $ e _{ \text{gw}} $ for each NR waveform. 
The different line styles highlight the cases with the worst maximum mismatch.
For each panel, the first row shows the $(2,2)$-mode mismatches $ \mathcal M_{22} $, and the second row the sky-and-polarization averaged, SNR-weighted mismatches $\overline{\mathcal{M}}_\mr{SNR}$ for inclination $\iota _{ \text{s}} = \pi/3$.
\emph{Botttom panel:}
The same as in the top panel, but for 15 highly eccentric NR waveforms with initial GW eccentricities $ e _{ \text{gw}} > 0.5 $.
}
\label{fig:mismatch_spaghetti_all_ecc}
\end{figure*}

\begin{figure*}%[h]
\centering
% First row, two figures
\subfloat{
\includegraphics[width=0.48\linewidth]{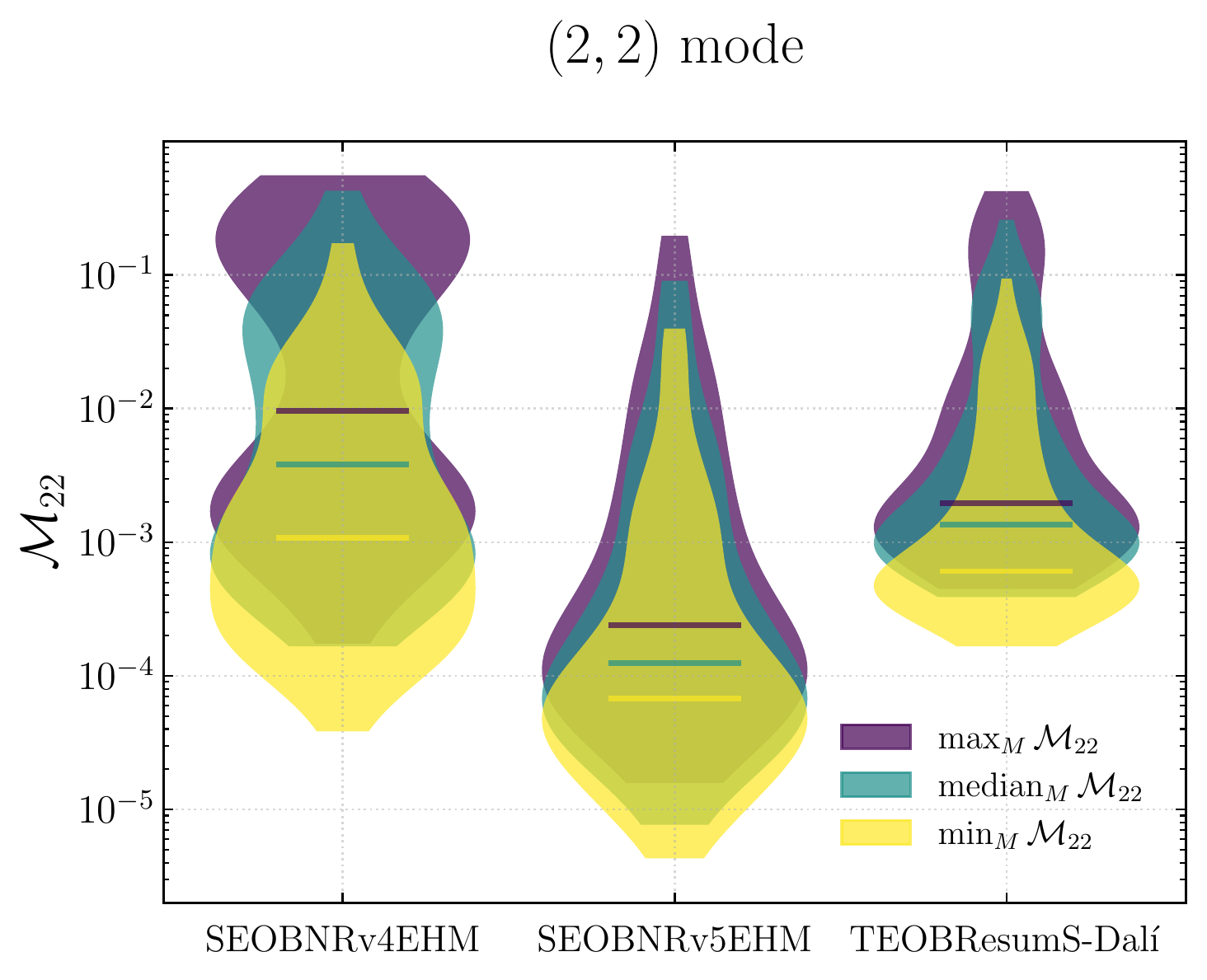}
}
\hfill
\subfloat{
\includegraphics[width=0.48\linewidth]{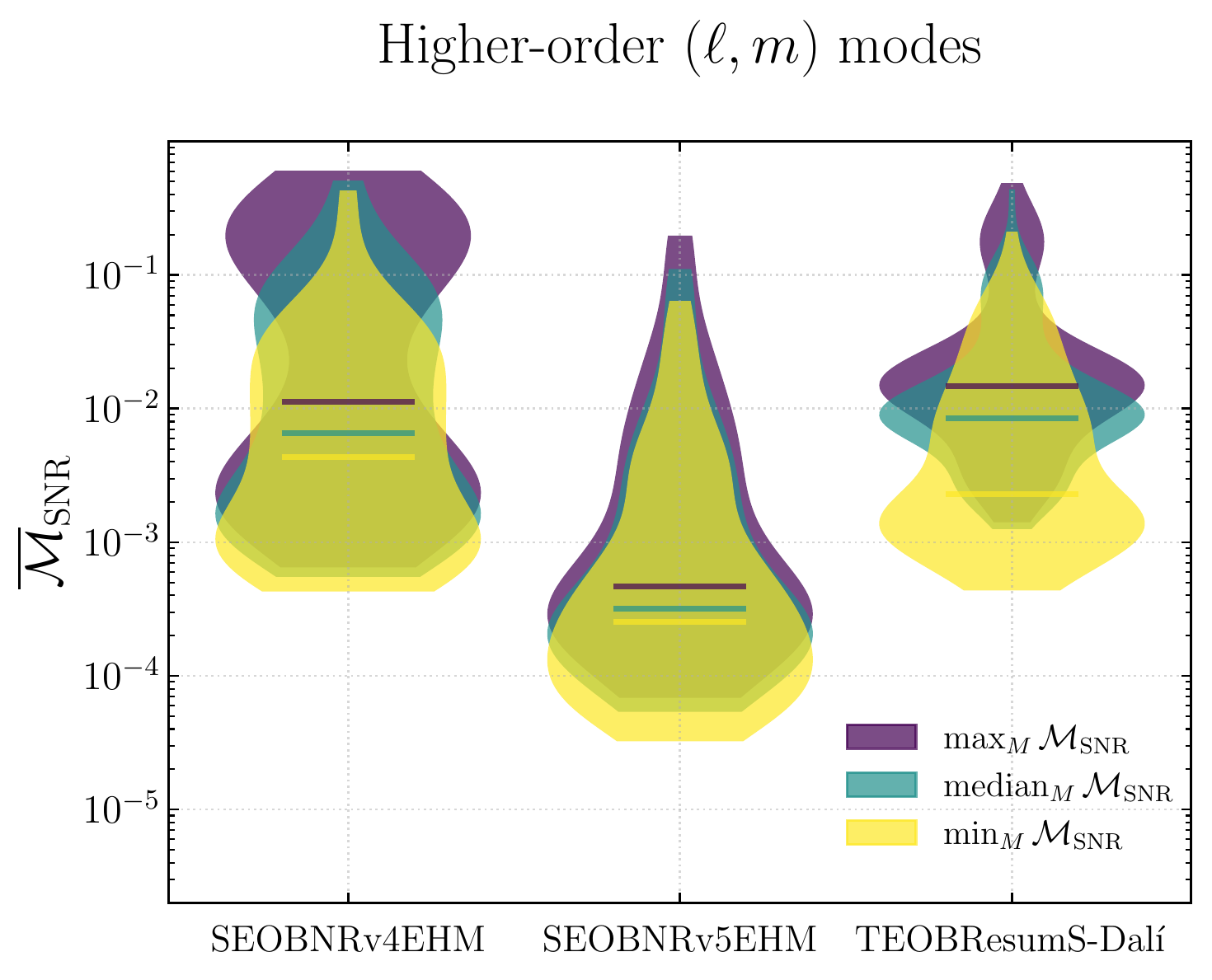}
}

% Second row, two figures
\hspace{2pt}
\subfloat{
\includegraphics[width=0.47\linewidth]{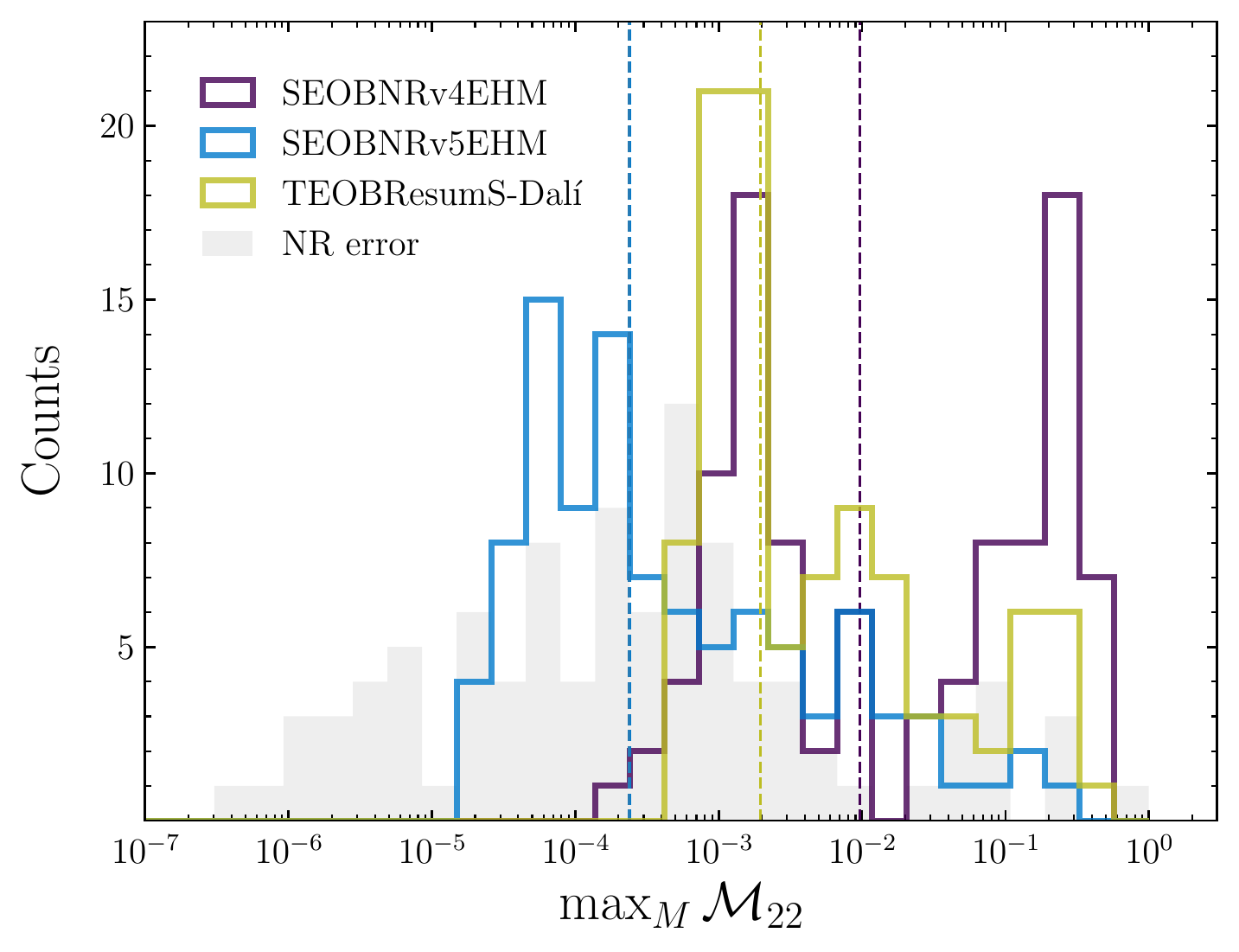}  
}
\hspace{14pt}
\subfloat{
\includegraphics[width=0.469\linewidth]{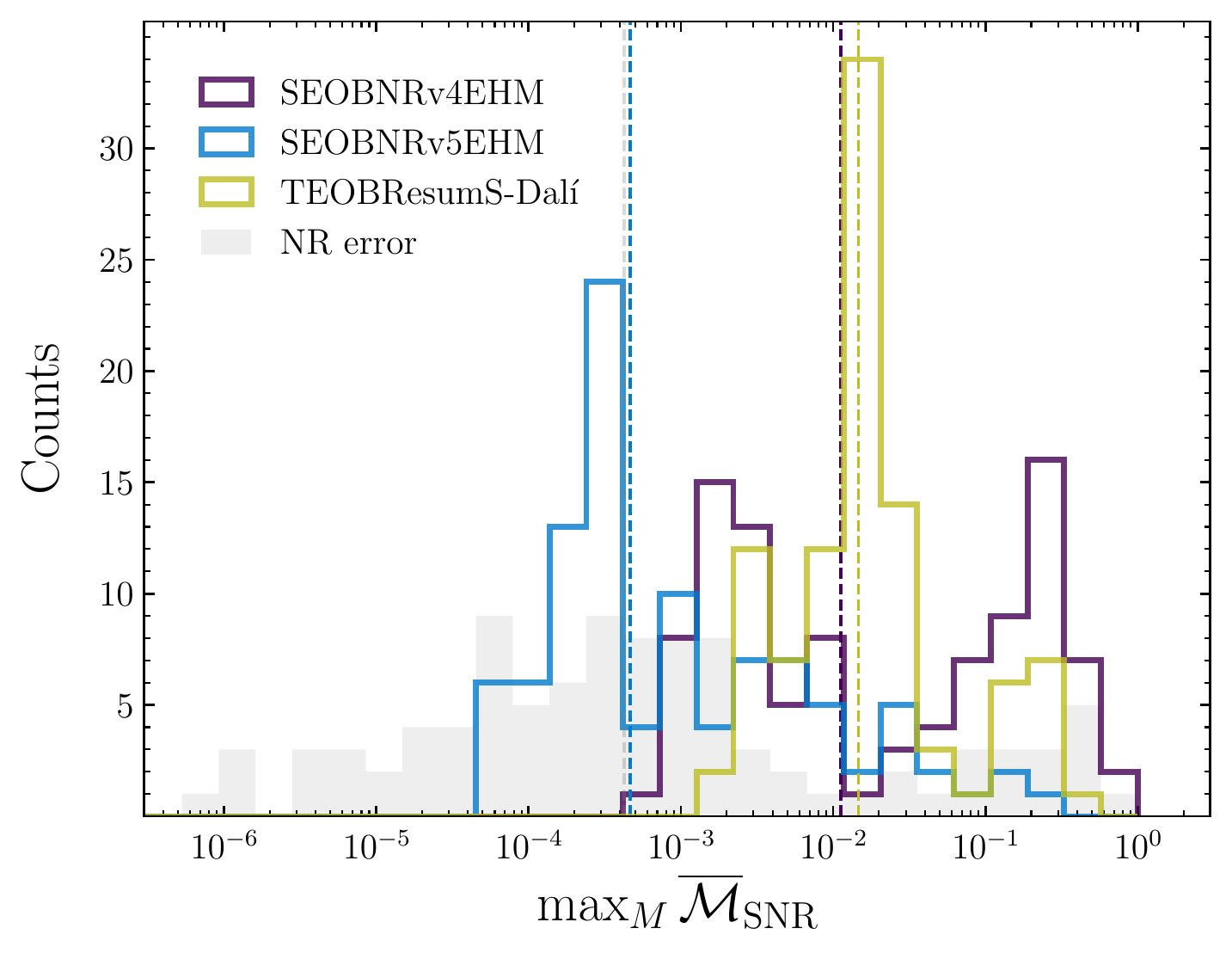}  
}
%\vspace{-5pt}
\caption{
\emph{Top panels:}
Distributions of the maximum (purple), median (green), and minimum (yellow) $(2,2)$-mode mismatches $ \mathcal M_{22} $ (left) and sky-and-polarization-averaged, SNR-weighted mismatches $ \overline{\mathcal M} _{ \text{SNR}} $ (right) over a range of total masses from $20$ and $200\, \solarmass$, for different eccentric, aligned-spin waveform models against the 99 eccentric NR waveforms employed in this work.
The horizontal lines represent the medians of the maximum (purple), median (green), and minimum (yellow) distributions.
\emph{Bottom panels:}
Histograms of the maximum $(2,2)$-mode mismatches (left) and maximum sky-and-polarization-averaged, SNR-weighted mismatches (right) for all the 99 eccentric NR waveforms against the considered models.
The NR error histograms are obtained from the waveform mismatches between NR simulations with the highest and second-highest resolutions.
The vertical dashed lines show the median of the corresponding distributions.
}
\label{fig:mismatch_distributions}
\end{figure*}

Having confirmed the reliability of our method for determining the best-fitting eccentric waveforms, we now present our results for the accuracy of different eccentric waveform models.
In particular, in the first row of the top panel of Fig.~\ref{fig:mismatch_spaghetti_all_ecc}, we show the $ (2,2) $-mode mismatches for the subset of 84 eccentric NR waveforms with initial GW eccentricities $ e _{ \text{gw}} <\nolinebreak 0.5 $, for the three eccentric, aligned-spin approximants \texttt{SEOBNRv4EHM}, \texttt{SEOBNRv5EHM}, and \texttt{TEOBResumS-Dal\'i}.
Analogous plots for the subset of 15 highly eccentric NR waveforms with initial GW eccentricities $ e _{ \text{gw}} >\nolinebreak 0.5 $ are shown in the first row of the bottom panel of Fig.~\ref{fig:mismatch_spaghetti_all_ecc}.
The mismatches are computed over a range of total masses between $20$ and $200\, \solarmass$ with the faithfulness function given in Eq.~\eqref{eq:faithfulness_ecc_22}.
In these plots, each curve quantifies the mismatch between a certain NR waveform and the corresponding best-fitting waveform produced by each of the models.
To facilitate comparisons with previous and future works, we show analogous plots for the $ 28 $ public SXS NR eccentric waveforms alone in Appendix~\ref{sec:public_ecc_NR_waveforms}.
Additionally, in Table~\ref{tab:simulation_data} of Appendix~\ref{sec:tables}, we list the optimum parameters for each model that produce the best-fitting waveform to each NR simulation.

In the left panels of Fig.~\ref{fig:mismatch_distributions}, we present plots that summarize the statistical information about the $ (2,2) $-mode mismatches for all the 99 eccentric NR simulations.
In the top-left panel, we show a plot with the distributions of the maximum, median, and minimum $ (2,2) $-mode mismatches over the same range of total masses $ M \in\nolinebreak {[20, 200]}  \solarmass $ and for the three eccentric, aligned-spin models \texttt{SEOBNRv4EHM}, \texttt{SEOBNRv5EHM}, and \texttt{TEOBResumS-Dal\'i}.
In the bottom-left panel, we show the histograms of maximum $ (2,2) $-mode mismatches, and also an \emph{estimate} of the NR error computed as the $ (2,2) $-mode mismatch between the waveforms of NR simulations with the highest and second-highest resolutions.\footnote{Comparing the waveforms of the NR simulations with the highest and second-highest resolutions gives a crude estimate of the NR error.
During this procedure, we find that the late inspiral can differ significantly between the two resolutions;
this explains the high mismatches ($ \gtrsim 10^{-2} $) for the NR errors observed in Fig.~\ref{fig:mismatch_distributions}.
We leave for the future a study of the NR errors and their impact on the quality of the waveforms.
Additionally, we note that only the simulation \texttt{SXS:BBH:0089} is omitted from the calculation of NR errors, as it has only a single resolution.
\label{fn:nr_error}}

\begin{figure}%[H]
%\centering
%\begin{subfigure}{\linewidth}
\includegraphics[width=1.03\linewidth]{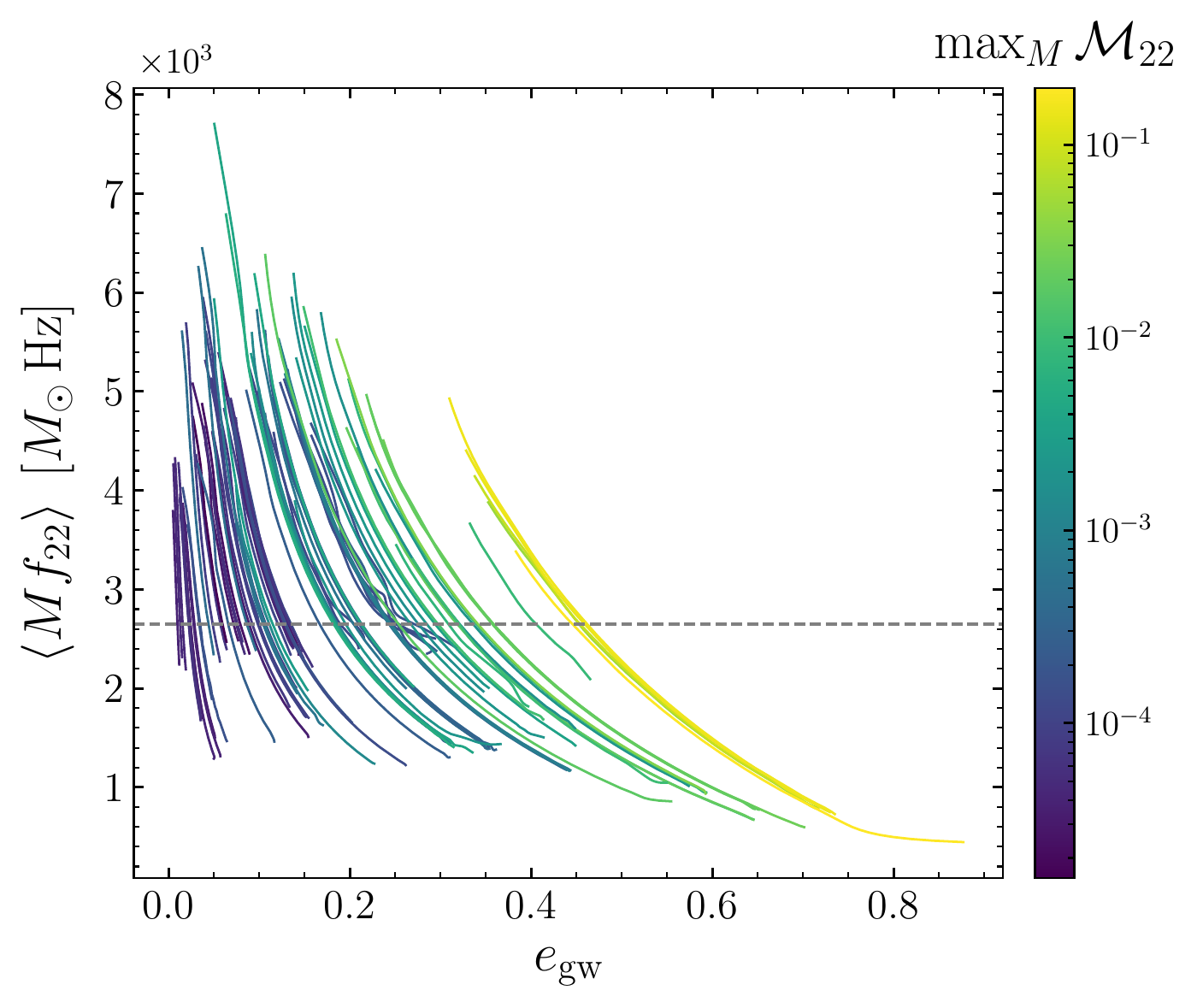}
\vspace{-20pt}
\caption{
Evolution of the mass-scaled, orbit-averaged $(2,2)$-mode frequency $ \langle M f_{22} \rangle $ as a function of the GW eccentricity $ e _{ \text{gw}} $ for the 99 eccentric NR waveforms employed in this work.
The color of each line is associated with the maximum $ (2,2) $-mode mismatch $ \mathcal M_{22} $ over an interval of total masses between $ 20 $ and $ 200 \, \solarmass$.
The dashed horizontal line indicates the value $ \langle M f_{22} \rangle = 2650 \, \solarmass \, \text{Hz} $, which is common to all the waveforms, and such that $ \langle f_{22} \rangle \approx 20 \, $Hz for a binary with total mass $ M = 132 \, \solarmass $.
}
\label{fig:mismatch_egw_frequency}
\end{figure}

An additional characterization of the parameter space of the 99 eccentric NR waveforms, including information about the $ (2,2) $-mode accuracy of \texttt{SEOBNRv5EHM}, is presented in Fig.~\ref{fig:mismatch_egw_frequency}.
Specifically, for each waveform, we show the evolution of its mass-scaled, orbit-averaged $ (2,2) $-mode frequency $ \langle M f _{ 22} \rangle $ as a function of its GW eccentricity $ e _{ \text{gw}} $,\footnote{
We note the presence of small oscillations in the curves of Fig.~\ref{fig:mismatch_egw_frequency}.
These oscillations are related to noise in the NR waveforms, which affects the interpolants used in the computation of the GW eccentricity.
}
and colored according to the value of the maximum $ (2,2) $-mode mismatch over the considered range of total masses, $ \max_M \mathcal M_{22} $, corresponding to the best-fitting \texttt{SEOBNRv5EHM} waveform.
In this way, one can estimate the physical frequencies associated with these waveforms just by rescaling with the binary's total mass.
As a reference, we place a dashed horizontal line that indicates the value $ \langle M f_{22} \rangle  = 2650 \, \solarmass \, \text{Hz}$, which is common to all the 99 eccentric NR waveforms;
for a total mass $ M = 132 \, \solarmass $, this line would correspond to $ \langle f_{22} \rangle \approx 20 \, $Hz.

Our results indicate that the overall $ (2,2) $-mode accuracy of \texttt{SEOBNRv5EHM} is about one order of magnitude better than the accuracies of \texttt{SEOBNRv4EHM} and \texttt{TEOBResumS-Dal\'i} (e.g., see the left panels in Fig.~\ref{fig:mismatch_distributions}).
In general, systems with high initial eccentricities represent the most challenging cases.
This makes sense since all the considered models employ approximations valid only for low eccentricities.
Furthermore, as with QC models, high mass ratios and high spin magnitudes also represent challenging systems to model.

We summarize our findings as follows:
\begin{itemize}
\item
For \texttt{SEOBNRv5EHM}, the bulk of $ (2,2) $-mode mismatches for initial eccentricities $ e _{ \text{gw}} \lesssim 0.3 $ lies approximately between $ 10^{-5} $ and $ 10^{-3} $.
For initial eccentricities $ 0.3 \lesssim\nolinebreak e _{ \text{gw}} \lesssim 0.5 $, \texttt{SEOBNRv5EHM} retains a good accuracy, with mismatches always below or very close to $ 10^{-2} $.
For initial eccentricities $ e _{ \text{gw}} \gtrsim 0.5 $, the mismatches lie approximately between $ 10^{-3} $ and $ 10^{-1} $, with the highest mismatches (3 cases) near the $ 20\% $ level for initial $ e _{ \text{gw}} \gtrsim 0.7 $.
Considering all the 99 eccentric NR waveforms, \texttt{SEOBNRv5EHM} achieves an overall $ (2,2) $-mode accuracy of $\sim 2 \times 10^{-4} = 0.02 \%$, as seen in the distribution for the maximum mismatches $ \max_M \mathcal M_{22} $ in the bottom left panel of Fig.~\ref{fig:mismatch_distributions}.
\item
For the previous-generation \texttt{SEOBNRv4EHM} model, the bulk of $ (2,2) $-mode mismatches for initial eccentricities $ e _{ \text{gw}} \lesssim 0.3 $ lies approximately between $ 10^{-4} $ and $ 10^{-2} $.
However, for initial eccentricities $ e _{ \text{gw}} > 0.3 $, we observe a degradation in accuracy, with mismatches going above $ 2 \times 10^{-2} $, and even reaching the $ \sim 60\% $ level for the most challenging systems.
This indicates a limitation of \texttt{SEOBNRv4EHM} for systems with moderate eccentricities, which is consistent with the fact that this model only employs eccentricity corrections in the waveform modes and not in the binary dynamics \cite{Ramos-Buades:2021adz}.
Considering all the 99 eccentric NR waveforms, \texttt{SEOBNRv4EHM} achieves an overall $ (2,2) $-mode accuracy of $\sim 10^{-2} = 1 \%$, as seen in the distribution for maximum mismatches $ \max_M \mathcal M_{22} $ in the bottom left panel of Fig.~\ref{fig:mismatch_distributions}.
\item
For the state-of-the-art \texttt{TEOBResumS-Dal\'i} model \cite{Nagar:2024dzj}, the bulk of $ (2,2) $-mode mismatches for initial eccentricities $ e _{ \text{gw}} \lesssim 0.3 $ lies approximately between $ 10^{-4} $ and $ 10^{-2} $.
For initial eccentricities $ 0.3 \lesssim\nolinebreak e _{ \text{gw}} \lesssim 0.5 $, the model has mismatches between $ 10^{-3} $ and $ 10^{-1} $.
For initial eccentricities $ e _{ \text{gw}} \gtrsim 0.5 $, the bulk of mismatches is concentrated between $ 10^{-2} $ and $ 10^{-1} $, but there are several cases with highest mismatch around or above the $ 20\% $ level for initial $ e _{ \text{gw}} \gtrsim 0.6 $.
Considering the 99 eccentric NR waveforms, \texttt{TEOBResumS-Dal\'i} achieves an overall $ (2,2) $-mode accuracy of $\sim 2 \times 10^{-3} = 0.2 \%$, as seen in the distribution for maximum mismatches $ \max_M \mathcal M_{22} $ in the bottom left panel of Fig.~\ref{fig:mismatch_distributions}, which is a factor 10 higher than for \texttt{SEOBNRv5EHM}.
\end{itemize}

We also note that there are some cases with low initial eccentricity but relatively high mismatch for the three models.
These high mismatches are associated with the quality of the eccentric NR waveforms or with an inefficient modeling of the merger-ringdown phase.
We refer the reader to Appendix~\ref{sec:public_ecc_NR_waveforms} for a discussion about these cases.

Furthermore, as a complementary study, we assess the impact of the 3PN eccentricity corrections to the modes and RR force on the model's accuracy.
In the top panel of Fig.~\ref{fig:mismatch_distributions_2PN}, we show histograms of maximum $ (2,2) $-mode mismatches for the set of 99 eccentric NR simulations employed in this work against the default \texttt{SEOBNRv5EHM} model (with 3PN corrections) and a version of \texttt{SEOBNRv5EHM} with only 2PN eccentricity corrections.
The prominent tails observed in the 2PN-version are associated with systems with high initial eccentricities ($ e _{ \text{gw}} \gtrsim 0.4 $).
The $ (2,2) $-mode mismatches as functions of the total mass are shown in Appendix~\ref{sec:mms_2PN}.

\begin{figure}%[h]
\centering
% First row, two figures
\subfloat{
\includegraphics[width=1\linewidth]{v5EHM_2PN_eccsims_allecc_histogram_22}
}

\vspace{-10pt}
\subfloat{
\includegraphics[width=1\linewidth]{v5EHM_2PN_eccsims_allecc_histogram_HoM}  
}
%\vspace{-5pt}
\caption{
Summary of mismatches for the 99 eccentric NR waveforms employed in this work, calculated for the \emph{default} \texttt{SEOBNRv5EHM} model (blue) with 3PN eccentricity corrections in the RR force and gravitational waveform modes, and also for \texttt{SEOBNRv5EHM} but with only 2PN eccentricity corrections (green).
\emph{Top panel:}
Histograms of the maximum $(2,2)$-mode mismatches over a range of total masses $ M \in [20, 200]\, \solarmass$.
\emph{Bottom panel}:
Histograms of the maximum sky-and-polarization-averaged, SNR-weighted mismatches over the same range of total masses.
The vertical dashed lines show the median of the corresponding distributions.
}
\label{fig:mismatch_distributions_2PN}
\end{figure}

Our results indicate that the 3PN eccentricity corrections are crucial to avoid high mismatches for high eccentricity waveforms ($ e _{ \text{gw}} \gtrsim 0.4 $).
This is expected since these configurations are associated with higher velocities close to periastron.
However, the 3PN eccentricity corrections do not significantly change the model's accuracy for low eccentricities, compared to the 2PN eccentricity corrections. 
As a complementary study, in Appendix~\ref{sec:forces_modes} we analyze the impact on the model's accuracy of using the same QC expressions for the RR force and/or waveform modes as in \texttt{SEOBNRv5HM}.
Overall, our results suggest that a significant source of improvement of \texttt{SEOBNRv5EHM} with respect to \texttt{SEOBNRv4EHM} and \texttt{TEOBResumS-Dal\'i} could be the accurate calibration to QC NR simulations inherited from the \texttt{SEOBNRv5HM} model, as well as the 3PN eccentricity corrections to the RR force.
Further work is needed to understand the impact on the model's performance of other ingredients (e.g., parametrization and factorization of the eccentricity corrections, or the initial conditions prescription).

Finally, even though \texttt{SEOBNRv5EHM} has the best accuracy also for high eccentricity waveforms (as seen, e.g.,~in Figs.~\ref{fig:22_mode_waveforms} and \ref{fig:mismatch_spaghetti_all_ecc}), the mismatches are around the $ 20\% $ value for the most extreme configurations ($ e _{ \text{gw}} \gtrsim 0.7$).
Therefore, our results indicate that eccentric waveform models still require further development to accurately treat these extreme systems. 
Statements about highly eccentric waveforms produced with these models should be taken with caution.

%%%%%%%%%%%%%%%%%%%%%%%%%%%
%%%%%%%%%%%%%%%%%%%%%%%%%%%
%%%%%%%%%%%%%%%%%%%%%%%%%%%

\subsubsection{Accuracy of the eccentric higher-order modes}

We quantify the accuracy of eccentric waveforms with HMs in terms of the sky-and-polarization-averaged, SNR-weighted faithfulness functions given in Eqs.~\eqref{eq:faithfulness_ecc_hom} and \eqref{eq:faithfulness_avg_snr}.
The corresponding mismatches, for an inclination angle $ \iota _{ \text{s}} = \pi/3 $, are shown in the second row of the top panel of Fig.~\ref{fig:mismatch_spaghetti_all_ecc} (for the subset of 84 eccentric NR waveforms with initial $ e _{ \text{gw}} < 0.5 $) and in the second row of the bottom panel of Fig.~\ref{fig:mismatch_spaghetti_all_ecc} (for the subset of 15 eccentric NR waveforms with initial $ e _{ \text{gw}} > 0.5 $).
The waveforms are constructed with all modes available for each model up to $ \ell = 4 $:
for \texttt{SEOBNRv5EHM}, we include the modes $( \ell, |m| ) = \{(2, 2),\, (3, 3),\, (2, 1),\, (4, 4),\, (3, 2),\, (4, 3)\}$, and for \texttt{SEOBNRv4EHM} and \texttt{TEOBResumS-Dal\'i}, we include the modes $( \ell, |m| ) = \{(2, 2),\, (3, 3),\, (2, 1),\, (4, 4)\}$.
The NR waveforms include all modes up to $ \ell = 4 $ with $ |m| > 0 $.

\begin{figure}%[H]
\includegraphics[width=\linewidth]{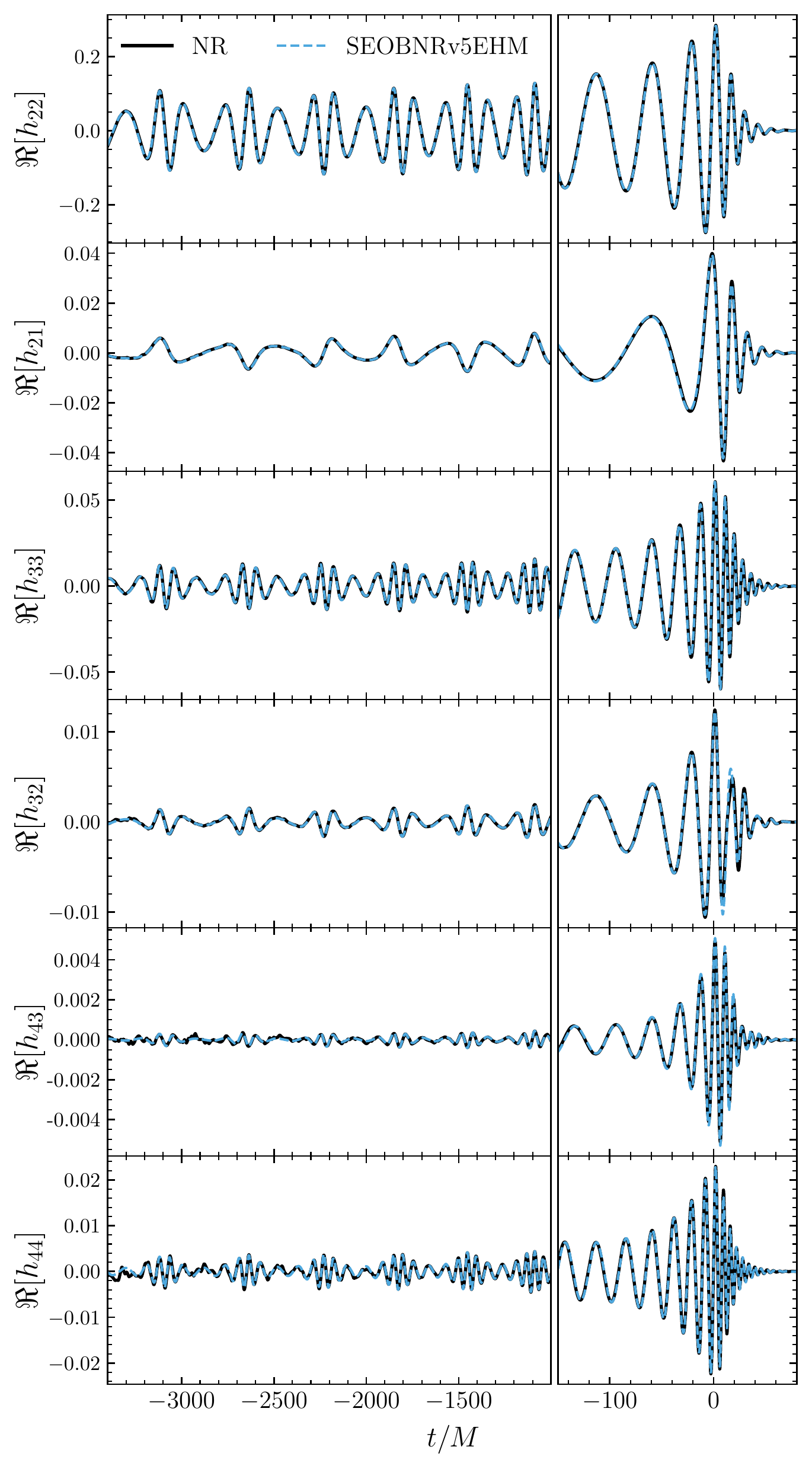}
\vspace{-20pt}
\caption{
Real part of different waveform modes for the nonspinning NR simulation \texttt{SXS:BBH:1374} with mass ratio $ q  =\nolinebreak 3 $ (black) and the best-fitting \texttt{SEOBNRv5EHM} waveform (blue), in geometric units.
The optimum values of the \texttt{SEOBNRv5EHM} waveform are $ e = 0.299 $ and $ \zeta = \pi $ at $\langle M \Omega \rangle = 0.0157$. 
The left panels show the inspiral part of the waveform, while the right panels focus on the merger-ringdown.
}
\label{fig:all_modes_NR_sim}
\end{figure}

As in the case of the $ (2,2) $-mode analysis, in the right panels of Fig.~\ref{fig:mismatch_distributions}, we present a summary of the statistical information about the sky-and-polarization-averaged, SNR-weighted mismatches $ \overline{\mathcal M} _{ \text{SNR}} $ for all the 99 eccentric NR simulations.
In the top-right panel, we show a plot with the distributions of the maximum, median, and minimum $ \overline{\mathcal M} _{ \text{SNR}} $ mismatches over the total masses range $ M \in [20, 200]\,\solarmass $ and for the three models \texttt{SEOBNRv4EHM}, \texttt{SEOBNRv5EHM}, and \texttt{TEOBResumS-Dal\'i}.
In the bottom right panel, we focus on the histogram of maximum $ \overline{\mathcal M} _{ \text{SNR}} $ mismatches, where we also show an estimate of the NR error computed as the mismatch between the waveforms (with HMs) of NR simulations with the highest and second-highest resolutions (see footnote \ref{fn:nr_error}).

The results for the mismatches of eccentric waveforms with HMs follow the same trend as for the $ (2,2) $-mode mismatches.
Namely, we see an overall improvement of about one order of magnitude for \texttt{SEOBNRv5EHM} with respect to \texttt{SEOBNRv4EHM} and \texttt{TEOBResumS-Dal\'i}, and we find again that the 3PN eccentricity corrections to the HMs and RR force are crucial to obtain better accuracy for high eccentricity systems ($ e _{ \text{gw}} \gtrsim 0.4 $).
This can be seen in the bottom panel of Fig.~\ref{fig:mismatch_distributions_2PN}, where we show histograms of maximum sky-and-polarization-averaged, SNR-weighted mismatches for the set of 99 eccentric NR simulations against the default \texttt{SEOBNRv5EHM} model (with 3PN corrections) and its 2PN version.

We additionally observe an increase in the mismatch with respect to the total mass for all models, as in the QC case (see, e.g., Fig.~\ref{fig:qc_mismatch_spaghetti_22} above, Fig. 7 of Ref.~\cite{Pompiliv5}, or Fig. 6 of Ref.~\cite{Ramos-Buades:2021adz}; the latter two figures were produced with Advanced LIGO PSD \cite{Barsotti:2018} and all NR modes up to $ \ell = 5 $).
This behavior reflects the fact that HMs are less well-modeled compared to the $ (2,2) $ mode.
We note that this degradation in accuracy is more marked in the \texttt{TEOBResumS-Dal\'i} model, as can be seen in the top-right panel of Fig.~\ref{fig:mismatch_distributions}, indicating that the HMs for some configurations are more challenging to model within the \texttt{TEOBResumS-Dal\'i} approximant.

As an illustrative example, in Fig.~\ref{fig:all_modes_NR_sim} we show the waveform modes for one NR simulation (\texttt{SXS:BBH:1374}) together with the corresponding best-fitting \texttt{SEOBNRv5EHM} waveform modes.
The rounded optimum values that produce the \texttt{SEOBNRv5EHM} waveform are $ e = 0.299 $ and $ \zeta = \pi $ at $\langle M \Omega \rangle = 0.0157$ (these values are given with full digits in the ancillary file to this work; see footnote \ref{fn:ancillary}).
In this figure, we observe a very good agreement between NR and \texttt{SEOBNRv5EHM} modes for a big portion of the waveform.
Only during the merger-ringdown phase do some HMs, such as the $ (3,2) $ mode, show a noticeable difference from the NR waveform.
This makes sense since \texttt{SEOBNRv5EHM} does not include eccentricity effects in the merger-ringdown phase, as is the case for all currently available eccentric models.

Combining the above results for eccentric systems with the very good performance of the model in the QC limit, we conclude that \texttt{SEOBNRv5EHM} is the most accurate eccentric waveform model currently available.

%%%%%%%%%%%%%%%%%%%%%%%%%%%
%%%%%%%%%%%%%%%%%%%%%%%%%%%

\subsection{Computational performance}
\label{sec:performance}

Typical applications of waveform models require the generation of millions of samples over different regions of parameter space.
Therefore, computational efficiency becomes an important and necessary property for any waveform model.

The new eccentric, aligned-spin \texttt{SEOBNRv5EHM} waveform model is implemented in Python within the high-performance and flexible \texttt{pySEOBNR} package \cite{Mihaylovv5}.
This flexibility allows us to develop an efficient parser code that we employ to translate the new mathematical expressions in Eqs.~\eqref{eq:EOMprStr}-\eqref{eq:eob_modes} into pure \texttt{Cython} code.
In this way, we obtain optimized, easier-to-evaluate versions of the long, 3PN accurate, expressions containing the eccentricity contributions to the RR force and modes, as well as efficient expressions for the new evolution equations for the Keplerian parameters.

\begin{figure}%[H]
\includegraphics[width=\linewidth]{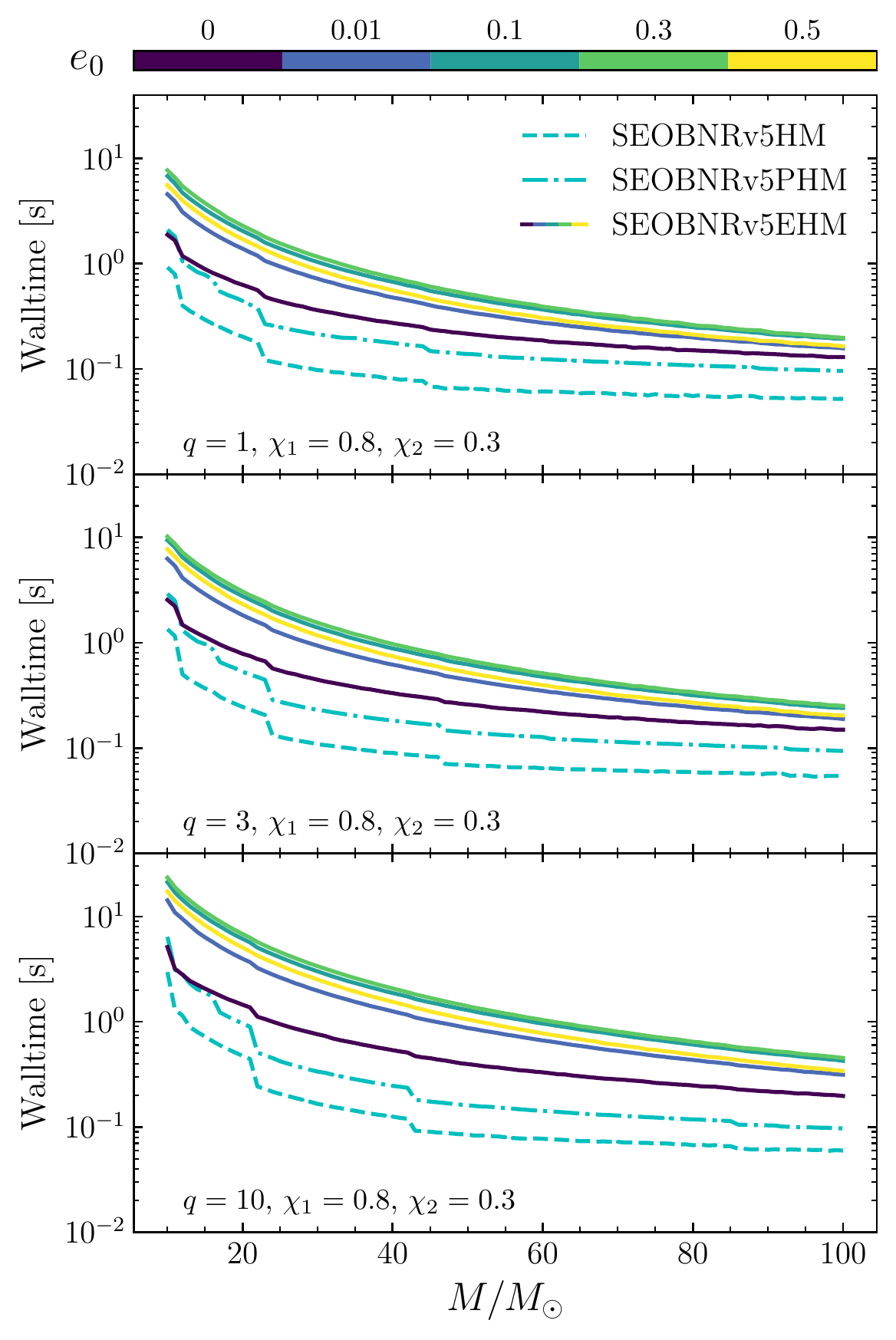}
\vspace{-20pt}
\caption{
Walltimes  for the models \texttt{SEOBNRv5HM} (dashed), \texttt{SEOBNRv5PHM} (dash-dotted), and \texttt{SEOBNRv5EHM} (solid), as functions of the total mass $ M \in [10, \, 100]\s \solarmass$.
The systems are characterized by a starting orbit-averaged frequency $ \langle f _{ \text{start}} \rangle = 10\, $Hz, dimensionless spin components $ \chi_{1} = 0.8 $ and $ \chi_{2} = 0.3 $, and three different mass ratios $ q \in \{1, 3, 10 \} $.
For \texttt{SEOBNRv5EHM}, the systems are initialized at apastron ($ \zeta_0 = \pi $) with different starting eccentricities $ e_0 \in \{0, 0.01, 0.1, 0.3, 0.5\} $, each represented by a different color.
%For all cases, the sampling rate is chosen such that the Nyquist criterion is satisfied for the $ \ell = 4 $ multipoles.
}
\label{fig:benchmarks}
\end{figure}

To assess the computational speed of \texttt{SEOBNRv5EHM}, we calculate the wall times required for waveform generation (time-domain polarizations) for various binary configurations typically expected for current LVK detectors, and we compare the results against the other models of the \texttt{SEOBNRv5} family.
Specifically, in Fig.~\ref{fig:benchmarks}, we show the waveform evaluation wall time as a function of the binary's total mass $ M \in [10, \, 100]\s \solarmass$ for three mass ratios $ q \in \{ 1, 3, 10\} $, and for three waveform approximants:
the QC aligned-spin \texttt{SEOBNRv5HM} model, the QC precessing-spin \texttt{SEOBNRv5PHM} model, and the eccentric aligned-spin \texttt{SEOBNRv5EHM} model.
The considered systems correspond to aligned-spin BBHs with dimensionless spin components $ \chi_{1} = 0.8 $ and $ \chi_{2} = 0.3 $, starting at an orbit-averaged frequency $ \langle f _{ \text{start}} \rangle = 10\, $Hz.
For \texttt{SEOBNRv5EHM}, the binaries are initialized with different eccentricities $ e_0 \in\nolinebreak \{0, 0.01, 0.1, 0.3, 0.5 \} $ at apastron, $ \zeta_0 = \pi $.
For all the models, we include all the modes up to $ \ell = 4 $, and choose a sampling rate such that the Nyquist criterion is satisfied for the $ \ell = 4 $ multipoles;
the rest of the settings are kept as default.
The benchmarks were performed with the \texttt{Hypatia} computer cluster at the Max Planck Institute for Gravitational Physics in Potsdam, on a compute node equipped with a dual-socket 64-core AMD EPYC (Rome) 7742 CPU.

In the zero eccentricity limit, we see from Fig.~\ref{fig:benchmarks} that \texttt{SEOBNRv5EHM} is half as fast as \texttt{SEOBNRv5HM} for total masses close to $ 10 \, \solarmass $.
The reason for this is that \texttt{SEOBNRv5EHM} evolves a background QC dynamics, which, as explained in Sec.~\ref{sec:model_waveform}, is employed i) to compute the NQC corrections needed to improve the agreement with NR waveforms toward merger, and ii) to infer the attachment time of the merger-ringdown phase.
Hence, it makes sense that the speed of \texttt{SEOBNRv5EHM} in the QC limit is at least two times the speed of \texttt{SEOBNRv5HM} for binaries with the same intrinsic parameters.

However, the difference in speed between \texttt{SEOBNRv5HM} and \texttt{SEOBNRv5EHM} increases considerably as the total mass increases due to the Nyquist criterion.
Selecting the sampling rate in accordance with this criterion explains the discontinuities in the curves shown in Fig.~\ref{fig:benchmarks}, since smaller sampling rates are associated with faster waveform generations (there are fewer points in the waveforms to resolve).
However, for \texttt{SEOBNRv5EHM}, the change in the sampling rate does not have a significant impact since the major cost, in the zero eccentricity limit, comes from the background QC evolution.
In fact, repeating the benchmarks shown in Fig.~\ref{fig:benchmarks} but without the application of the Nyquist criterion (thus, using a fixed sampling rate for all cases) leads to the result that \texttt{SEOBNRv5EHM} is about half as fast as \texttt{SEOBNRv5HM} for all the total masses.
Nevertheless, the results obtained using the Nyquist criterion are more representative of inference applications.

We also observe in Fig.~\ref{fig:benchmarks} an increase of evaluation times in \texttt{SEOBNRv5EHM} as the eccentricity is augmented, until a certain point at which a further increase of eccentricity decreases the evaluation time.
To explain this behavior, we remember that, at a fixed starting orbit-averaged frequency, an increment of eccentricity is associated with a decrease in waveform length.
This is because high eccentricity systems lose more energy and angular momentum at each periastron passage.
From this reasoning alone, we would expect a decrease in waveform evaluation time as the eccentricity increases.
However, evaluating Eqs.~\eqref{eq:EOMprStr}--\eqref{eq:eob_modes} at each time during the numerical integration of the EOM for a nonzero eccentricity is an expensive process since the integrator needs to resolve the orbital time scales associated to eccentric modulations.
Therefore, there is a competition between the numerical integrator resolving the eccentricity timescales and the waveforms becoming shorter as the eccentricity increases.
This explains the increased evaluation times of \texttt{SEOBNRv5EHM} for systems with fixed starting frequency and $ e > 0 $, compared to systems with $ e = 0 $ for which the eccentricity contributions completely disappear from the binary's EOM by construction.

We also note that systems with $ e > 0 $ have smooth curves in Fig.~\ref{fig:benchmarks}, without large discontinuities.
This means that changing the sampling rate in accordance with the Nyquist criterion does not significantly affect systems with $ e > 0 $, since the numerical integration of the eccentric EOM (and the numerical integration of the background QC dynamics) dominates the computational time of the waveform generation.

Another observation from Fig.~\ref{fig:benchmarks} is that the eccentric, aligned-spin \texttt{SEOBNRv5EHM} model has a computational speed $ \sim 2 - 10 $ times the one from the QC, precessing-spin \texttt{SEOBNRv5PHM} model \cite{RamosBuadesv5}, depending on the mass ratio and the value of eccentricity.
This indicates that \texttt{SEOBNRv5EHM} can be integrated into commonly used software for GW analysis under relatively similar settings as \texttt{SEOBNRv5PHM}.
As a particular case, we provide details about the application of \texttt{SEOBNRv5EHM} to Bayesian parameter-estimation studies in Sec.~\ref{sec:model_applications}.
There, in Table~\ref{tab:peruntime}, we show the settings and evaluation times of parameter-estimation runs for two LVK GW events for different waveform approximants, including \texttt{SEOBNRv5EHM} and \texttt{SEOBNRv5PHM}.

\begin{figure*}%[ht]
\includegraphics[width=\textwidth]
{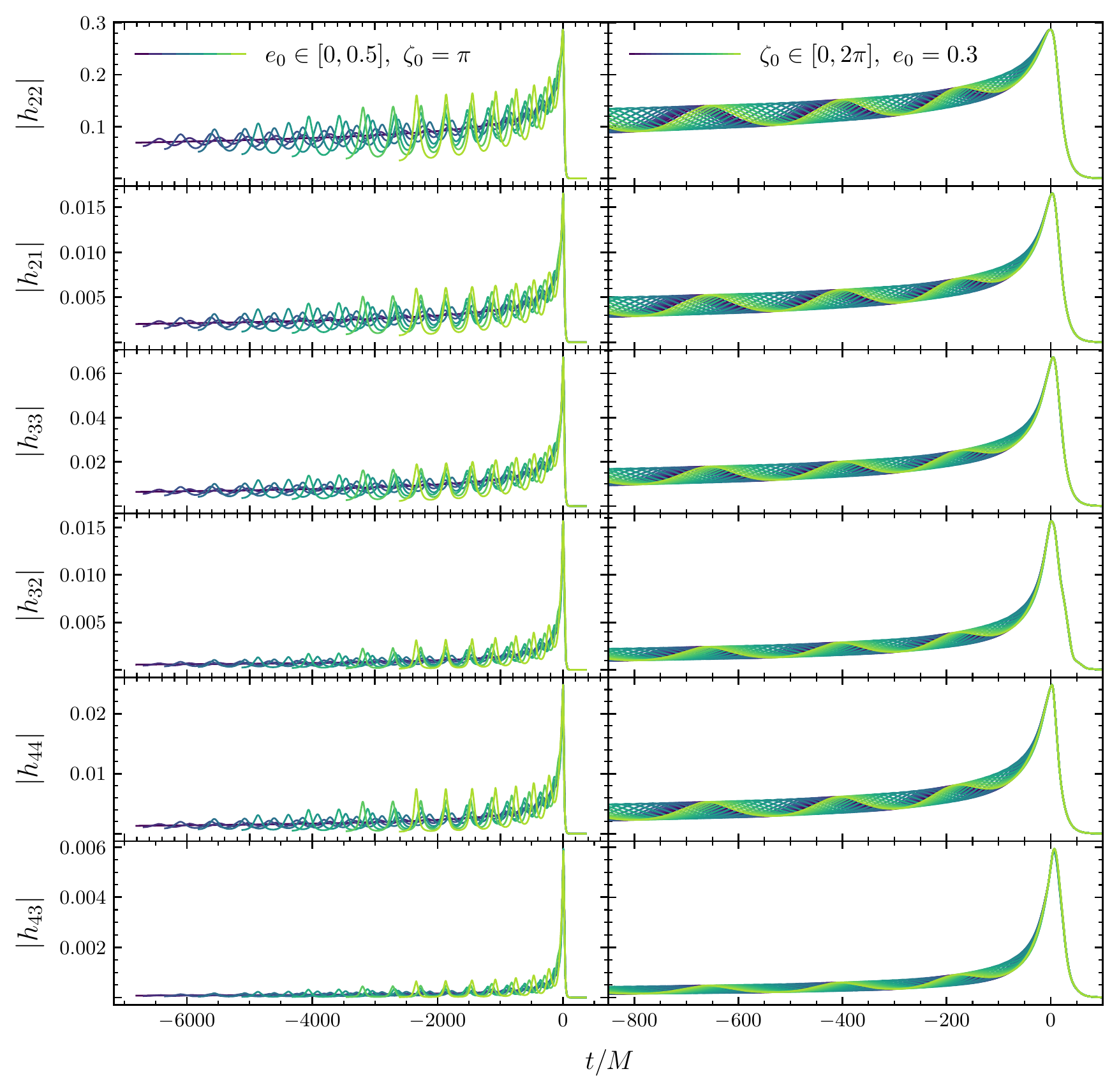}
\vspace{-20pt}
\caption{
Uniform variation of eccentricity (left panels) and relativistic anomaly (right panels) for the different \texttt{SEOBNRv5EHM} waveform modes, in geometric units.
Each of these systems is characterized by mass ratio $ q = 3 $, dimensionless spin components $ \chi_{1} = 0.5 $ and $ \chi_{2} = -0.5 $, and starting frequency $ \langle f _{ \text{start}} \rangle = 20\, $Hz for a total mass $ M = 50 \, \solarmass $.
The systems in the left panels are initialized at apastron ($ \zeta_0 = \pi $), and the eccentricity is varied uniformly from $ e_0 = 0 $ (dark blue) to $ e_0 = 0.5 $ (yellow).
The systems in the right panels are initialized with eccentricity $ e_0 = 0.3 $, and the relativistic anomaly is varied uniformly from $ \zeta_0 = 0 $ (dark blue) to $ \zeta_0 = 2 \pi $ (yellow).
In the right panels, we show the end of the inspirals to make more visible the predicted behavior with respect to the relativistic anomaly variation.
}
\label{fig:smoothness_e_z}
\end{figure*}

Finally, we note that we are not including in Fig.~\ref{fig:benchmarks} the wall times for other eccentric models.
An appropriate comparison would require calculating evaluation times for the same physical system, which is complicated due to the different definitions of eccentricity.
In practice, however, one is interested in the time scales required for computationally expensive analyses, such as parameter inference.
Thus, in Table~\ref{tab:peruntime}, we also show the run-times of the \texttt{SEOBNRv4EHM\_opt} model~\cite{Ramos-Buades:2023yhy} (which is an optimized version of \texttt{SEOBNRv4EHM} \cite{Ramos-Buades:2021adz}), and we observe that \texttt{SEOBNRv5EHM} is faster than its predecessor.

%%%%%%%%%%%%%%%%%%%%%%%%%%%
%%%%%%%%%%%%%%%%%%%%%%%%%%%

\subsection{Robustness across parameter space}
\label{sec:robustness}

Assessing the robustness of any waveform model across parameter space is an important process required to get confidence in its predictions.
Here, we demonstrate the robustness of \texttt{SEOBNRv5EHM} by i) identifying the region in parameter space where the model generates physically sensible eccentric waveforms, and ii) by testing the waveforms' behavior under variations of the eccentric input parameters.

For QC models, the determination of a valid region of parameter space is simplified by the fact that QC binaries evolve from weak-field to strong-field configurations.
In contrast, eccentric binaries could continuously switch between weak-field and strong-field configurations, depending on the values of the eccentricity, frequency, and radial anomaly.
This poses a challenge to eccentric models since various underlying approximations become inaccurate for strong-field, high-velocity regimes, such as those encountered near each periastron passage of highly eccentric binaries.

Determining EOB initial conditions for binaries in strong-field, high-velocity configurations can be particularly challenging.
To relate the input values of eccentricity $ e $, radial anomaly $ \zeta $, and starting frequency $ x = \langle M \Omega \rangle ^{2/3} $ to the initial values of the EOB variables, we use PN expressions valid under the postadiabatic approximation (see Sec.~\ref{sec:ICs}).
Thus, trying to initialize a binary in a strong-field, high-velocity state could lead to a different solution that does not physically correspond to the given input values.
To overcome this issue, in \texttt{SEOBNRv5EHM} we employ a set of secular-evolution equations for $ (e, \zeta, x) $, given by Eqs.~\eqref{eq:e}, \eqref{eq:zeta}, and \eqref{eq:dotx}, which is evolved backward in time whenever the starting EOB separation of the binary is less than $ 10 M $, as described in Sec.~\ref{sec:ICs}.
This way, we obtain past values of $ (e, \zeta, x) $ that correspond to the same physical system but in an earlier configuration, which allows an accurate determination of EOB initial conditions.
Therefore, the inclusion of the set of secular evolution equations \eqref{eq:e}, \eqref{eq:zeta}, and \eqref{eq:dotx} allows us to extend the parameter space of \texttt{SEOBNRv5EHM} toward higher eccentricities and total masses for all values of the relativistic anomaly $ \zeta \in [0, 2 \pi] $.

After testing the robustness of waveform generation and visually inspecting sample waveforms across different regions of parameter space, we recommend that the eccentric, aligned-spin waveform model \texttt{SEOBNRv5EHM} is used in the parameter space determined by
mass ratios $ q \in [1, 20] $,
dimensionless spin components $ \chi_{1, \, 2} \in [-0.999, 0.999] $,
eccentricities $ e \in [0, 0.45] $,
and
relativistic anomalies $ \zeta \in [0, 2 \pi] $,
for total masses $ \geq 10 \, \solarmass $
at $ \langle f _{ \text{start}} \rangle = 20\,$Hz.
This represents a safe (conservative) region of parameter space where the generated waveforms are physically sensible.

We note that going outside this conservative region of parameter space could cause the appearance of unphysical features in the waveforms.
In particular, the coupled set of EOB plus the Keplerian evolution equations \eqref{eq:EOMprStr} employed in \texttt{SEOBNRv5EHM} is overdetermined (seven equations for four degrees of freedom), and it can be desynchronized if the binary sustains its dynamics under conditions of high velocities, decreasing the accuracy of the PN approximation.
This is the case for \emph{some} binaries with very high eccentricities $ (e \gtrsim 0.7) $, or with moderate eccentricities $ (e \gtrsim 0.5) $ that evolve during many cycles (e.g., systems with total mass $ \sim 5 \, \solarmass $, systems with mass ratios $ q \gg 20 $, etc.).
Unfortunately, characterizing the problematic region of parameter space is not an easy task due to the complexity of eccentric systems.
Hence, the safest option is to employ the model in the conservative part of the parameter space where there are no detected issues.
We refer the reader to Appendix \ref{sec:desynchronization}, where we discuss in detail this desynchronization issue; in particular, we show some examples and explain the origin of the problem. 

However, note that \texttt{SEOBNRv5EHM} \emph{can} be applied outside this conservative region --- 
for example, at high eccentricities, \texttt{SEOBNRv5EHM} still keeps the best accuracy among the models considered in this work (see the bottom panel of Fig.~\ref{fig:mismatch_spaghetti_all_ecc}).
In general, going outside the conservative region of parameter space should be fine, but the model must be used with caution, e.g., by visually inspecting some sample waveforms.

As an example, we show in Fig.~\ref{fig:smoothness_e_z} the smooth behavior of the amplitude of all the \texttt{SEOBNRv5EHM} waveform modes, $( \ell, |m| ) = \{(2, 2),\, (3, 3),\, (2, 1),\, (4, 4),\, (3, 2),\, (4, 3)\}$, under uniform variations of the eccentric parameters $ e \in [0, \s 0.5] $ (left panels) and $ \zeta \in [0, \s 2\pi] $ (right panels) at a fixed starting frequency $ \langle f _{ \text{start}} \rangle = 20 \, \text{Hz} $, for a system with total mass $ M = 50 \, \solarmass $ and dimensionless spin components $ \chi_{1} = 0.5 $ and $ \chi_{2} = -0.5 $.
We note that all systems have the same behavior at merger for each of the waveform modes.
This is a consequence of the QC merger-ringdown assumption employed in \texttt{SEOBNRv5EHM}, which should be valid as long as the binaries have enough time to circularize their orbit.
Additionally, one can observe in the right panels (although very slightly) the mode-mixing in the merger-ringdown of the $ (3,2) $ and $ (4,3) $ modes.

The results discussed in this section form part of the many tests and consistency checks performed during the successful review of the eccentric \texttt{SEOBNRv5EHM} waveform model by the LVK Collaboration.

%%%%%%%%%%%%%%%%%%%%%%%%%%%
%%%%%%%%%%%%%%%%%%%%%%%%%%%
%%%%%%%%%%%%%%%%%%%%%%%%%%%

\section{Applications to Bayesian inference}
\label{sec:model_applications}

One of the main applications of waveform models is the Bayesian inference of source parameters from the observed GWs.
In this section, we demonstrate the applicability of the eccentric, aligned-spin \texttt{SEOBNRv5EHM} waveform model by performing a set of parameter-estimation studies.
First, we introduce the methods and codes used for parameter estimation.
Then, we investigate the model's accuracy by performing a series of synthetic NR signal injections into zero noise.\footnote{
This represents an idealized scenario in which the background data is entirely free from noise or environmental disturbances.
}
Finally, we analyze two real GW events detected by the LVK Collaboration (GW150914 and GW190521), and we compare our results with those from the literature.

%%%%%%%%%%%%%%%%%%%%%%%%%%%
%%%%%%%%%%%%%%%%%%%%%%%%%%%

\subsection{Methodology for parameter estimation}\label{sec:PEsetup}

For the Bayesian inference studies presented here, we employ a very similar setup as in Ref.~\cite{Ramos-Buades:2023yhy}.
The runs are performed using the Bayesian inference Python packages  \texttt{parallel} {\tt Bilby} \cite{Smith:2019ucc},
and \texttt{Bilby} \cite{Ashton:2018jfp,Romero-Shaw:2020owr} called henceforth {\tt serial} {\tt  Bilby}.\footnote{
We employ the \texttt{parallel} {\tt Bilby} code from the public repository \url{https://git.ligo.org/lscsoft/parallel\_bilby} with git hash \texttt{b56d25b87b3b33b33a91a8410ae3a6c2a5c92a2e}, which corresponds to the version \texttt{2.0.2}, while for \texttt{Bilby} we use the code from the repository \url{https://git.ligo.org/lscsoft/parallel\_bilby} with git hash \texttt{bad30b7cfd732d16b62322e45db5bb5f9ebc4f3c.}}
Both codes incorporate the nested sampler \texttt{dynesty} \cite{Speagle:2019ivv}. 
Based on previous experience with \texttt{parallel/serial} {\tt Bilby} \cite{RamosBuadesv5,Ramos-Buades:2023yhy}, we use a number of autocorrelation times $\mathrm{nact}=30$ and a number of live points $\mathrm{nlive}=2048$ for \texttt{parallel} {\tt Bilby}, while we use a number of accepted steps  $\mathrm{naccept}=60$ and a number of live points $\mathrm{nlive}=1000$ for {\tt Bilby}; the remaining sampling parameters are set to their default values, unless otherwise specified.
Additionally, we employ distance marginalization as implemented in \texttt{Bilby}.
If the model is restricted to the $ (\ell,|m|)=(2,2) $ mode, we activate the phase marginalization option to further reduce the computational cost.

For the choice of priors, we follow broadly Refs.~\cite{Veitch:2014wba,LIGOScientific:2018mvr,RamosBuadesv5,Ramos-Buades:2023yhy}.
We choose priors on inverse mass ratio $1/q$ and chirp mass $\mathcal{M}$ such that we have a uniform prior on component masses.
The priors on initial eccentricity $e_0$ and relativistic anomaly $\zeta_0 \in [0,2\pi ]$ are chosen to be uniform.
To facilitate the comparison with QC, precessing-spin results, the priors on the dimensionless spin components $ \chi_{1} $ and $ \chi_{2} $ are chosen such that they correspond to the projections of a uniform and isotropic spin distribution along a perpendicular direction to the binary's orbital plane \cite{Veitch:2014wba}.
The prior on the luminosity distance $d_L$ is chosen to be proportional to $d_L^2$, unless otherwise specified.
The rest of the priors are set according to Appendix C of Ref.~\cite{LIGOScientific:2018mvr}.
The specific values of the prior boundaries for the different parameters vary depending on the application, and we specify them in subsequent sections. 

In general, comparing the eccentric parameters from any two waveforms requires additional postprocessing due to the gauge dependence and nonuniqueness of these parameters in general relativity.
In the next sections, we present our results using a common definition of eccentric parameters to facilitate direct comparisons against results from other eccentric waveform models.
Particularly, we employ the definitions of GW eccentricity $ e _{ \text{gw}} $ and GW mean anomaly $ l _{ \text{gw}} $, as introduced in Ref.~\cite{Ramos-Buades:2022lgf}, and we use the highly efficient implementation of such definitions in the \texttt{gw\_eccentricity} Python package \cite{Shaikh:2023ypz} (see footnote \ref{fn:e_gw}).
To present our results in terms of these eccentric parameters, we follow the same strategy as in Sec.~III~D of Ref.~\cite{Ramos-Buades:2023yhy}, which consists in evaluating the waveform for each sample of the posterior distributions and applying the \texttt{gw\_eccentricity} package to measure $e_\text{gw}$ and $l_\text{gw}$ at a desired point in the evolution of the system.
To ease the \texttt{gw\_eccentricity} postprocessing, we implement an optional setting in the \texttt{SEOBNRv5EHM} model to evolve backward in time the equations of motion~\eqref{eq:EOMprStr} of the binary, and we integrate $2000\,M$ backward for each parameter-estimation run.

%%%%%%%%%%%%%%%%%%%%%%%%%%%
%%%%%%%%%%%%%%%%%%%%%%%%%%%

\subsection{Numerical-relativity injections}\label{sec:NRInj}

In Sec.~\ref{sec:comparison_nr_eccentric}, we studied the accuracy of the \texttt{SEOBNRv5EHM} model by computing waveform mismatches against a dataset of eccentric NR waveforms \cite{Hinder:2017sxy,Boyle:2019kee,Ramos-Buades:2022lgf}.
Here, we further investigate the model's accuracy by performing zero-noise injections of eccentric NR waveforms and using the model to recover the parameters of the injected waveforms.
For these analyses, we include all the modes up to $ \ell \leq 8$ for the injected NR signals, and for the recovery we use only the $ (\ell, |m|)=(2,2) $ modes of \texttt{SEOBNRv5EHM} to reduce computational time.
We label this restricted model as \texttt{SEOBNRv5E}.

We consider a set of 3 public eccentric NR simulations \texttt{SXS:BBH:1355}, \texttt{SXS:BBH:1359}, and \texttt{SXS:BBH:1363}, which correspond to equal-mass, nonspinning BH binaries with initial eccentricities of $0.07$, $0.13$, and $0.25$, respectively (these eccentricities are measured from the orbital frequency at first periastron passage; see Table I of Ref.~\cite{Ramos-Buades:2021adz} for details).
For these injections, we choose a total mass
$M=70 \, \solarmass$,
an inclination between the system's orbital angular momentum and the line of sight
$\iota=0 \,$ rad, coalescence phase $\varphi =0 \,$rad, and luminosity distance $d_L=2307\,$Mpc.
This configuration produces a three-detector network  matched-filtered SNR of
$\rho_{\text{mf}}^\text{N} = 20$
for LIGO Hanford/Livingston and Virgo, when using the zero-detuned Advanced LIGO PSD and the Virgo PSD at design sensitivity.
We note that the choice of equal-mass, nonspinning NR signals with zero inclination ($\iota=0 \,$rad) justifies the use of \texttt{SEOBNRv5E} (no higher-order modes) to increase computational efficiency.

\setlength{\extrarowheight}{8pt}
\begin{table}%[h!]
    \centering
    \begin{tabular}{ c  c  c  c c  }
 \hline
 \hline
 Parameter & \makecell[cc]{Injected \\ value} & \makecell[cc]{ \small SXS:1355} & \makecell[cc]{ \small SXS:1359} & \makecell[cc]{ \small SXS:1363}  \\
 \hline
 \centering
$M/\solarmass$ &  ${70.0}$ &   ${71.05}^{+2.62}_{-2.35} $ & ${70.77}^{+2.59}_{-2.44} $ & ${71.13}^{+3.53}_{-3.25} $ \\
 %\hline
 $\mathcal{M}/\solarmass$ &  ${30.47}$ & ${30.52}^{+1.03}_{-0.97} $ & ${30.43}^{+1.09}_{-1.09} $ & ${30.61}^{+1.53}_{-1.44} $ \\
 %\hline
 $1/q$ & ${1.0}$ &  ${0.8}^{+0.16}_{-0.19} $ & ${0.8}^{+0.16}_{-0.19} $ & ${0.81}^{+0.15}_{-0.18} $ \\
 %\hline
 $\chi_{\text{eff}}$ & $0.0$ & ${0.02}^{+0.09}_{-0.09} $ & ${0.02}^{+0.1}_{-0.11} $ & ${0.03}^{+0.12}_{-0.12} $\\
 %\hline
 $e_0$ & - &  ${0.05}^{+0.04}_{-0.04} $ & ${0.13}^{+0.03}_{-0.04} $ & ${0.24}^{+0.03}_{-0.03} $ \\
 %\hline
 $\zeta_0$ & - &  ${2.27}^{+1.14}_{-1.10} $ & ${1.27}^{+1.83}_{-0.9} $ & ${4.01}^{+0.83}_{-0.71} $ \\
 %\hline
 %\hline
$\iota$ & $0.0$ &   ${0.62}^{+0.48}_{-0.37} $ & ${0.62}^{+0.48}_{-0.38} $ & ${0.62}^{+0.49}_{-0.38} $ \\
 %\hline
 $d_L$ & ${2307}$  & ${1835}^{+376}_{-569} $ & ${1827}^{+381}_{-564} $ & ${1910}^{+411}_{-593} $ \\
 %\hline
$\varphi$ & $0.0$ & ${3.14}^{+2.52}_{-2.51} $ & ${3.14}^{+2.52}_{-2.48} $ & ${3.13}^{+2.52}_{-2.52} $ \\
 %\hline
 $\rho^{\text{N}}_{\text{mf}}$  & $20.0$ &  ${19.07}^{+0.09}_{-0.14} $ & ${19.05}^{+0.08}_{-0.14} $ & ${18.99}^{+0.09}_{-0.15} $ \\ [0.1cm]
\hline
\multirow{2}{*}{$e_{\text{gw}}$} & Injected  & ${0.07}$  &  ${0.13}$  &  ${0.25}$   \\
& Measured  & ${0.06}^{+0.04}_{-0.04} $ & ${0.13}^{+0.03}_{-0.04} $ & ${0.24}^{+0.03}_{-0.03} $\\ 
\multirow{2}{*}{$l_{\text{gw}}$} & Injected  & ${1.96}$  &  ${0.81}$  &  ${4.27}$   \\
& Measured  &   ${2.10}^{+1.20}_{-1.06} $ & ${1.13}^{+4.59}_{-0.84} $ & ${4.1}^{+0.88}_{-0.91} $ \\  [0.1cm]
 \hline
 \hline
    \end{tabular}
 \caption{
Injected, median values, and $90\%$ credible intervals of the posterior distributions for three synthetic NR injections corresponding to equal-mass, nonspinning BBHs with different initial eccentricities, and recovered with \texttt{SEOBNRv5E}.
The binary parameters correspond to the total mass $M$, chirp mass $\mathcal{M}$, inverse mass ratio $1/q$, effective spin parameter $\chi_{\text{eff}}$, initial eccentricity $e_0$, initial relativistic anomaly $\zeta_0$, angle between the total angular momentum and the line of sight $\iota$, luminosity distance $d_L$, coalescence phase $\varphi$, and the network matched-filtered SNR for LIGO-Hanford/Livingston and Virgo detectors $\rho^{\text{N}}_{\text{mf}}$.
We also report the injected and measured GW eccentricity $e_{\text{gw}}$ and mean anomaly $l_{\text{gw}}$, at a reference frequency of $ 20$ Hz.
}
\label{tab:nr_injection}
\end{table}

\begin{figure*}%[!] 
\includegraphics[width=0.71\columnwidth]{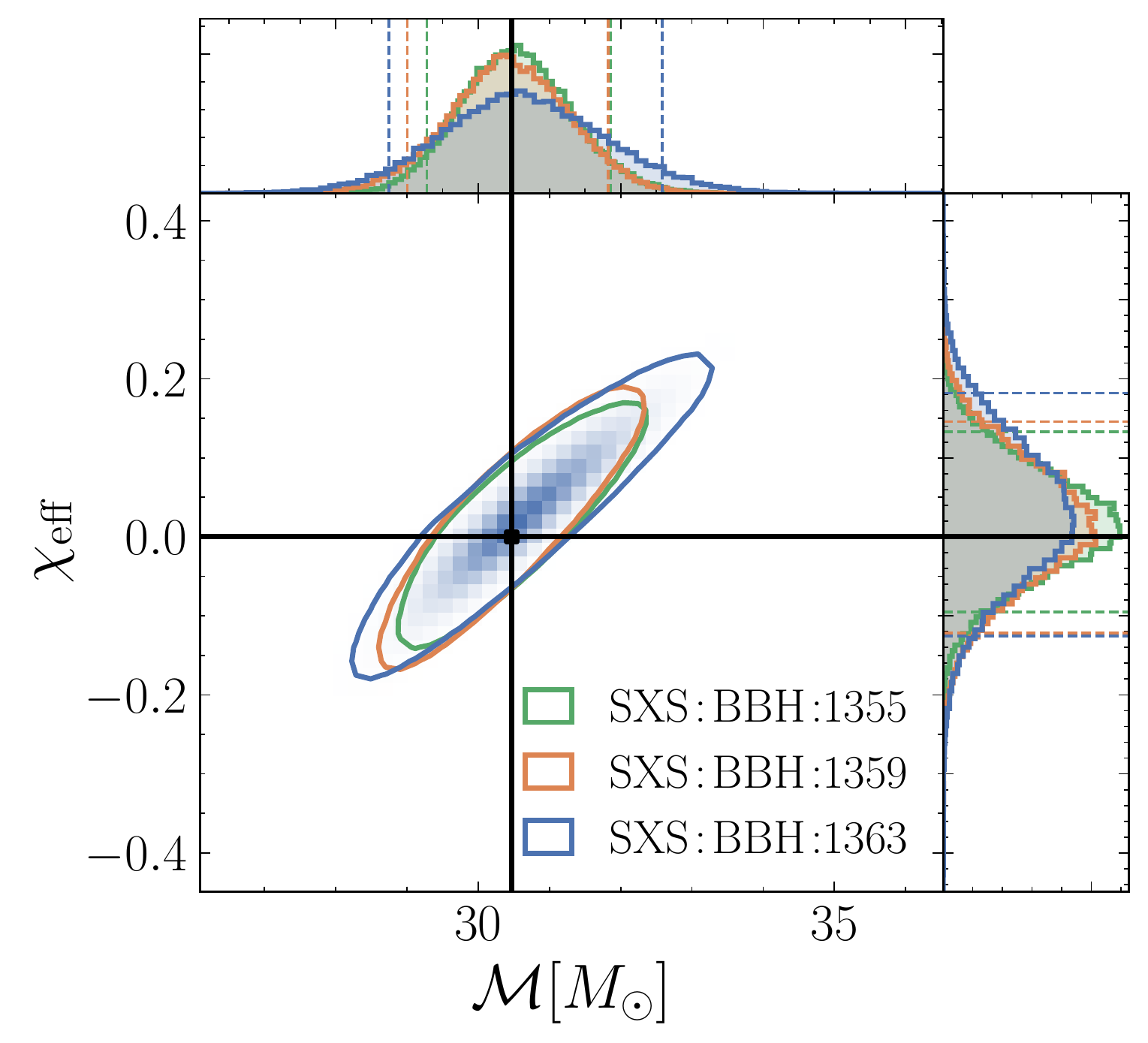}
\hfill
\includegraphics[width=0.665\columnwidth]{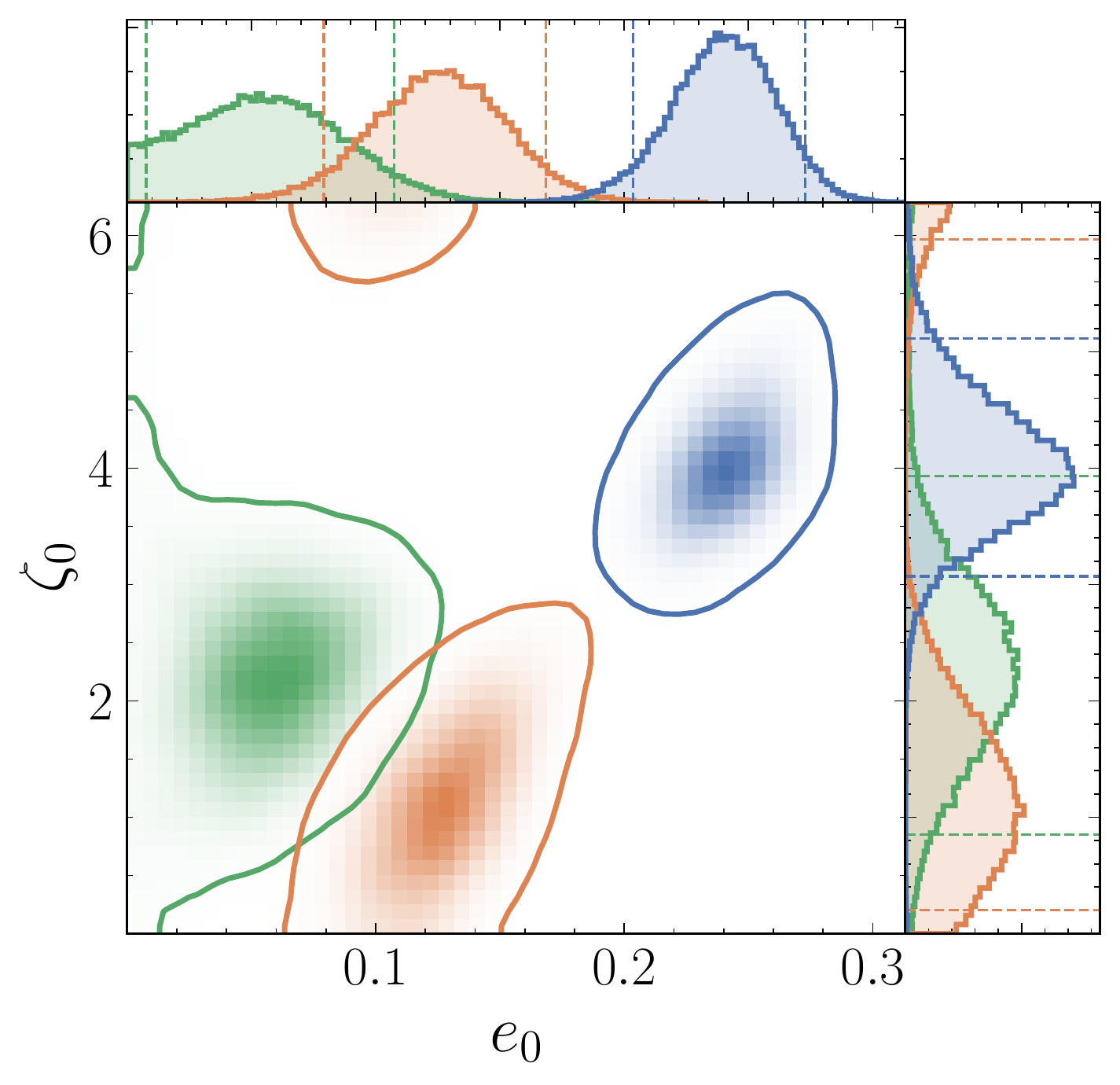}
\hfill
\includegraphics[width=0.67\columnwidth]{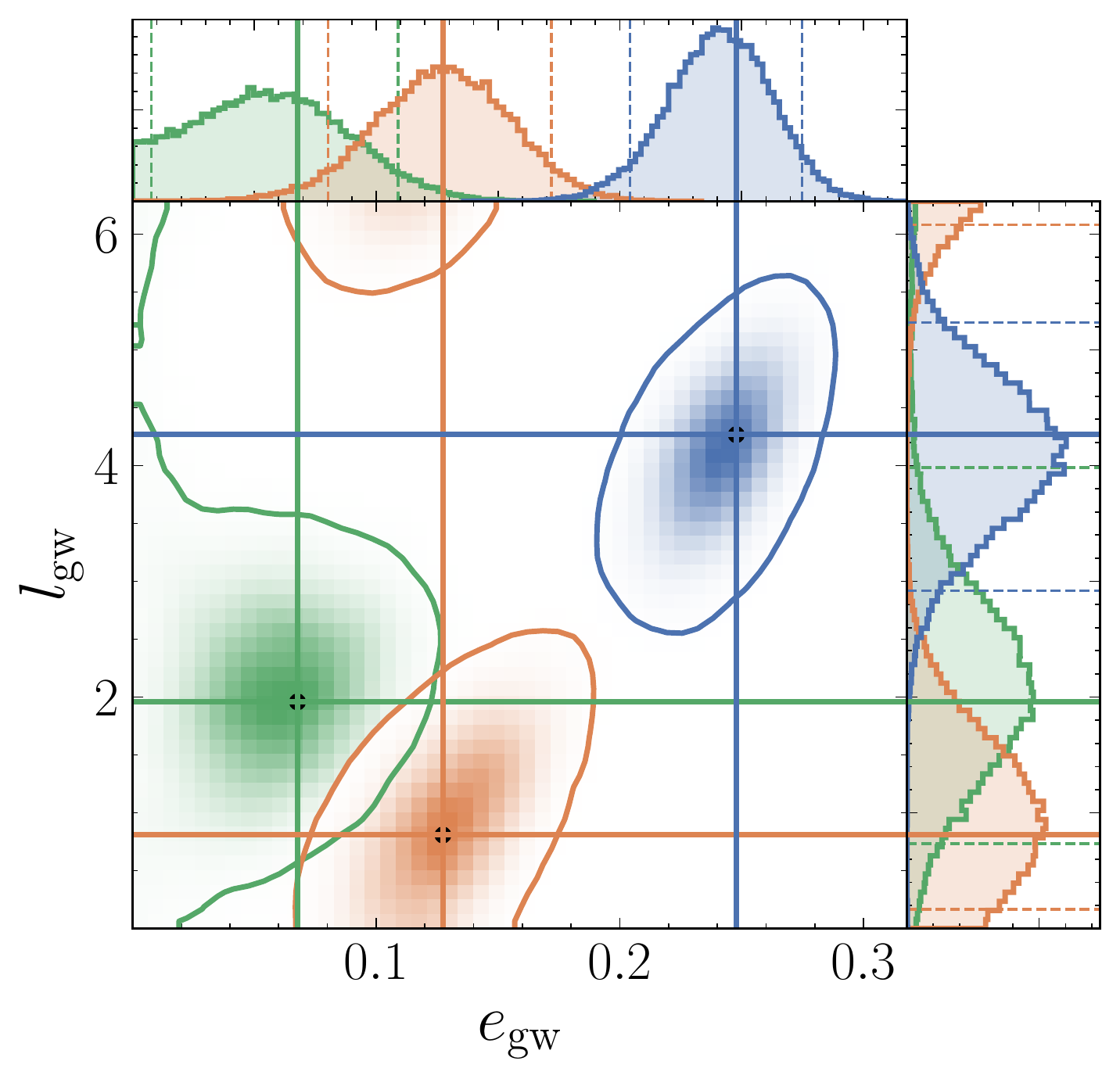}  
\caption{
Marginalized 1D and 2D posterior distributions obtained with the \texttt{SEOBNRv5E} model for some relevant parameters for three equal-mass, nonspinning NR synthetic BBH signals with total mass $70 \, \solarmass$.
The signal NR waveforms
\texttt{SXS:BBH:1355},
\texttt{SXS:BBH:1359},
and \texttt{SXS:BBH:1363}
belong to the public SXS catalog and have GW eccentricities
$e_\text{gw} \in \{0.07, 0.13, 0.25\}$
and GW mean anomalies
$l_\text{gw} \in \{ 1.96, 0.81, 4.27 \}$
defined at $20\,$Hz, respectively.
The rest of the parameters are specified in Table \ref{tab:nr_injection}.
In the 2D posteriors, the solid contours represent the $90\%$ credible intervals, and the black dots indicate the true parameters of the injected signal.
In the 1D posteriors, the $90\%$ credible intervals are represented by dashed lines, and the true parameters of the injected signal are indicated by black solid lines.
\textit{Left panel:} Posteriors of the chirp mass $ \mathcal M $ and effective spin parameter $ \chi _{ \text{eff}} $.
\textit{Middle panel:} Posteriors of the initial eccentricity $ e_0 $ and initial relativistic anomaly $ \zeta_0 $ at $20\,$Hz.
\textit{Right panel:} Posteriors of the GW eccentricity $ e _{ \text{gw}} $ and GW mean anomaly $ l _{ \text{gw}} $ measured at $20\,$Hz.
}
\label{fig:nrInj}
\end{figure*}

We choose priors on inverse mass ratio
$1/q \in [0.05, 1]$
and chirp mass
$\mathcal{M} \in [5, 100] \,\solarmass$ such that the induced priors on component masses are uniform.
The priors on the magnitudes of the dimensionless spin components have ranges
$|\chi_{1, \s 2}| \in [0,0.99]$.
The prior on the initial eccentricity $ e_0 $ is taken to be uniform with range
$e_0 \in [0, 0.3]$,
except for the run corresponding to the
\texttt{SXS:BBH:1363}
waveform for which the range is set to be
$e_0 \in [0,0.5]$
to avoid railing of the posterior against the upper bound of the prior.
The prior on the relativistic anomaly $ \zeta_0 $ is chosen to be uniform, with range
$\zeta_0 \in [0,2 \pi]$.
These parameters are defined at a reference frequency for starting the waveform generation $f_\text{ref} = 20\,$Hz.

We summarize in Fig.~\ref{fig:nrInj} the results of the injection studies. 
For each NR injection, we report the marginalized 1D and 2D posteriors for the chirp mass $\mathcal{M}$ and the effective spin parameter $\chi_{\text{eff}}$, the initial eccentricity $ e_0 $ and initial relativistic anomaly $ \zeta_0 $ at $20\,$Hz, and the GW eccentricity $ e _{ \text{gw}} $ and GW mean anomaly $ l _{ \text{gw}} $ measured at $20\,$Hz.
The injected parameters and the median of the inferred posterior distributions with the $90\%$ credible intervals are shown in Table \ref{tab:nr_injection}.

Our results demonstrate that \texttt{SEOBNRv5E} recovers accurately the binary's intrinsic parameters $\{ M, \,\mathcal{M},\, 1/q,\, \chi_{\text{eff}},\, e _{ \text{gw}},\,  l _{ \text{gw}} \}$ for all the injections within the $90\%$ credible intervals.
The small biases in the luminosity distance $ d_L $ and $\iota $ parameter can be explained by the different multipolar content in NR (up to $\ell \leq 8$) and  \texttt{SEOBNR5E} (only the $(\ell,|m|)=(2,2)$ modes) waveforms.
These biases were also observed with the \texttt{SEOBNRv4E} model in Ref.~\cite{Ramos-Buades:2023yhy}.

For all the NR injections, the eccentricity posterior obtained with the \texttt{SEOBNRv5E} model is Gaussian and unimodal, even for the highest eccentricity injection
(\texttt{SXS:BBH:1363}).
This was not the case for the previous generation \texttt{SEOBNRv4E} model (see Fig.~8 of Ref.~\cite{Ramos-Buades:2023yhy}).
This is a consequence of the better description of the initial EOB conditions implemented in the \texttt{SEOBNRv5EHM} model, as well as the more accurate description of the eccentric dynamics.

The posterior distributions for the GW eccentricity $ e _{ \text{gw}} $ and GW mean anomaly $ l _{ \text{gw}} $  measured at $ 20 \,$Hz are shown in the right panel of Fig.~\ref{fig:nrInj}.
We find that the \texttt{SEOBNRv5E} model recovers accurately the NR injected values within the $90 \%$ credible intervals.
Furthermore, we observe that the $(e _{ \text{gw}},\s l _{ \text{gw}})$ posteriors are very similar to the $(e_0, \s\zeta_0)$ posteriors.
This indicates that the eccentric input parameters of the \texttt{SEOBNRv5EHM} model are a well-behaved representation of the eccentric modulations in the waveform for the considered cases.

These studies demonstrate that the \texttt{SEOBNRv5E} model recovers very accurately the intrinsic parameters of the injected NR waveforms.
This is consistent with the low unfaithfulness values of \texttt{SEOBNRv5EHM} against NR waveforms, as reported in Sec.~\ref{sec:comparison_nr_eccentric}.
Further studies of the model's accuracy will require a larger coverage of parameter space including also different SNRs.
We leave for future work investigating the waveform systematic errors of the \texttt{SEOBNRv5EHM} model and its biases against NR waveforms.

%%%%%%%%%%%%%%%%%%%%%%%%%%%
%%%%%%%%%%%%%%%%%%%%%%%%%%%

\subsection{Analysis of GW events}\label{sec:GWevents}

In this section, we employ the \texttt{SEOBNRv5EHM} model to analyze two GW events observed by the LIGO and Virgo detectors during the first and third observing runs \cite{LIGOScientific:2018mvr,LIGOScientific:2021usb,LIGOScientific:2021djp}: GW150914 \cite{LIGOScientific:2016aoc} and GW190521 \cite{LIGOScientific:2020iuh}.
We employ strain data from the Gravitational Wave Open Source Catalog (GWOSC) \cite{LIGOScientific:2019lzm}, and the released PSD and calibration envelopes included in the Gravitational Wave Transient Catalog GWTC-2.1 \cite{LIGOScientific:2021usb}.
For the analysis, we employ the parameter estimation settings described in Sec.~\ref{sec:PEsetup}.
Additionally, we compare the results for the eccentric, aligned-spin \texttt{SEOBNRv5EHM} model with samples from the eccentric, aligned-spin \texttt{SEOBNRv4EHM\_opt} model presented in Ref.~\cite{Ramos-Buades:2023yhy} (this model is an optimized version of \texttt{SEOBNRv4EHM} \cite{Ramos-Buades:2021adz}), and the QC, precessing-spin \texttt{SEOBNRv5PHM} model from Ref.~\cite{RamosBuadesv5}.

%%%%%%%%%%%%%%%%%%%%%%%%%%%
  
\subsubsection*{GW150914}
GW150914 was the first observation of GWs from a BBH coalescence with one of the highest SNRs ($\sim \!23.7$) observed during the first three observing runs  \cite{LIGOScientific:2016aoc,LIGOScientific:2021usb} of the LVK Collaboration.
Its intrinsic parameters are consistent with a QC, nonspinning binary with comparable masses \cite{LIGOScientific:2016ebw, Romero-Shaw:2019itr,Bonino:2022hkj,Iglesias:2022xfc}.

For our analysis, we choose priors on inverse mass ratio $1/q \in [0.05, 1]$ and chirp mass $\mathcal{M} \in [20,50] \,\solarmass$ such that the induced priors on component masses are uniform.
We use uniform priors on the initial eccentricity $ e_0 $ and initial relativistic anomaly $ \zeta_0 $, with ranges $e_0 \in [0,0.3]$ and $\zeta_0 \in [0, 2\pi]$. These parameters are defined at a starting frequency for the waveform generation of $  f _{ \text{ref}}  = 10\,$Hz.
The other priors are chosen as in Sec.~\ref{sec:NRInj}, and the sampler settings are summarized in Table \ref{tab:peruntime}.
These choices coincide with the settings employed in Ref.~\cite{Ramos-Buades:2023yhy} for the \texttt{SEOBNRv4EHM\_opt} model.

\begin{figure*}%[htpb!] 
\includegraphics[width=0.71\columnwidth]{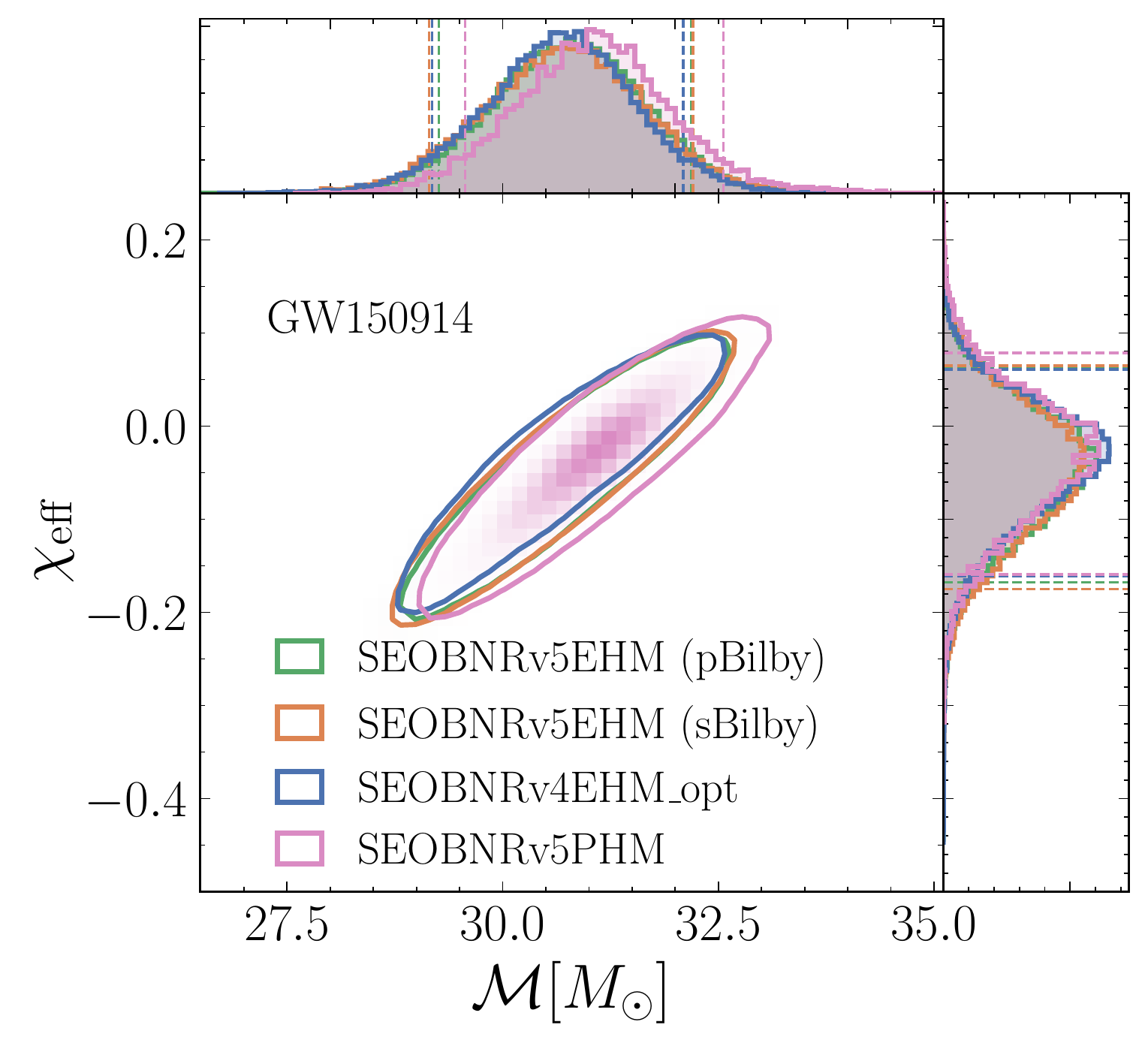} 
\includegraphics[width=0.68\columnwidth]{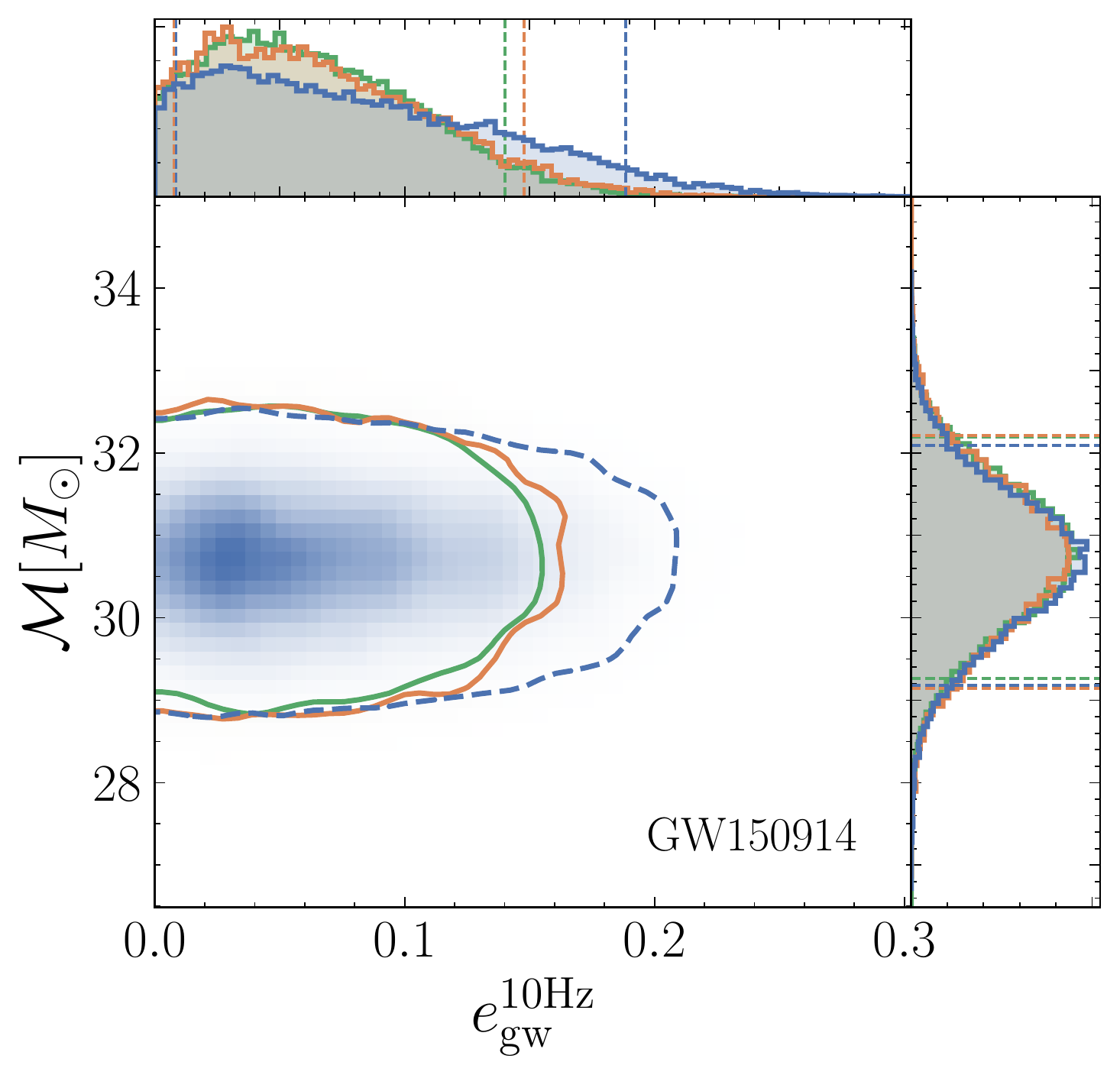}  
\includegraphics[width=0.66\columnwidth]{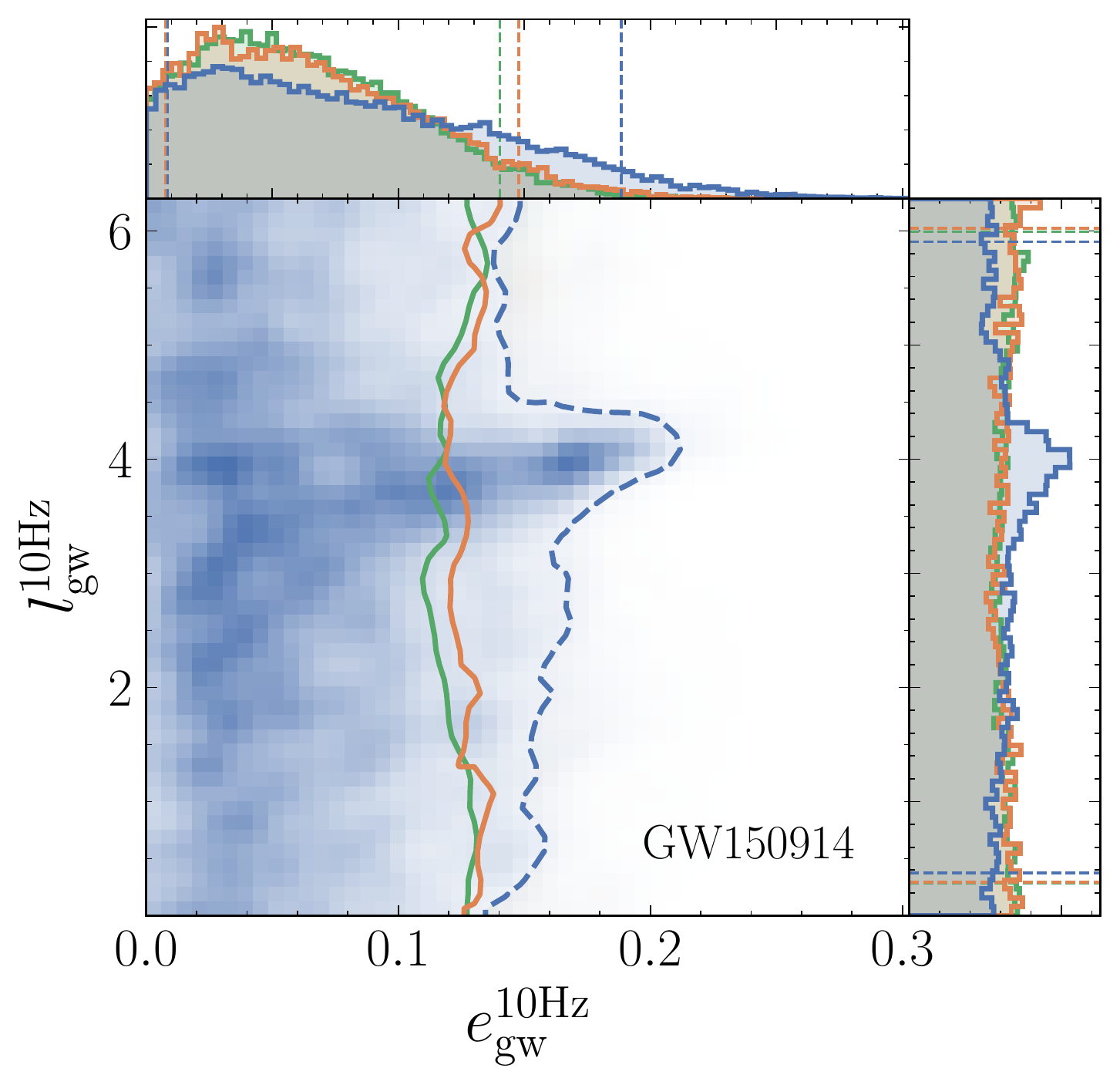} \\
\includegraphics[width=0.71\columnwidth]{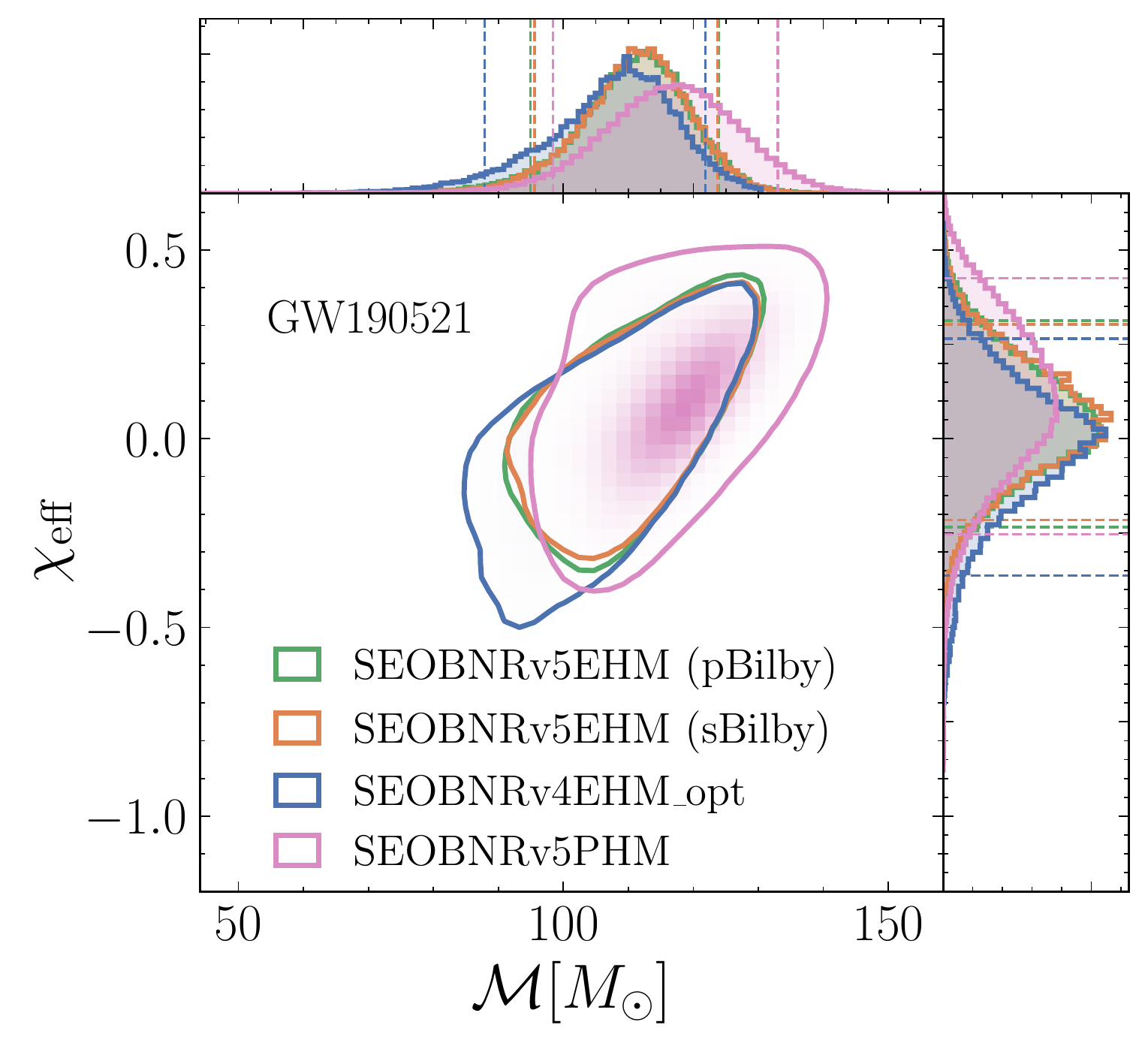} 
\includegraphics[width=0.68\columnwidth]{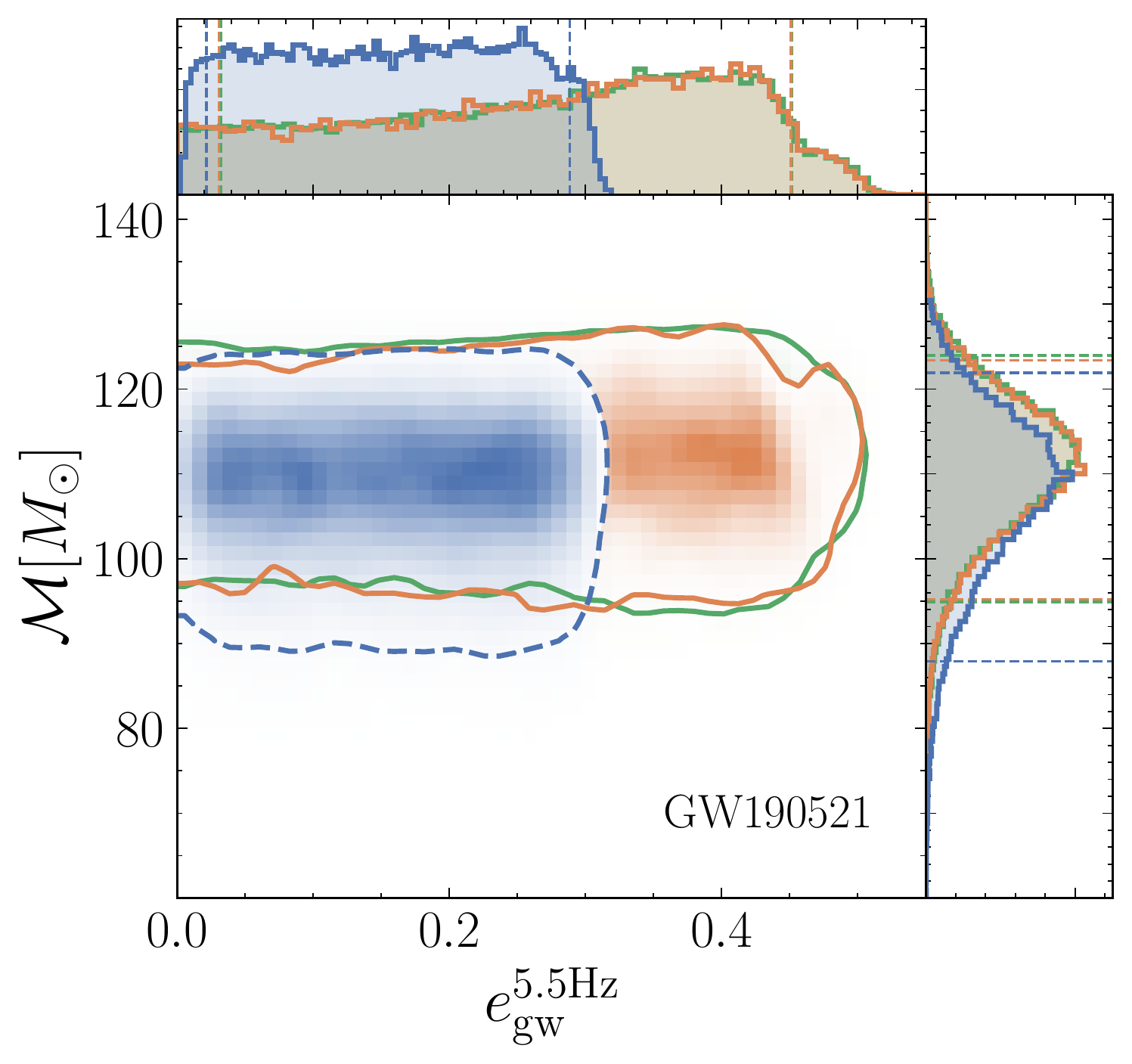} 
\includegraphics[width=0.66\columnwidth]{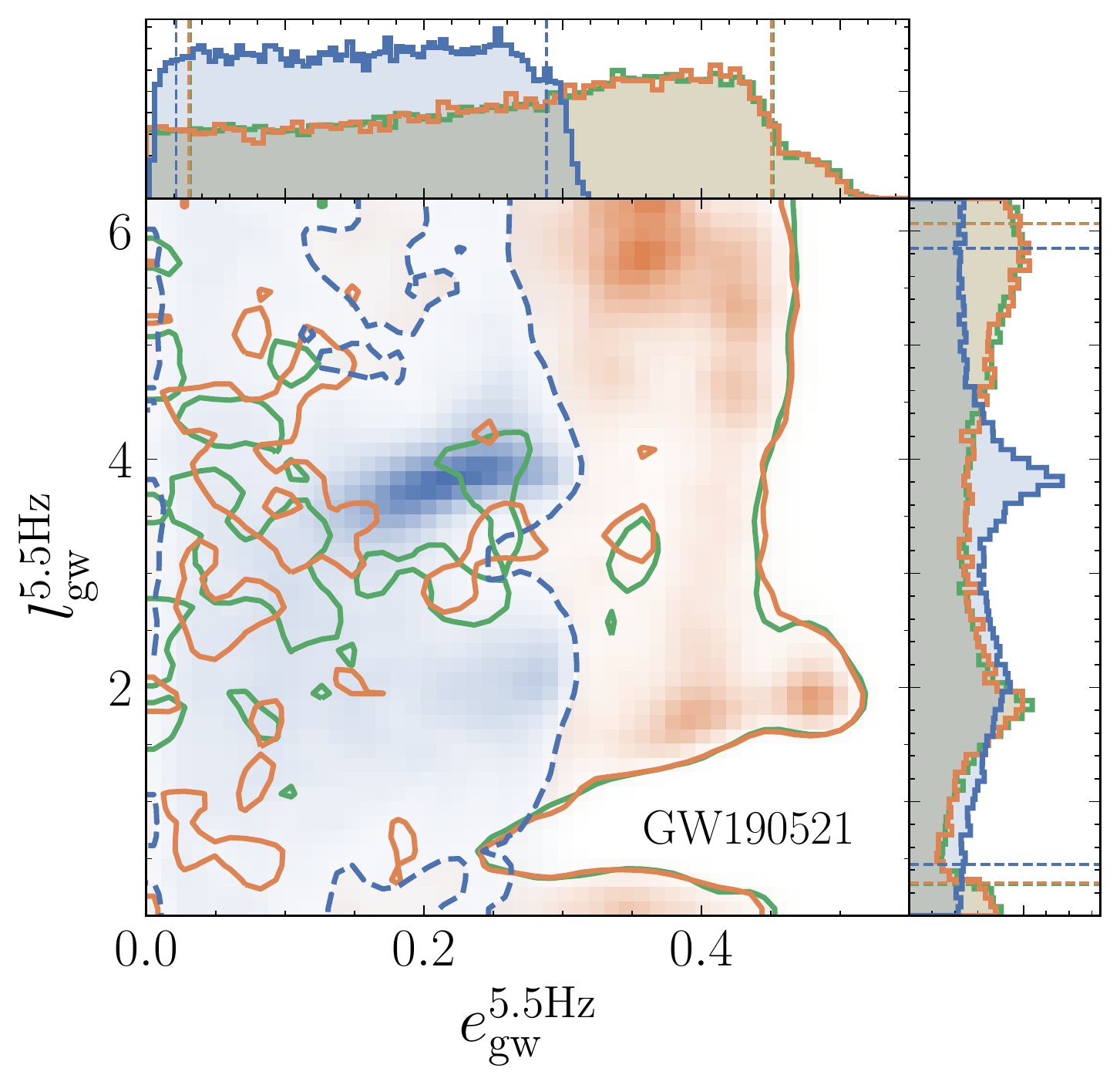}  \\ 
\vspace{-0.3cm}
\caption{
Marginalized 1D and 2D posterior distributions corresponding to the real events GW150914 and GW190521 obtained with the eccentric, aligned-spin \texttt{SEOBNRv5EHM} model, for the effective spin parameter $ \chi _{ \text{eff}} $, chirp mass $ \mathcal M $, GW eccentricity $ e _{ \text{gw}} $, and GW mean anomaly $ l _{ \text{gw}} $. 
We also show the results for the eccentric, aligned-spin \texttt{SEOBNRv4EHM\_opt} model and the QC, precessing-spin \texttt{SEOBNRv5PHM} model from Refs.~\cite{Ramos-Buades:2023yhy,RamosBuadesv5}. For {\tt SEOBNRv5EHM}, we indicate whether the run is performed with {\tt parallel} {\tt Bilby} (pBilby), or  {\tt serial} {\tt Bilby} (sBilby). The reference frequency for starting the waveform generation for GW150914 is chosen to be $ f_{\text{ref}}  = 10\,$Hz, while for GW190521 is $ f_{\text{ref}}  = 5.5\,$Hz, so that all the waveform modes are in band at the likelihood minimum frequency ($f_{\text{min}}= 11\,$Hz).}
\label{fig:gwEvents}
\end{figure*}

In the top row of Fig.~\ref{fig:gwEvents}, we display the marginalized 1D and 2D posterior distributions for the chirp mass $ \mathcal M $, effective-spin parameter $ \chi _{ \text{eff}} $, GW eccentricity $ e _{ \text{gw}} $, and GW mean anomaly $ l _{ \text{gw}} $ corresponding to GW150914, measured by \texttt{SEOBNRv5EHM} and other approximants. For \texttt{SEOBNRv5EHM}, we show results with two different samplers, {\tt parallel} {\tt Bilby} (pBilby) and {\tt serial} {\tt Bilby} (sBilby). In Table \ref{tab:gwEvents}, we show the median values and the $90 \%$ credible intervals of the posterior distributions for the different binary parameters.
For comparisons, we include in our analysis the samples of the eccentric, aligned-spin \texttt{SEOBNRv4EHM\_opt} model from Ref.~\cite{Ramos-Buades:2023yhy}, and the samples of the QC, precessing-spin \texttt{SEOBNRv5PHM} model from Ref.~\cite{RamosBuadesv5}.

We find that the noneccentric binary parameters like chirp mass, effective spin, and mass ratio, measured by \texttt{SEOBNRv5EHM} are consistent with the ones measured by \texttt{SEOBNRv4EHM\_opt} \cite{Ramos-Buades:2023yhy}. We also observe broad agreement between the QC, precessing-spin \texttt{SEOBNRv5PHM} model and the eccentric models, as GW150914 is consistent with a nonspinning binary \cite{LIGOScientific:2016ebw}.
The slight shift in chirp mass observed in the eccentric models with respect to the QC model is consistent with the different physical effects included in the waveform approximants. Regarding sampler systematics, we observe consistent results for \texttt{SEOBNRv5EHM} runs produced with  {\tt parallel} {\tt Bilby} and {\tt serial} {\tt Bilby}.

Although both \texttt{SEOBNRv5EHM pBilby} (\texttt{SEOBNRv5EHM sBilby}) and \texttt{SEOBNRv4EHM\_opt} have median values of eccentricity distinct from zero, $e^{10\,\text{Hz}}_{\text{gw}}=  {0.06}^{+0.06}_{-0.05} $ $\big( e^{10\,\text{Hz}}_{\text{gw}}= {0.06}^{+0.07}_{-0.05} \big) $
and $e^{10\,\text{Hz}}_{\text{gw}}= {0.07}^{+0.09}_{-0.06}$, respectively, the posterior distributions have strong support in the zero eccentricity region, which is in agreement with other analyses of GW150914 with eccentric waveforms \cite{LIGOScientific:2016ebw,Romero-Shaw:2019itr,Bonino:2022hkj,Iglesias:2022xfc}. Furthermore, with \texttt{SEOBNRv5EHM} we obtain a tighter posterior constraint in the eccentricity towards the zero value. This demonstrates that the increased accuracy of the \texttt{SEOBNRv5EHM} model translates into more accurate measurements of the eccentric parameters.

Finally, we also compute the log-{10} Bayes factor between the eccentric, aligned-spin (EAS) and the QC, aligned-spin (QCAS) hypothesis, $\log_{10} \mathcal{B}_\text{EAS/QCAS}$, by producing a run using the QC, aligned-spin \texttt{SEOBNRv5HM} model \cite{Pompiliv5}.
The main parameters obtained with this run, including $\log_{10} \mathcal{B}_\text{EAS/QCAS}$ for the \texttt{SEOBNRv5EHM} and  \texttt{SEOBNRv4EHM\_opt} models, are shown in Table \ref{tab:gwEvents}.
The  $\log_{10} \mathcal{B}_\text{EAS/QCAS}$ values are  $-0.34^{+0.10}_{-0.10} $  and  $-0.35^{+0.09}_{-0.09} $ for \texttt{SEOBNRv5EHM} and  \texttt{SEOBNRv4EHM\_opt}, respectively.\footnote{
We note that the nominal values of the Bayes factors obtained with serial and parallel Bilby for \texttt{SEOBNRv5EHM} differ slightly, although they are consistent within the error bars, and differences might be explained due to different implementations of the \texttt{dynesty} nested sampler~\cite{Speagle:2019ivv}.
}
This indicates a slight preference for the QC hypothesis and is in agreement with the results reported in Table III of Ref.~\cite{Gupte:2024jfe}, where the $\log_{10} \mathcal{B}_\text{EAS/QCAS}$ between the \texttt{SEOBNRv4EHM} and \texttt{SEOBNRv4HM} models is found to be $-0.44$.

\begin{table}%[h!]
\centering
\resizebox{\columnwidth}{!}{%
\begin{tabular}{l c c c c c c c }
 \hline
 \hline
  \makecell[cc]{Event} & \makecell[cc]{Model \\ (sampler)} & \multicolumn{2}{c}{\makecell[cc]{Data \\ settings}} &\multicolumn{2}{c}{\makecell[cc]{Sampler \\ settings}} & \makecell[cc]{Computing \\ resources} & Run-time  \\
 \hline
 & &  \makecell[cc]{srate  \\ (Hz)} & \makecell[cc]{$f_\text{ref}$ \\(Hz)} & \makecell[cc]{naccept/\\nact} & nlive & cores$\times$nodes &   \\
 \hline
   \multirow{6}{*}{ \makecell[ccc]{  GW150914	}%\\ \fontsize{7.5pt}{7pt}\selectfont ($\langle f_{\text{start}} \rangle=10\,$Hz) } 
   }  &   \makecell[cc]{\fontsize{7.5pt}{7pt}\selectfont  \texttt{SEOBNRv5EHM} \\ \fontsize{7.5pt}{7pt}\selectfont  (\texttt{pBilby})} & 4096  & 10 & 30 & 2048 & $32\times 10$ & 2d 9h\\
                           & \makecell[cc]{\fontsize{7.5pt}{7pt}\selectfont  \texttt{SEOBNRv5EHM} \\ \fontsize{7.5pt}{7pt}\selectfont  (\texttt{sBilby})}   & 4096 & 10 & 60 & 1000 & $64\times 1$ & 3d 20h  \\[0.08cm]
                           & \makecell[cc]{\fontsize{7.5pt}{7pt}\selectfont  \texttt{SEOBNRv4EHM\_opt} \\ \fontsize{7.5pt}{7pt}\selectfont  (\texttt{pBilby})}   & 4096 & 10 & 30 & 2048 & $32\times 10$ & 2d 21h  \\[0.08cm]
                           & \makecell[cc]{\fontsize{7.5pt}{7pt}\selectfont  \texttt{SEOBNRv5PHM} \\ \fontsize{7.5pt}{7pt}\selectfont  (\texttt{sBilby})}   & 2048 & 20 & 60 & 1000 & $64\times 1$ & 1d 17h  \\[0.08cm]
\hline
 \multirow{5}{*}{ \makecell[ccc]{  GW190521	%\\ \fontsize{7.5pt}{7pt}\selectfont ($\langle f_{\text{start}} \rangle=5.5\,$Hz) 
 } }  & \makecell[cc]{\fontsize{7.5pt}{7pt}\selectfont  \texttt{SEOBNRv5EHM} \\ \fontsize{7.5pt}{7pt}\selectfont  (\texttt{pBilby})} & 4096  & 5.5 & 30 & 2048 & $32\times 16$ & 1d 2h\\
                           & \makecell[cc]{\fontsize{7.5pt}{7pt}\selectfont  \texttt{SEOBNRv5EHM} \\ \fontsize{7.5pt}{7pt}\selectfont  (\texttt{sBilby})}   & 4096 & 5.5 & 60 & 1000 & $64\times 1$ & 1d 13h  \\[0.08cm]
                           & \makecell[cc]{\fontsize{7.5pt}{7pt}\selectfont  \texttt{SEOBNRv4EHM\_opt} \\ \fontsize{7.5pt}{7pt}\selectfont  (\texttt{pBilby})}   & 4096 & 5.5 & 30 & 2048 & $32\times 16$ & 1d 6h  \\[0.08cm]
                           & \makecell[cc]{\fontsize{7.5pt}{7pt}\selectfont  \texttt{SEOBNRv5PHM} \\ \fontsize{7.5pt}{7pt}\selectfont  (\texttt{pBilby})}   & 2048 & 5.5 & 30 & 8192 & $64\times 8$ & 3d 4h  \\[0.08cm]
 \hline
\end{tabular}
}
\caption{
Settings and evaluation times for the different runs on two real GW events (GW150914 and GW190521) corresponding to the eccentric, aligned-spin \texttt{SEOBNRv5EHM} model.
We also include the run-times for the eccentric, aligned-spin \texttt{SEOBNRv4EHM\_opt} model from Ref.~\cite{Ramos-Buades:2023yhy}, and the run times for the QC, precessing-spin \texttt{SEOBNRv5PHM} model from Ref.~\cite{RamosBuadesv5}.
We indicate the sampler for each run, either {\tt parallel} {\tt Bilby} (pBilby) \cite{Smith:2019ucc} or {\tt serial} {\tt Bilby} (sBilby)  \cite{Ashton:2018jfp,Romero-Shaw:2020owr}.
Sampling rate ($\text{srate}$) and reference frequency for starting the waveform generation ($f_\text{ref}$) are specified in the data settings, while the number of live points ($\text{nlive}$), the number of autocorrelation times ($\text{nact}$, for pBilby), and the number of accepted steps ($\text{naccept}$, for sBilby) are specified in the sampler settings.
We use a segment length of $ 8\,$s for all the runs in this table.
The time reported is wall time, while the total computational cost in CPU hours can be obtained by multiplying this time by the reported number of CPU cores employed.
}
\label{tab:peruntime}
\end{table}
%

% Table of GW events 
\setlength{\extrarowheight}{8pt}
\begin{table*}%[!]
    \centering 
\resizebox{\linewidth}{!}{
%\begin{adjustbox}{width=\linewidth,center}
%\scalebox{0.8}{
\begin{tabular*}{\textwidth}{c@{\extracolsep{\fill}} c c c c c c c  c  c   c c }
 \hline
 \hline
Event & Model & $M/\solarmass$ & $\mathcal{M}/\solarmass$ & $1/q$ &  $\chi_{\text{eff}}$ & $e_{\text{gw}}$ &  $l_{\text{gw}}$ &  $d_L$   & $ \log_{10} \mathcal{B}$\scalebox{0.75}{$_\text{EAS/QCAS}$}   \\[0.05cm]
 \hline 
 \centering
 \multirow{6}{*}{ \makecell[cc]{  GW150914	
 %\\ \fontsize{7.5pt}{7pt}\selectfont ($\langle f_{\text{start}} \rangle=10\,$Hz) 
 } }  & \makecell[cc]{\fontsize{7.5pt}{7pt}\selectfont  \texttt{SEOBNRv5EHM} \\ \fontsize{7.5pt}{7pt}\selectfont  (\texttt{pBilby})}   &   ${70.99}^{+2.53}_{-2.63} $ & ${30.76}^{+1.11}_{-1.17} $ & ${0.89}^{+0.09}_{-0.14} $ & ${-0.04}^{+0.08}_{-0.1} $ & ${0.06}^{+0.06}_{-0.05} $ & ${3.17}^{+2.48}_{-2.55} $ & ${475}^{+113}_{-118} $ &  $-0.34^{+0.10}_{-0.10} $ \\
                           &  \makecell[cc]{\fontsize{7.5pt}{7pt}\selectfont  \texttt{SEOBNRv5EHM} \\ \fontsize{7.5pt}{7pt}\selectfont  (\texttt{sBilby})}  & ${70.9}^{+2.62}_{-2.8} $ & ${30.72}^{+1.15}_{-1.24} $ & ${0.88}^{+0.09}_{-0.14} $ & ${-0.05}^{+0.09}_{-0.1} $ & ${0.06}^{+0.07}_{-0.05} $ & ${3.17}^{+2.49}_{-2.54} $ & ${480}^{+116}_{-125} $ & $-0.57^{+0.13}_{-0.13} $  \\
                           &  \makecell[cc]{\fontsize{7.5pt}{7pt}\selectfont  \texttt{SEOBNRv4EHM\_opt} \\ \fontsize{7.5pt}{7pt}\selectfont  (\texttt{pBilby})}   &${70.8}^{+2.43}_{-2.6} $ & ${30.68}^{+1.07}_{-1.16} $ & ${0.89}^{+0.09}_{-0.14} $ & ${-0.04}^{+0.08}_{-0.09} $ & ${0.07}^{+0.09}_{-0.06} $ & ${3.12}^{+2.53}_{-2.5} $ & ${445}^{+115}_{-125} $ &  $-0.35^{+0.09}_{-0.09} $  \\
                           & \makecell[cc]{\fontsize{7.5pt}{7pt}\selectfont  \texttt{SEOBNRv5HM} \\ \fontsize{7.5pt}{7pt}\selectfont  (\texttt{pBilby})}  &   ${71.37}^{+2.38}_{-2.41} $ & ${30.92}^{+1.05}_{-1.06} $ & ${0.88}^{+0.09}_{-0.13} $ & ${-0.03}^{+0.08}_{-0.09} $ & - & - & ${497}^{+105}_{-127} $ & -   \\[0.08cm]
                           & \makecell[cc]{\fontsize{7.5pt}{7pt}\selectfont  \texttt{SEOBNRv5PHM} \\ \fontsize{7.5pt}{7pt}\selectfont  (\texttt{sBilby})}  &   ${71.64}^{+2.61}_{-2.56} $ & ${31.05}^{+1.14}_{-1.14} $ & ${0.89}^{+0.09}_{-0.14} $ & ${-0.03}^{+0.08}_{-0.1} $ & -  & -  & ${493}^{+113}_{-122} $ & -   \\[0.08cm]

%%%%%%%%%%%%%%%%%%%%%%%%%%%%%%%%%%%%%%%%%%%%%%%%%%%%%%%%%%%%%%%%%%%%%%%%%%%%%%%%%%%%%%%%%
\hline
%%%%%%%%%%%%%%%%%%%%%%%%%%%%%%%%%%%%%%%%%%%%%%%%%%%%%%%%%%%%%%%%%%%%%%%%%%%%%%%%%%%%%%%%%
  \multirow{6}{*}{ \makecell[cc]{  GW190521	
  %\\ \fontsize{7.5pt}{7pt}\selectfont ($\langle f_{\text{start}} \rangle=5.5\,$Hz) 
  } }  & \makecell[cc]{\fontsize{7.5pt}{7pt}\selectfont  \texttt{SEOBNRv5EHM} \\ \fontsize{7.5pt}{7pt}\selectfont  (\texttt{pBilby})}  & ${260.45}^{+20.72}_{-19.79} $ & ${111.37}^{+9.8}_{-11.93} $ & ${0.73}^{+0.21}_{-0.22} $ & ${0.04}^{+0.21}_{-0.21} $ & ${0.3}^{+0.16}_{-0.23} $ & ${3.12}^{+2.54}_{-2.51} $ & ${4773}^{+1240}_{-1225} $ & $-0.42^{+0.08}_{-0.08} $  \\
   &  \makecell[cc]{\fontsize{7.5pt}{7pt}\selectfont  \texttt{SEOBNRv5EHM} \\ \fontsize{7.5pt}{7pt}\selectfont  (\texttt{sBilby})}   & ${260.71}^{+20.29}_{-19.52} $ & ${111.42}^{+9.68}_{-11.66} $ & ${0.73}^{+0.21}_{-0.21} $ & ${0.05}^{+0.2}_{-0.2} $ & ${0.29}^{+0.16}_{-0.23} $ & ${3.14}^{+2.54}_{-2.52} $ & ${4786}^{+1261}_{-1230} $ & $-0.36^{+0.11}_{-0.11} $ \\
                           &  \makecell[cc]{\fontsize{7.5pt}{7pt}\selectfont  \texttt{SEOBNRv4EHM\_opt} \\ \fontsize{7.5pt}{7pt}\selectfont  (\texttt{pBilby})}  & ${253.97}^{+21.48}_{-24.94} $ & ${108.4}^{+10.56}_{-15.27} $ & ${0.72}^{+0.22}_{-0.24} $ & ${-0.01}^{+0.21}_{-0.26} $ & ${0.15}^{+0.12}_{-0.12} $ & ${3.09}^{+2.57}_{-2.47} $ & ${4172}^{+1262}_{-1286} $ & $-0.61^{+0.08}_{-0.08} $   \\
                           & \makecell[cc]{\fontsize{7.5pt}{7pt}\selectfont  \texttt{SEOBNRv5HM} \\ \fontsize{7.5pt}{7pt}\selectfont  (\texttt{pBilby})}  &  ${259.97}^{+20.52}_{-20.84} $ & ${111.15}^{+9.73}_{-12.78} $ & ${0.73}^{+0.21}_{-0.22} $ & ${0.03}^{+0.21}_{-0.21} $ & - & - & ${4840}^{+1290}_{-1297} $ & -  \\[0.08cm]                            
                           & \makecell[cc]{\fontsize{7.5pt}{7pt}\selectfont  \texttt{SEOBNRv5PHM} \\ \fontsize{7.5pt}{7pt}\selectfont  (\texttt{pBilby})}  &  ${279.06}^{+36.2}_{-28.84} $ & ${116.73}^{+12.75}_{-13.84} $ & ${0.67}^{+0.24}_{-0.3} $ & ${0.1}^{+0.26}_{-0.27} $ & -  & -  & ${4194}^{+1539}_{-1651} $ & -  \\[0.08cm]                            
 \hline
 \hline
    \end{tabular*} 
   } 
%\end{adjustbox}
 \caption{
Median and $90\%$ credible intervals corresponding to GW150914 and GW190521 measured with the eccentric, aligned-spin \texttt{SEOBNRv5EHM} model for the total mass $ M $, chirp mass $ \mathcal M $, inverse mass ratio $ 1/q $, effective spin parameter $ \chi _{ \text{eff}} $,  GW eccentricity $ e _{ \text{gw}} $ and GW mean anomaly $ l_{ \text{gw}} $ measured at the reference frequency $ f_{\text{ref}} $ for starting the waveform generation ($f_{\text{ref}}  = 10\,$Hz for GW150914, and $f_{\text{ref}}  =\nolinebreak 5.5\,$Hz for GW190521, so that all the waveform modes are in band at the likelihood minimum frequency $f_{\text{min}}= 11\,$Hz), luminosity distance $ d_L $, and log-10 Bayes factor between the eccentric, aligned-spin and the QC, aligned-spin hypothesis, $\log_{10} \mathcal{B}_\text{EAS/QCAS}$ (for each event, we performed a run using the QC, aligned-spin \texttt{SEOBNRv5HM} model \cite{Pompiliv5} to calculate $\log_{10} \mathcal{B}_\text{EAS/QCAS}$).
Additionally, we include the results corresponding to the eccentric, aligned-spin \texttt{SEOBNRv4EHM\_opt} model and the QC, precessing-spin \texttt{SEOBNRv5PHM} model from Refs.~\cite{Ramos-Buades:2023yhy,RamosBuadesv5}.
We indicate the sampler for each run, either {\tt parallel} {\tt Bilby} (pBilby) \cite{Smith:2019ucc} or {\tt serial} {\tt Bilby} (sBilby)  \cite{Ashton:2018jfp,Romero-Shaw:2020owr}.
%The reference frequency for starting the waveform generation for GW150914 is chosen to be $ f_{\text{ref}}  = 10\,$Hz, while for GW190521 is $ f_{\text{ref}}  = 5.5\,$Hz, so that all the waveform modes are in band at the likelihood minimum frequency ($f_{\text{min}}= 11\,$Hz).
}
\label{tab:gwEvents}
\end{table*}

%%%%%%%%%%%%%%%%%%%%%%%%%%%

\subsubsection*{GW190521}

GW190521 is a merger-ringdown-dominated signal, with only 4 cycles in the frequency band of the detectors \cite{LIGOScientific:2020iuh}.
If this signal corresponds to a QC BBH coalescence, then the associated total mass of the binary would be $ \sim 151 \, \solarmass $, thus becoming one of the most massive BBHs detected.
However, subsequent studies have attributed to this event different origins, such as a head-on collision of exotic compact objects \cite{CalderonBustillo:2020fyi}, a nonspinning hyperbolic capture \cite{Gamba:2021gap}, and an eccentric binary merger \cite{Gayathri:2020coq,Romero-Shaw:2020thy}, although other recent studies do not find clear evidence for eccentricity  \cite{Gupte:2024jfe,Iglesias:2022xfc,Bonino:2022hkj,Ramos-Buades:2023yhy}.

We analyze GW190521 with \texttt{SEOBNRv5EHM} with {\tt parallel} {\tt Bilby} (pBilby) and {\tt serial} {\tt Bilby} (sBilby) using uniform priors on initial eccentricity $e_0$ and initial relativistic anomaly $\zeta_0$, with ranges
$e_0 \in [0,0.5]$
and
$\zeta_0\in[0,2\pi]$.
We employ priors on inverse mass ratio
$1/q \in [0.05,1]$
and chirp mass
$\mathcal{M}\in [60,200] \, \solarmass$
such that the induced priors on component masses are uniform.
The rest of the priors are chosen as in the analysis of GW150914, except for the prior on the luminosity distance which is chosen to be uniform in comoving volume instead of $\propto d_L^2$, to match the settings of Ref.~\cite{Ramos-Buades:2023yhy}.
The reference frequency for starting the waveform generation is $  f _{ \text{ref}}  = 5.5\,$Hz. 
The sampler settings are specified as in Ref.~\cite{Ramos-Buades:2023yhy},
and they are summarized in Table \ref{tab:peruntime}.

In the bottom row of Fig.~\ref{fig:gwEvents}, we show the marginalized 1D and 2D posterior distributions for chirp mass $ \mathcal M $, effective spin parameter $ \chi _{ \text{eff}} $, GW eccentricity $ e _{ \text{gw}} $, and GW mean anomaly $ l _{ \text{gw}} $ corresponding to GW190521, measured by \texttt{SEOBNRv5EHM} and other waveform approximants.
The median values and the $90 \%$ credible intervals are reported in Table \ref{tab:gwEvents}. 
For comparisons, we include in our analysis the samples of the eccentric, aligned-spin \texttt{SEOBNRv4EHM\_opt} model from Ref.~\cite{Ramos-Buades:2023yhy}, and the samples of the QC, precessing-spin \texttt{SEOBNRv5PHM} model from Ref.~\cite{RamosBuadesv5}.

For this event, we observe larger differences in some noneccentric parameters measured by \texttt{SEOBNRv5EHM} with respect to the QC, precessing-spin \texttt{SEOBNRv5PHM} model.
For example, this is the case for the chirp mass and the total mass.
Other parameters, such as the effective spin and the mass ratio, have consistent median values.
Additionally, we see a remarkable agreement in the noneccentric parameters measured by the eccentric, aligned-spin \texttt{SEOBNRv5EHM} and \texttt{SEOBNRv4EHM\_opt} models.
This indicates that the differences in the posteriors between the eccentric and QC models are due to the absence of spin precession effects in both 
eccentric models. Furthermore, between both \texttt{SEOBNRv5EHM} runs performed with different samplers, we observe good agreement, indicating that sampler systematics is a subdominant effect.

For the eccentric parameters, we find median values of the GW eccentricity of $e^\text{5.5Hz}_\text{ gw}={0.30}^{+0.16}_{-0.23}$ for \texttt{SEOBNRv5EHM} (pBilby), $e^\text{ 5.5Hz}_\text{ gw}={0.29}^{+0.16}_{-0.23} $ for \texttt{SEOBNRv5EHM} (sBilby), and $e^\text{ 5.5Hz}_\text{ gw}={0.15}^{+0.12}_{-0.12}$ for \texttt{SEOBNRv4EHM\_opt}, which seems to indicate that \texttt{SEOBNRv5EHM} has a larger preference for the eccentric hypothesis than  \texttt{SEOBNRv4EHM\_opt}. However, this is an artifact of the different priors used for the analysis.
Due to its limited accuracy, the upper limit in the eccentricity prior for \texttt{SEOBNRv4EHM\_opt} goes up to 0.3, while for \texttt{SEOBNRv5EHM} it goes up to 0.5.
A visual inspection of the eccentricity posteriors in the bottom row of Fig.~\ref{fig:gwEvents} reveals that the eccentricity posterior for \texttt{SEOBNRv5EHM} is mostly uninformative and prior dominated.
Consistently with Ref.~\cite{Ramos-Buades:2023yhy}, we find large uncertainty in the $90 \%$ credible intervals of the eccentricity posterior which, combined with an uninformative posterior distribution of the GW mean anomaly, indicates that for GW190521 the eccentricity parameter is poorly constrained. 

Furthermore, we also compute the log-10 Bayes factor between the eccentric, aligned-spin and the QC, aligned-spin hypothesis, $\log_{10} \mathcal{B}_\text{EAS/QCAS}$, by producing a run using the QC, aligned-spin \texttt{SEOBNRv5HM} model \cite{Pompiliv5}.
The main parameters obtained with this run for the \texttt{SEOBNRv5EHM} and  \texttt{SEOBNRv4EHM\_opt} models are shown in Table \ref{tab:gwEvents}.
The corresponding $\log_{10} \mathcal{B}_\text{EAS/QCAS}$ values are $-0.42^{+0.08}_{-0.08} $  and  $-0.61^{+0.08}_{-0.08} $  for \texttt{SEOBNRv5EHM} and  \texttt{SEOBNRv4EHM\_opt}, respectively.
These values indicate a slight preference for the QC hypothesis.
Comparing to Table III of Ref.~\cite{Gupte:2024jfe}, where the $\log_{10} \mathcal{B}_\text{EAS/QCAS}$ value between the \texttt{SEOBNRv4EHM} and \texttt{SEOBNRv4HM} models is found to be $-0.04$, we find slightly more negative values which can be explained because we are computing the log-10 Bayes factor between different models  \texttt{SEOBNRv5EHM}/\texttt{SEOBNRv5HM} and \texttt{SEOBNRv4EHM\_opt}/\texttt{SEOBNRv5HM} with respect to Ref.~\cite{Gupte:2024jfe}, and also because Ref.~\cite{Gupte:2024jfe} is using a different stochastic sampler, namely \texttt{DINGO} \cite{dax2021group, Dax:2021tsq, Green:2020dnx}.
Apart from these small quantitative differences, the comparison of the Bayes factors indicates a preference for the noneccentric hypothesis, which is in agreement with the results of Refs.~\cite{Ramos-Buades:2023yhy,Gupte:2024jfe}.

The poor constraint on the eccentricity posteriors can be explained by the extremely short duration of the signal and the lack of spin-precession effects (which can lead to a complicated signal morphology) in \texttt{SEOBNRv5EHM}.
Additionally, the assumption of quasicircularity at merger-ringdown employed by \texttt{SEOBNRv5EHM} (and other eccentric waveforms) could also lead to some unexpected biases due to unmodeled eccentricity effects.
Hence, to accurately measure eccentricity in high-mass systems, such as GW190521, eccentric waveform models must be extended to include spin-precession effects and eccentricity effects in the merger-ringdown part of the waveform.
We leave such extensions of the model for future work.

Finally, we focus on the efficiency of the \texttt{SEOBNRv5EHM} model reported in Table \ref{tab:peruntime}.
We observe a reduction in run-times with respect to the previous generation model \texttt{SEOBNRv4EHM\_opt} due to an efficient implementation of \texttt{SEOBNRv5EHM} in the Python infrastructure \texttt{pySEOBNR} \cite{Mihaylovv5}.

For instance, the analysis of GW150914 with {\tt parallel} {\tt Bilby} and \texttt{SEOBNRv5EHM} was performed in 2 days, while it took 3 days for \texttt{SEOBNRv4EHM\_opt}.
For {\tt serial} {\tt Bilby} and a single node with 64 cores, we see an increase in wall time ($\sim 3$ days and $20$ hours), but still being competitive for parameter estimation with similar wall times as the QC, precessing-spin model \texttt{SEOBNRv5PHM} (see Table III in Ref. \cite{RamosBuadesv5}). 
% is still consisespecially if one wants to have all the HMs in band at the starting frequency of the likelihood calculation.
%As GW150914 is consistent with a face-on GW signal, one could set the starting frequency equal to the likelihood frequency without incurring biases, since the HMs are highly suppressed for this orientation; this was the choice made for the \texttt{SEOBNRv5PHM} run \cite{RamosBuadesv5}.
%Thus, we also produce a run with only the $(\ell,|m|)=(2,2)$ modes and a starting frequency of $20\,$Hz, and we observe that the runtime reduces to $\sim 5$ days (see Table \ref{tab:peruntime}).  

For GW190521, we see that the run-times of the  {\tt parallel} {\tt Bilby} and  {\tt serial} {\tt Bilby} runs of \texttt{SEOBNRv5EHM} become comparable, and are about 2 times faster than the QC, precessing-spin \texttt{SEOBNRv5PHM} model (this is likely related to the fact that the $\text{nlive}$ value employed with \texttt{SEOBNRv5PHM} is larger than the one employed with \texttt{SEOBNRv5EHM}).
This shows that one can obtain results on the order of hours/days using \texttt{parallel/serial} {\tt Bilby}, which would make \texttt{SEOBNRv5EHM} a standard tool for parameter estimation for current and future GW events. 
Taking advantage of this efficiency, we are systematically analyzing other GW events with \texttt{SEOBNRv5EHM}, and we plan to present the results in future work.

%%%%%%%%%%%%%%%%%%%%%%%%%%%
%%%%%%%%%%%%%%%%%%%%%%%%%%%
%%%%%%%%%%%%%%%%%%%%%%%%%%%

\section{Conclusions}
\label{sec:conclusions}

The observation of GWs emitted by BBHs with non-negligible orbital eccentricity is well in sight, given the large number of events that existing GW detectors will observe in the next years.
Such a detection will serve as a smoking gun for the BBH dynamical formation channel, and ignoring eccentricity effects on GW analyses could lead to biases in parameter estimation and false violations of general relativity. 

In this work, we developed \texttt{SEOBNRv5EHM}: a new, time-domain, multipolar waveform model for eccentric BBHs with aligned spins.
It models the gravitational waveform modes $( \ell, |m| ) = \{(2, 2),\, (3, 3),\, (2, 1),\, (4, 4),\, (3, 2),\, (4, 3)\}$ and is applicable to bound orbits. The model is constructed upon the state-of-the-art, QC waveform model \texttt{SEOBNRv5HM} \cite{Pompiliv5}. \texttt{SEOBNRv5EHM} is the first EOB waveform model to include \emph{third} PN order information in the nonspinning eccentricity corrections to the EOB radiation-reaction force and to the waveform modes, as derived in the companion paper \cite{Gamboa:2024imd}.

Here, we i) outlined the main theoretical and technical ingredients of \texttt{SEOBNRv5EHM}, ii) validated its accuracy by comparing its waveforms against 441 QC and 99 eccentric NR waveforms from the SXS Collaboration, iii) quantified its robustness and speed, and iv) showed its applicability in different inference studies, namely, injection-recovery studies and analysis of two real GW events: GW150914 and GW190521.

Our results show that \texttt{SEOBNRv5EHM} is the most accurate eccentric model available.
In the zero eccentricity limit, it has a very good agreement with the accurate QC \texttt{SEOBNRv5HM} model; a faithful QC limit is an essential property for any eccentric model to avoid biases in parameter estimation due to a residual eccentricity in the waveforms \cite{Bonino:2022hkj, Ramos-Buades:2023yhy}.
For eccentric waveforms, \texttt{SEOBNRv5EHM} has a median unfaithfulness, for the (2,2) mode, of $ \sim 0.02 \% $ when compared against the set of NR eccentric waveforms employed in this work.
This is an accuracy improvement of about one order of magnitude with respect to the previous-generation \texttt{SEOBNRv4EHM} model \cite{Ramos-Buades:2021adz}, and the state-of-the-art eccentric, aligned-spin \texttt{TEOBResumS-Dal\'i} model \cite{Nagar:2024dzj}.

\texttt{SEOBNRv5EHM} is also robust across a substantial region of the parameter space, which for 
eccentric binaries is larger and more complex than that of QC binaries.
We determined a safe (conservative) region of parameter space, where the generated waveforms are physically sensible.
This region is characterized by
mass ratios $ q \in [1, 20] $,
dimensionless spin components $ \chi_{1, \, 2} \in [-0.999, 0.999] $,
eccentricities $ e \in [0, 0.45] $,
and
relativistic anomalies $ \zeta \in [0, 2 \pi] $,
for total masses $ \geq 10 \, \solarmass $
at $ \langle f _{ \text{start}} \rangle = 20\,$Hz.
Exploring outside this region of parameter space is generally acceptable, but care must be taken to ensure that the generated waveforms look physically sane, especially in challenging regions of parameter space: high eccentricities, high mass ratios, and high total masses.

Another key characteristic of the \texttt{SEOBNRv5EHM} model is its computational speed.
Typical applications of waveform models require millions of evaluations; hence, computational efficiency becomes a necessity for any waveform model.
\texttt{SEOBNRv5EHM} is efficiently implemented into the optimized Python infrastructure \texttt{pySEOBNR} \cite{Mihaylovv5}.
This enables a competitive speed for common applications of waveform models, as demonstrated by a set of waveform evaluation benchmarks conducted in this work.
\texttt{SEOBNRv5EHM} is also the first IMR eccentric waveform model to complete a review process in the LVK Collaboration.

The accuracy, robustness, and speed of \texttt{SEOBNRv5EHM} are ideal for a wide range of applications, such as Bayesian inference of source parameters, GW searches, population studies, waveform systematic analyses, improvement of other waveform models, etc.
Here, we demonstrated the applicability of \texttt{SEOBNRv5EHM} to inference studies. 
In particular, we performed a series of synthetic NR signal injections into zero noise to assess the model's ability to recover the parameters of the injected signals.
In all cases, \texttt{SEOBNRv5EHM} is able to recover the injected signals, in consistency with the very low unfaithfulness of \texttt{SEOBNRv5EHM} when compared against NR waveforms. 
Additionally, we analyzed public strain data from the LVK Collaboration corresponding to the real GW events GW150914 and GW190521.
We obtained consistency with previous results in the literature, but we showed the extended ability of \texttt{SEOBNRv5EHM} to provide tighter constraints on the inferred posterior distributions of the BHs' parameters.
Furthermore, we plan a systematic analysis of previous GW events reported by the LVK Collaboration, and also the use of \texttt{SEOBNRv5EHM} into the machine-learning inference code \texttt{DINGO} \cite{dax2021group, Dax:2021tsq, Green:2020dnx, Gupte:2024jfe}.

We note that there are several potential improvements for \texttt{SEOBNRv5EHM}.
For example, the agreement with eccentric NR waveforms could be increased by the inclusion of eccentricity effects in the merger-ringdown phase, the calibration to eccentric NR simulations, and the enhancement of the EOB Hamiltonian with post-Minkowskian results (e.g.~Ref.~\cite{Buonanno:2024byg}).
Another important option is to revisit the parametrization of the EOM employed in \texttt{SEOBNRv5EHM}.
Such parametrization was crucial in recovering the same accuracy as \texttt{SEOBNRv5HM} in the zero eccentricity limit and in achieving a very good unfaithfulness against NR waveforms.
However, it could lead to unphysical effects in the waveforms if the model is applied in challenging regions of the binary parameter space.

This research is an intermediate but fundamental step toward the development of an accurate eccentric waveform model valid for generic spin orientations.
Such a model will be important in assessing systematic errors (or biases) in parameter estimation.
However, due to the complexity of eccentric, spin-precessing binaries, only a few waveform models have been developed, so far.
Specifically, Refs.~\cite{Klein:2018ybm, Klein:2021jtd, Arredondo:2024nsl} have constructed inspiral-only models and, just very recently, Refs.~\cite{Liu:2023ldr, Gamba:2024cvy} started the development of complete IMR eccentric, spin-precessing waveform models. 
In this regard, the extension of \texttt{SEOBNRv5EHM} to generic spin orientations will be one of our main tasks for the near future.

%%%%%%%%%%%%%%%%%%%%%%%%%%%
%%%%%%%%%%%%%%%%%%%%%%%%%%%
%%%%%%%%%%%%%%%%%%%%%%%%%%%

\section*{Acknowledgments}

We thank Guglielmo Faggioli, Nihar Gupte, Benjamin Leather, Oliver Long, Philip Lynch, Maarten van de Meent, and Peter James Nee for helpful discussions.

It is with great appreciation that we thank Geraint Pratten, Md Arif Shaikh, Sylvain Marsat, Eleanor Hamilton, Shaun Nicholas Swain, and Yifan Wang for performing the LIGO-Virgo-KAGRA review of the \texttt{SEOBNRv5EHM} model.
We also thank Rossella Gamba for her support in using the \texttt{TEOBResumS-Dal\'i} code.

The computational work for this manuscript was carried out on the \texttt{Hypatia} computer cluster at the Max Planck Institute for Gravitational Physics in Potsdam.
Numerical-relativity simulations were performed at the Max Planck Computing and Data Facility on the Urania HPC system of the "Astrophysical and Cosmological Relativity” division.
M.K.'s work is supported by Perimeter Institute for Theoretical Physics. Research at Perimeter Institute is supported in part by the Government of Canada through the Department of Innovation, Science and Economic Development and by the Province of Ontario through the Ministry of Colleges and Universities.
A. Ramos-Buades is supported by the Veni research programme which is (partly) financed by the Dutch Research Council (NWO) under the grant VI.Veni.222.396; acknowledges support from the Spanish Agencia Estatal de Investigación grant PID2022-138626NB-I00 funded by MICIU/AEI/10.13039/501100011033 and the ERDF/EU; is supported by the Spanish Ministerio de Ciencia, Innovación y Universidades (Beatriz Galindo, BG23/00056) and co-financed by UIB.
M.B and L.K. are supported by  National Science Foundation Grants No.~PHY-2407742, No.~PHY-2207342, and No.~OAC-2209655, and by the Sherman Fairchild Foundation at Cornell.
This research has made use of data or software obtained from the Gravitational Wave Open Science Center (gwosc.org), a service of the LIGO Scientific Collaboration, the Virgo Collaboration, and KAGRA. This material is based upon work supported by NSF's LIGO Laboratory which is a major facility fully funded by the National Science Foundation, as well as the Science and Technology Facilities Council (STFC) of the United Kingdom, the Max-Planck-Society (MPS), and the State of Niedersachsen/Germany for support of the construction of Advanced LIGO and construction and operation of the GEO600 detector. Additional support for Advanced LIGO was provided by the Australian Research Council. Virgo is funded, through the European Gravitational Observatory (EGO), by the French Centre National de Recherche Scientifique (CNRS), the Italian Istituto Nazionale di Fisica Nucleare (INFN) and the Dutch Nikhef, with contributions by institutions from Belgium, Germany, Greece, Hungary, Ireland, Japan, Monaco, Poland, Portugal, Spain. KAGRA is supported by Ministry of Education, Culture, Sports, Science and Technology (MEXT), Japan Society for the Promotion of Science (JSPS) in Japan; National Research Foundation (NRF) and Ministry of Science and ICT (MSIT) in Korea; Academia Sinica (AS) and National Science and Technology Council (NSTC) in Taiwan.

%%%%%%%%%%%%%%%%%%%%%%%%%%%
%%%%%%%%%%%%%%%%%%%%%%%%%%%
%%%%%%%%%%%%%%%%%%%%%%%%%%%

\appendix

%\onecolumngrid

%%%%%%%%%%%%%%%%%%%%%%%%%%%
%%%%%%%%%%%%%%%%%%%%%%%%%%%

\section{
Assessing the applicability of a set of secular evolution equations in setting initial conditions for eccentric orbits
}
\label{sec:secular}

In this appendix, we discuss the use of a set of secular evolution equations for the Keplerian eccentric parameters to determine the initial conditions of eccentric binaries in challenging configurations.

To compute the initial conditions of eccentric binaries, we make use of the postadiabatic approximation~\cite{Buonanno:2000ef,Damour:2002vi,Buonanno:2005xu,Damour:2012ky}, where it is assumed that the binary evolves through a sequence of elliptic conservative orbits which slowly inspiral inwards due to GW emission.
This approximation will be less accurate for configurations in which the binary has a small relative separation and large relative velocity, as is the case for periastron passages of highly eccentric binaries, or for binaries which are close to merger.
For these challenging configurations, the mapping between the model input parameters $ (\langle M \Omega \rangle_0, e_0, \zeta_0) $ and the corresponding EOB initial values $ (r_0, p_{r_* 0}, p_{\phi,0} ) $ may become increasingly inaccurate, depending on how extreme is the configuration associated with $ (\langle M \Omega \rangle_0, e_0, \zeta_0) $.
More specifically and as mentioned previously, these configurations would correspond to systems initialized close to merger (high values of the starting orbit-averaged frequency $ \langle M \Omega \rangle_0 $), or close to a periastron passage $ (\zeta_0 \approx 0) $ for high values of the initial eccentricity $ e_0 $.

\begin{figure}%[h!]
%\centering
%\vspace{170pt}
\includegraphics[width=\linewidth]
{histogram_22_bwd_sec}
%\vspace{-10pt}
\caption{
Normalized histograms of the $ (2,2) $-mode mismatches between systems with and without backward secular evolution, for binary configurations which have a starting separation $ r_0 \leq 10 M $.
The blue histogram corresponds to a test involving $ 10^5 $ waveform evaluations over the parameter space
$ q = 3 $, $ \chi_1 = 0.5 $, $ \chi_2 =\nolinebreak -0.5 $, $ M \in \nolinebreak {[10, 200]} \solarmass $, $ e \in [0, 0.6] $, and $\zeta \in [0, 2 \pi]$ at $ \langle f _{ \text{start}} \rangle =\nolinebreak 10\,\text{Hz}$.
The yellow histogram corresponds to a broader test with $10^6$ evaluations spanning
$ q \in [1, 20] $, $ \chi_{1,2} \in {[-0.999, 0.999]} $, $ M \in \nolinebreak {[10, 200]} \solarmass $, $ e \in [0, 0.6] $, and $\zeta \in [0, 2 \pi]$ at $ \langle f _{ \text{start}} \rangle = 10\,\text{Hz}$.
}
\label{fig:histogram_bwd}
\end{figure}

To overcome this problem, if the starting separation of the binary satisfies $ r_0 < 10 M $, then in \texttt{SEOBNRv5EHM} we evolve backward in time a set of \textit{secular evolution equations} until $ r_0 = 10 M $.
Thus, we get new values for $ (\langle M\Omega \rangle, e, \zeta) $, which are then safely employed with the postadiabatic prescription for initial conditions, as described in Sec.~\ref{sec:ICs}.
The set of secular evolution equations is given by Eqs.~\eqref{eq:e}, \eqref{eq:zeta}, and \eqref{eq:dotx}, which we repeat here for completeness,
\begin{subequations}
\label{eq:secular_evo_app}
\begin{align}
\dot{e} &= -\frac{\nu e x^4}{M} \left[\frac{\left(121 e^2+304\right)}{15 \left(1-e^2\right)^{5/2}} + \text{3PN expansion}\right], \label{eq:e_app}
\\
\dot{\zeta} &= \frac{x^{3/2}}{M}\left[\frac{ (1+e \cos\zeta)^2}{\left(1-e^2\right)^{3/2}} + \text{3PN expansion}\right], \label{eq:zeta_app}
\\
\dot{x} &= \frac{2  \nu x^5}{3 M} \left[\frac{96 +292 e^2 + 37 e^4}{5 \left(1-e^2\right)^{7/2}} + \text{3PN expansion} \right]. \label{eq:dotx_app}
\end{align}
\end{subequations}

These are three ordinary differential equations for the Keplerian variables $ (e, \zeta, x = \langle M \Omega \rangle^{2/3}) $.
The variables $ e $ and $ x $ evolve secularly according to the energy and angular momentum losses due to GW emission, while $ \zeta $ evolves on the orbital time scale and contains postadiabatic contributions consistent with the emission of GWs.\footnote{
Strictly speaking, Eq.~\eqref{eq:zeta_app} is not a secular evolution equation, since $ \zeta $ evolves on the orbital timescale.
For ease of notation, we will refer to the set of equations \eqref{eq:secular_evo_app} as ``secular evolution equations''.
}
We refer the reader to the companion paper~\cite{Gamboa:2024imd} for the derivation of these equations.

To quantify the impact of using this method, we perform random waveform evaluations in a given region of parameter space using and not using the set of secular evolution equations.
More specifically, we perform two tests:
\begin{enumerate}
\item
We set $ q = 3 $, $ \chi_1 = 0.5 $, $ \chi_2 = -0.5 $, $ M \in \nolinebreak {[10, 200]} \solarmass $, $ e \in [0, 0.6] $, and $\zeta \in [0, 2 \pi]$ at $ \langle f _{ \text{start}} \rangle = 10\,\text{Hz}$, with $ 10^5 $ waveform evaluations.
\item
We set $ q \in [1, 20] $, $ \chi_{1,2} \in {[-0.999, 0.999]} $, $ M \in \nolinebreak {[10, 200]} \solarmass $, $ e \in [0, 0.6] $, and $\zeta \in [0, 2 \pi]$ at $ \langle f _{ \text{start}} \rangle = 10\,\text{Hz}$, with $ 10^6 $ waveform evaluations.
\end{enumerate}

For both tests, around 13\% of the cases have $ r_0 \leq 10M $, and hence undergo a backward evolution.
To quantify the waveform differences, we regenerate the same cases with the backward evolution deactivated, and compute the corresponding mismatches.
Figure~\ref{fig:histogram_bwd} shows the distribution of $(2,2)$-mode mismatches for the two tests; both histograms peak around or above the $ 10\% $ level, which indicates significant waveform differences.
To understand these differences, we show in Fig.~\ref{fig:wf_bwd_sec} the cases with the highest mismatch between the system undergoing a backward evolution (blue) and the one that does not (green), both having the same input parameters (in particular, same reference values of eccentricity $ e _{ \text{ref}} $, relativistic anomaly $ \zeta _{ \text{ref}} $, and starting frequency $ \langle M \Omega \rangle _{ \text{ref}} $).\footnote{
The fact that we are not evolving backward and forward the same set of equations explains why the duration of the systems undergoing a backward evolution in Fig.~\ref{fig:wf_bwd_sec} is shorter than the duration of the other systems.
Intuitively, one would expect the opposite behavior, but this depends on how similar the dynamics predicted by both sets of equations are.
}
The left panels show the maximum mismatch case ($ \mathcal M _{22} =  50.3\% $) for the first test, while the right panels show the maximum mismatch case ($ \mathcal M _{22} =  80.7\% $) for the second test;
the top panels show the $ (2,2) $ mode amplitude and frequency, along with the corresponding orbit-averaged frequencies (dashed curves); additionally, the bottom panel shows the evolution of the Keplerian eccentricity from Eq.~\eqref{eq:e} (solid curves), and the GW eccentricity $ e _{ \text{GW}} $ (dashed curves) computed from the \texttt{gw\_eccentricity} package (see footnote~\ref{fn:e_gw}).
In these plots, the vertical lines indicate the reference time at which the input values are specified.
Note that the reference time coincides with the starting time only for systems that do not undergo backward integration (green curves).

\begin{figure*}%[h!]
%\centering
%\vspace{170pt}
\includegraphics[width=0.49\linewidth]
{wf_bwd_sec_sample_single}
\hfill
\includegraphics[width=0.49\linewidth]
{wf_bwd_sec_sample_all}

%\vspace{-10pt}
\caption{
Amplitude and frequency of the $ (2,2) $ mode, as well as the eccentricity evolution of systems which have the highest mismatch when comparing the waveforms affected and not affected by the backward secular evolution (see Fig.~\ref{fig:histogram_bwd}).
Color blue indicates systems that undergo a backward evolution, while green represents those that do not.
The vertical lines indicate the time at which the reference values $ (e _{ \text{ref}}, \zeta _{ \text{ref}}, \langle f \rangle _{ \text{ref}} = 10\, \text{Hz}) $ are specified.
The binary parameters and the $ (2,2) $-mode mismatch between the blue and green waveforms are specified at the top of the panels.
The bottom panels display the eccentricity evolution predicted by Eq.~\eqref{eq:e} (solid curves), and the GW eccentricity computed with the \texttt{gw\_eccentricity} Python package (dashed curves).
}
\label{fig:wf_bwd_sec}
\end{figure*}

Focusing on the left panels of Fig.~\ref{fig:wf_bwd_sec}, we observe that this is a system with a high eccentricity at few orbits before merger.
Thus, we expect an inaccurate mapping between the reference values and the initial EOB variables.
This is precisely observed in the middle panel, where the starting orbit-averaged frequency of the system not undergoing a backward evolution (beginning of the green dashed curve) is not equal to the given reference frequency (dotted horizontal line);
in other words, the green dashed curve does not intersect the dotted horizontal line at the start of the evolution, meaning that this system starts with a notoriously different frequency than the reference value.
In contrast, for the system that undergoes a backward evolution, we observe that the orbit-averaged frequency (blue dashed curve) at the reference time is very close to the desired reference frequency.
Therefore, using the secular evolution equations allows us to generate a system that is more faithful to the given input values.
As for the eccentricity evolution, we note that both systems have the appropriate value at the reference time.
The discrepancies between $ e $ and $ e _{ \text{gw}} $ arise since these correspond to different definitions of eccentricity; hence, there is no reason for them to match.
Regarding the right panels of Fig.~\ref{fig:wf_bwd_sec}, we note that there is no discernible problem with any of the systems, and both have the expected reference values.
In this case, the high mismatch observed is due to the length of the waveforms, which allows for a significant dephasing between both systems.
The rest of the cases with high mismatch in Fig.~\ref{fig:histogram_bwd} have an analogous behavior to the examples of Fig.~\ref{fig:wf_bwd_sec}.

These results indicate that evolving backward in time the set of secular evolution equations~\eqref{eq:secular_evo_app} allows us to obtain a system which has the expected reference values, overcoming the issues that arise when the reference values correspond to a challenging binary configuration.
Future work could focus on finding a more robust method to determine EOB initial conditions that accurately match the given reference values.

\begin{figure*}%[!]
\includegraphics[width=\linewidth]
{v4EHM_v5EHM_Dali_eccsims_public_spaghetti_22_HoM_iota_1_05}
\vspace{-20pt}
\caption{
Waveform mismatches of the 28 publicly-available SXS eccentric NR waveforms against different eccentric, aligned-spin waveform models: \texttt{SEOBNRv4EHM} (first column), \texttt{SEOBNRv5EHM} (second column), and \texttt{TEOBResumS-Dal\'i} (third column), calculated over a range of total masses $ M \in [20, 200] \,\solarmass$.
The color of each curve indicates the initial value of the GW eccentricity $ e _{ \text{gw}} $ for each NR waveform. 
We highlight the curves associated with four particular NR waveforms with high mismatches (\texttt{SXS:BBH:0089}, \texttt{SXS:BBH:1149}, \texttt{SXS:BBH:1169}, and \texttt{SXS:BBH:1363}), and we discuss them in the main text.
The first row shows the $(2,2)$-mode mismatches $ \mathcal M_{22} $, and the second row the sky-and-polarization averaged, SNR-weighted mismatches $\overline{\mathcal{M}}_\mr{SNR}$ for inclination $\iota _{ \text{s}} = \pi/3$.
}
\label{fig:mismatch_spaghetti_public}
\end{figure*}
%

%%%%%%%%%%%%%%%%%%%%%%%%%%%
%%%%%%%%%%%%%%%%%%%%%%%%%%%

\section{
Mismatches against publicly available eccentric NR waveforms
}
\label{sec:public_ecc_NR_waveforms}

%\twocolumngrid

In this appendix, we obtain results for the waveform mismatches against the 28 publicly available eccentric NR waveforms from the SXS Collaboration \cite{SXS:catalog,Hinder:2017sxy, Boyle:2019kee} of the three eccentric, aligned-spin waveform models:
\texttt{SEOBNRv4EHM} \cite{Ramos-Buades:2021adz}, \texttt{SEOBNRv5EHM} (presented in this work), and \texttt{TEOBResumS-Dal\'i}~\cite{Nagar:2024dzj} (see footnote \ref{fn:dali} for the details about the specific version employed).
Thus, this appendix serves as a point of reference for past and future works that employ this set of public eccentric NR waveforms.
Additionally, we present a simple analysis of four particular cases with large mismatches for the three models.

\begin{figure*}%[h!]
%\centering
%\vspace{170pt}
\subfloat{\includegraphics[width=0.49\linewidth]{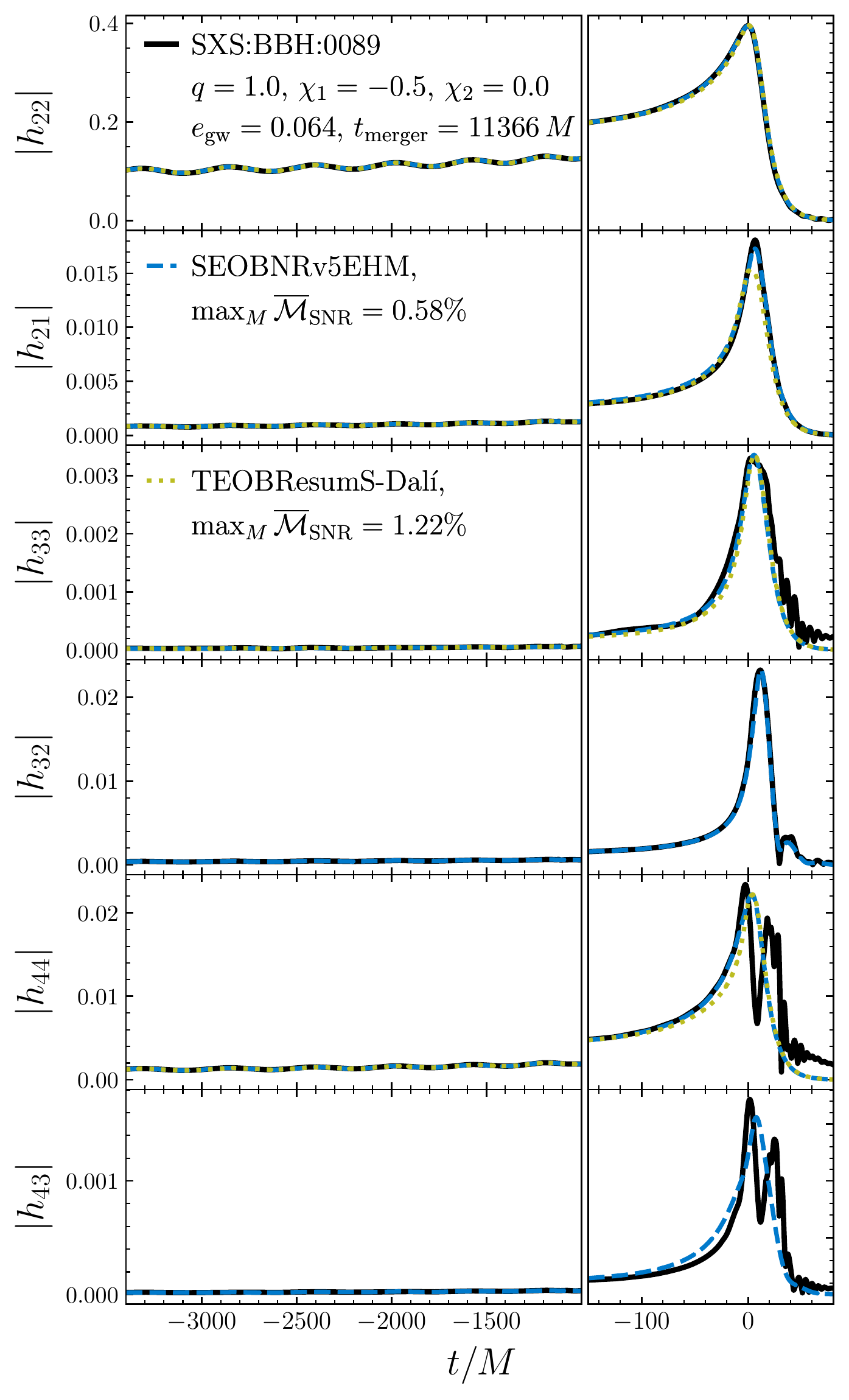}}
\hfill
\subfloat{
\raisebox{142.5pt}{
\includegraphics[width=0.49\linewidth]{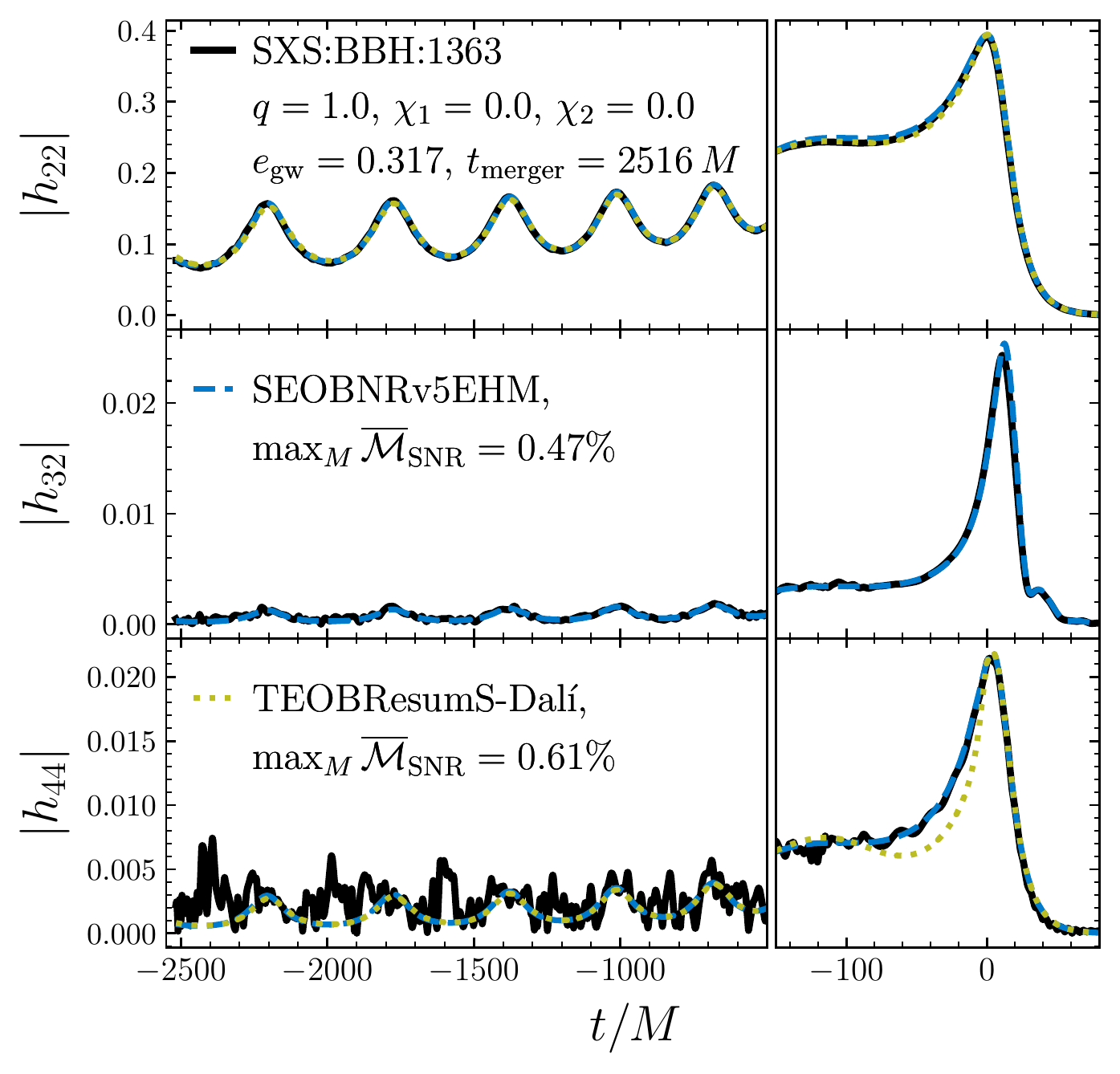}}}

%\vspace{-10pt}
\caption{
Amplitude of different waveform modes corresponding to two, publicly available SXS eccentric NR waveforms (\texttt{SXS:BBH:0089} and \texttt{SXS:BBH:1363}, represented by black curves), together with the best-fitting waveforms from the state-of-the-art, eccentric, aligned-spin waveform models \texttt{SEOBNRv5EHM} (blue, dashed) and \texttt{TEOBResumS-Dal\'i} (olive, dashed), in geometric units.
In each plot, we indicate the binary parameters: mass ratio $ q $, dimensionless spin components $ \chi_1 $ and $ \chi_2 $, initial value of the GW eccentricity $ e _{ \text{gw}} $, and time to merger $ t _{ \text{merger}} $.
In addition, we display the value of the maximum sky-and-polarization averaged, SNR-weighted mismatch $\overline{\mathcal{M}}_\mr{SNR}$ calculated over a range of total masses between $ 20 $ and $ 200 \, \solarmass $ for \texttt{SEOBNRv5EHM} and \texttt{TEOBResumS-Dal\'i}.
Note that there are no curves for the $ (3,2) $ and $ (4,3) $ modes for the \texttt{TEOBResumS-Dal\'i} model, as the reviewed version of this approximant does not produce these modes (see footnote~\ref{fn:dali}).
For \texttt{SXS:BBH:1363}, we only show the even-$ m $ modes, as this is an equal-mass, nonspinning simulation for which odd-$m$ modes are zero.
}
\label{fig:particular_waveforms_1}
\end{figure*}

In Fig.~\ref{fig:mismatch_spaghetti_public}, we show the results for the $ (2,2) $-mode mismatch $ \mathcal M_{22} $ (first row) and for the sky-and-polarization averaged, SNR-weighted mismatch $ \overline{\mathcal{M}}_\mr{SNR}$ (second row) of the 28 publicly available eccentric NR waveforms against \texttt{SEOBNRv4EHM} (first column), \texttt{SEOBNRv5EHM} (second column), and \texttt{TEOBResumS-Dal\'i} (third column).
The mismatches are computed over a range of total masses between $20$ and $200\, \solarmass$ with the faithfulness functions given in Eq.~\eqref{eq:faithfulness_ecc_22} (for $ \mathcal M_{22} $), and in Eqs.~\eqref{eq:faithfulness_ecc_hom} and \eqref{eq:faithfulness_avg_snr} (for $\overline{\mathcal{M}}_\mr{SNR}$ with $ \iota _{ \text{s}} =\nolinebreak \pi/3$).
When including HMs, we employ all modes available for each model up to $ \ell = 4 $:
for \texttt{SEOBNRv5EHM}, we include the modes $( \ell, |m| ) = \{(2, 2),\, (3, 3),\, (2, 1),\, (4, 4),\, (3, 2),\, (4, 3)\}$, and for \texttt{SEOBNRv4EHM} and \texttt{TEOBResumS-Dal\'i} we include the modes $( \ell, |m| ) = \{(2, 2),\, (3, 3),\, (2, 1),\, (4, 4)\}$.
The NR waveforms include all modes up to $ \ell = 4 $ with $ |m| > 0 $. 
We color each curve depending on the initial value of the GW eccentricity $ e _{ \text{gw}} $ as defined in Refs.~\cite{Ramos-Buades:2022lgf,Shaikh:2023ypz} (see footnote \ref{fn:e_gw} for the details about the specific version employed).

\begin{figure*}%[h]
%\ContinuedFloat
%\centering
%\vspace{170pt}
\subfloat{\includegraphics[width=0.49\linewidth]{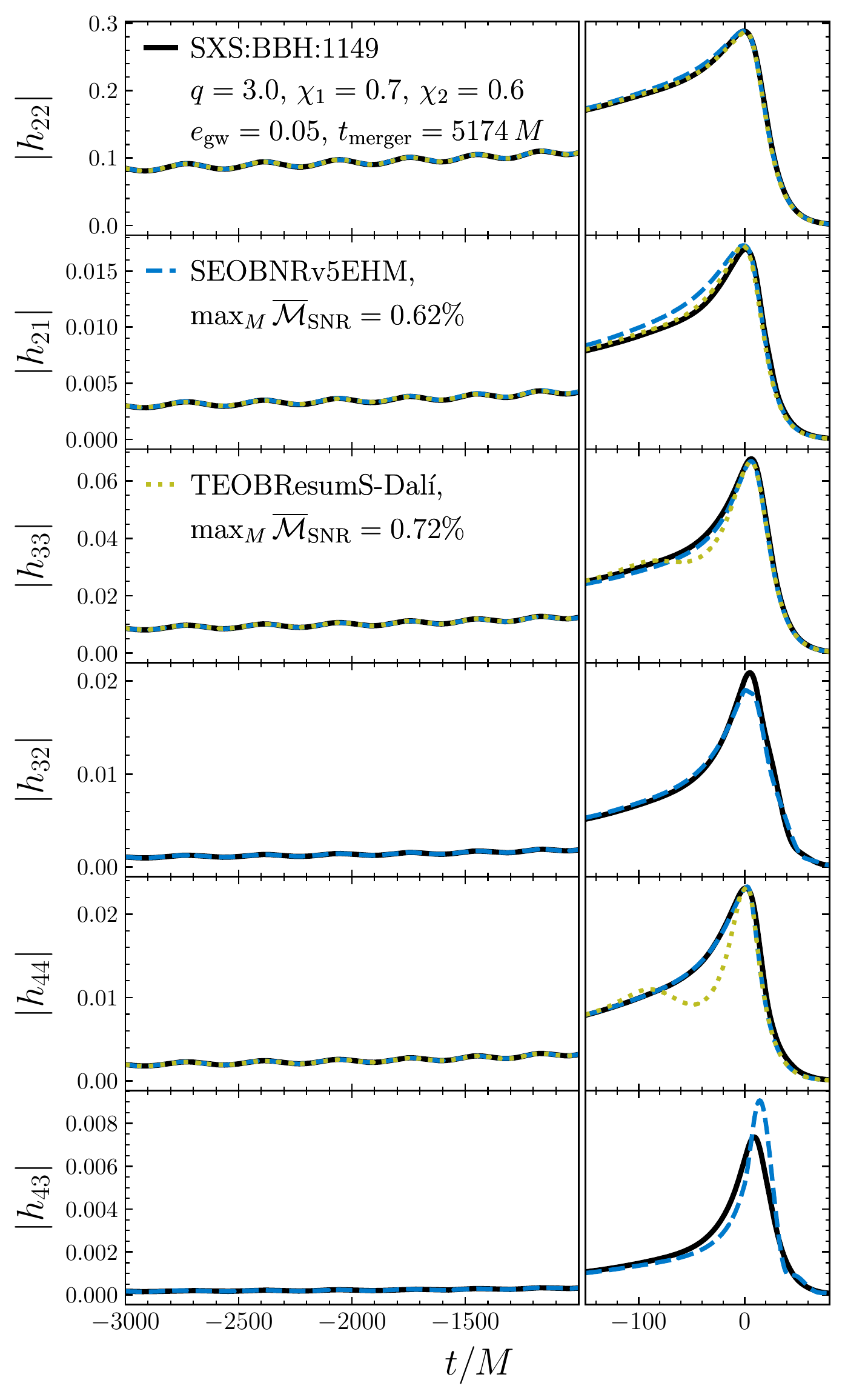}}
%\vspace{-10pt}
\hfill
\subfloat{\includegraphics[width=0.49\linewidth]{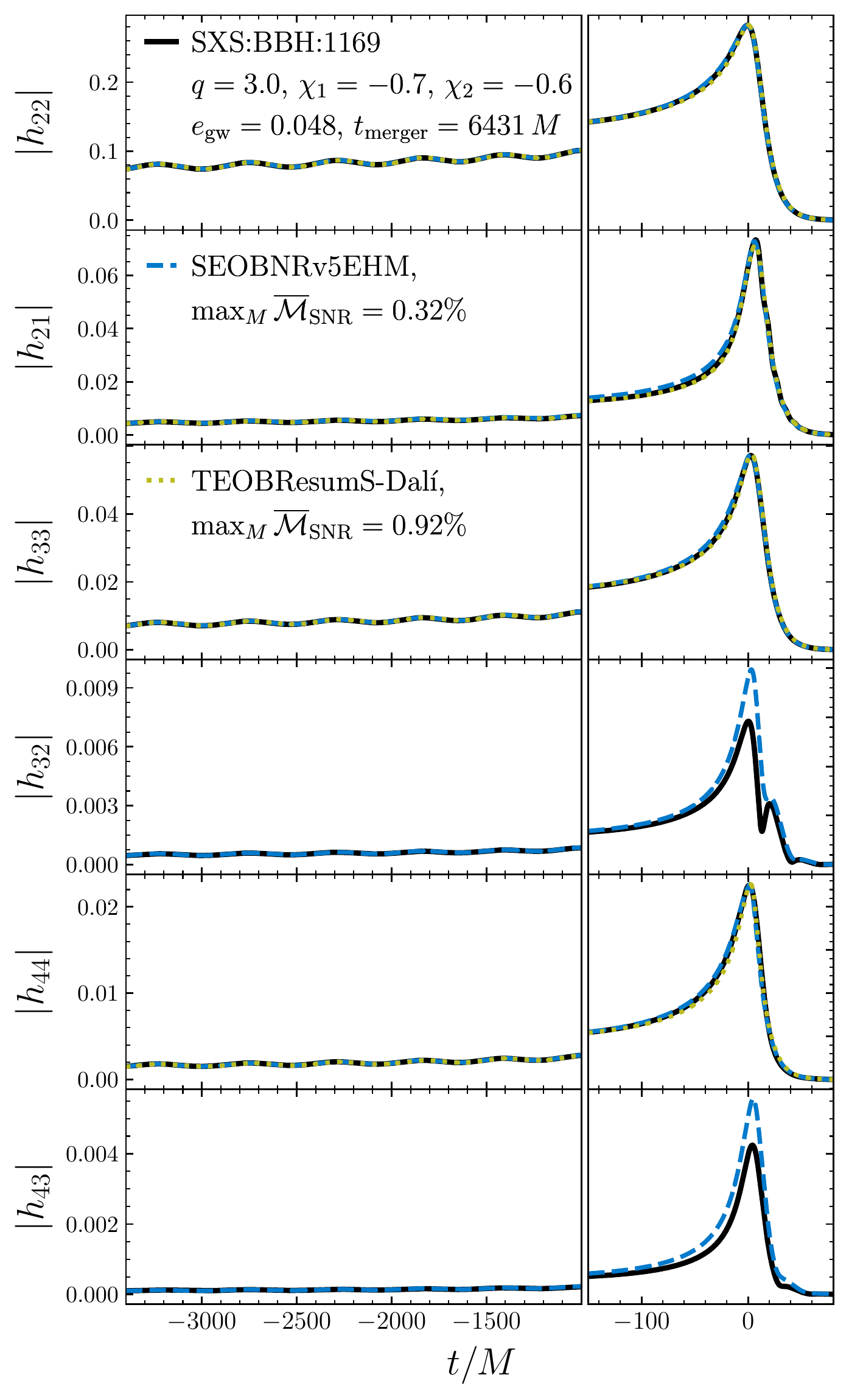}}

%\vspace{-10pt}
\caption{
Same as in Fig.~\ref{fig:particular_waveforms_1} but for the publicly available, SXS eccentric NR waveforms \texttt{SXS:BBH:1149} and \texttt{SXS:BBH:1169} (black), together with the best-fitting waveforms from the state-of-the-art, eccentric, aligned-spin waveform models \texttt{SEOBNRv5EHM} (blue, dashed) and \texttt{TEOBResumS-Dal\'i} (olive, dashed), in geometric units.
Note that there are no curves for the $ (3,2) $ and $ (4,3) $ modes for the \texttt{TEOBResumS-Dal\'i} model, as the reviewed version of this approximant does not produce these modes (see footnote~\ref{fn:dali}).
}
\label{fig:particular_waveforms_2}
\end{figure*}

Similar to our main results for a set of 99 eccentric NR simulations (see Fig.~\ref{fig:mismatch_spaghetti_all_ecc}), \texttt{SEOBNRv5EHM} demonstrates an overall improvement of one order of magnitude with respect to the previous-generation \texttt{SEOBNRv4EHM} model, and the state-of-the-art \texttt{TEOBResumS-Dal\'i} model (see Sec.~\ref{sec:comparison_nr_eccentric} for a discussion about the origin of this improved accuracy).
In addition, we note that we obtain a better accuracy for the $ (2,2) $ mode of \texttt{TEOBResumS-Dal\'i}, compared to the one reported in Ref.~\cite{Nagar:2024dzj}.
This is mainly because our method to find the waveform's parameters that correspond to an eccentric signal employs an \emph{optimization} over the parameters of the model, as described in Sec.~\ref{sec:faithfulness_ecc} (see also Sec.~\ref{sec:inner_product} for more details about the mismatch calculation).

For \texttt{SEOBNRv5EHM}, we note that there are four NR waveforms for which the mismatch $\overline{\mathcal{M}}_\mr{SNR}$ is relatively high (see the middle panel in the second row of Fig.~\ref{fig:mismatch_spaghetti_public}).
These waveforms correspond to the simulations \texttt{SXS:BBH:0089}, % ($ q = 1 $, $ \chi_1 = -0.5 $, $ \chi_2 = 0 $),
\texttt{SXS:BBH:1149}, % ($ q = 3 $, $ \chi_1 = 0.7 $, $ \chi_2 = 0.6 $),
\texttt{SXS:BBH:1169}, % ($ q = 3 $, $ \chi_1 = -0.7 $, $ \chi_2 = -0.6 $),
and
\texttt{SXS:BBH:1363}. % ($ q = 1 $, $ \chi_1 = 0 $, $ \chi_2 = 0 $).
In Figs.~\ref{fig:particular_waveforms_1} and \ref{fig:particular_waveforms_2}, we plot the amplitude of the different modes corresponding to these simulations, and we show the binary parameters of the corresponding NR waveform: mass ratio $ q $, dimensionless spin components $ \chi_1 $ and $ \chi_2 $, initial value of the GW eccentricity $ e _{ \text{gw}} $, and time to merger $  t _{ \text{merger}} $.
Additionally, we include the amplitudes of the best-fitting waveforms for the two state-of-the-art models \texttt{SEOBNRv5EHM} and \texttt{TEOBResumS-Dal\'i}, along with the associated value of the sky-and-polarization averaged, SNR-weighted mismatch $\overline{\mathcal{M}}_\mr{SNR}$ maximized over the considered range of total masses.

The high mismatches for these four eccentric NR waveforms are associated with inaccuracies in the modeling of the merger-ringdown phase and the quality of the NR waveforms.
In Fig.~\ref{fig:particular_waveforms_1}, we observe that the waveform corresponding to the simulation \texttt{SXS:BBH:0089} has prominent numerical noise in the merger-ringdown.
Similarly, the waveform associated with the simulation \texttt{SXS:BBH:1363} contains plenty of numerical noise throughout the entire, relatively short simulation ($ t _{ \text{merger}} = 2516 \, M $).
In contrast, the waveforms corresponding to \texttt{SXS:BBH:1149} and \texttt{SXS:BBH:1169} in Fig.~\ref{fig:particular_waveforms_2} have no noticeable numerical noise, but their merger-ringdown is considerably different from the one predicted by the \texttt{SEOBNRv5EHM} and \texttt{TEOBResumS-Dal\'i} models.

From this inspection, we should expect i) an increasing mismatch with respect to the total mass for the NR waveforms \texttt{SXS:BBH:0089}, \texttt{SXS:BBH:1149}, and \texttt{SXS:BBH:1169}, since the problems are restricted to the merger-ringdown phase, and ii) a decreasing mismatch with respect to the total mass for \texttt{SXS:BBH:1363}, since the numerical noise is affecting the inspiral part of the waveform.
This is precisely what we observe, and it is more noticeable for \texttt{SEOBNRv5EHM} thanks to its high faithfulness against the other NR waveforms.

%%%%%%%%%%%%%%%%%%%%%%%%%%%
%%%%%%%%%%%%%%%%%%%%%%%%%%%
%\clearpage

\section{Improvement due to the 3PN eccentricity corrections to the EOB RR force and waveform modes}
\label{sec:mms_2PN}

\begin{figure*}%[H]
\includegraphics[width=\linewidth]
{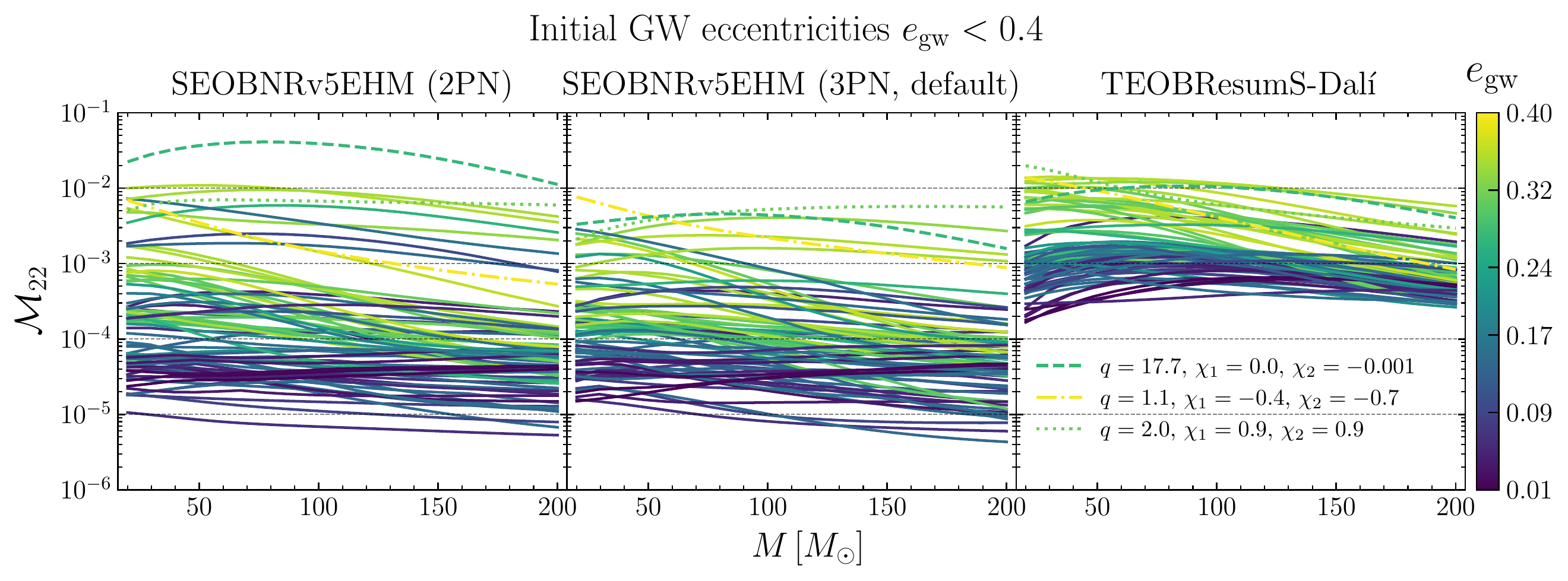}

\vspace{1pt}
\includegraphics[width=\linewidth]
{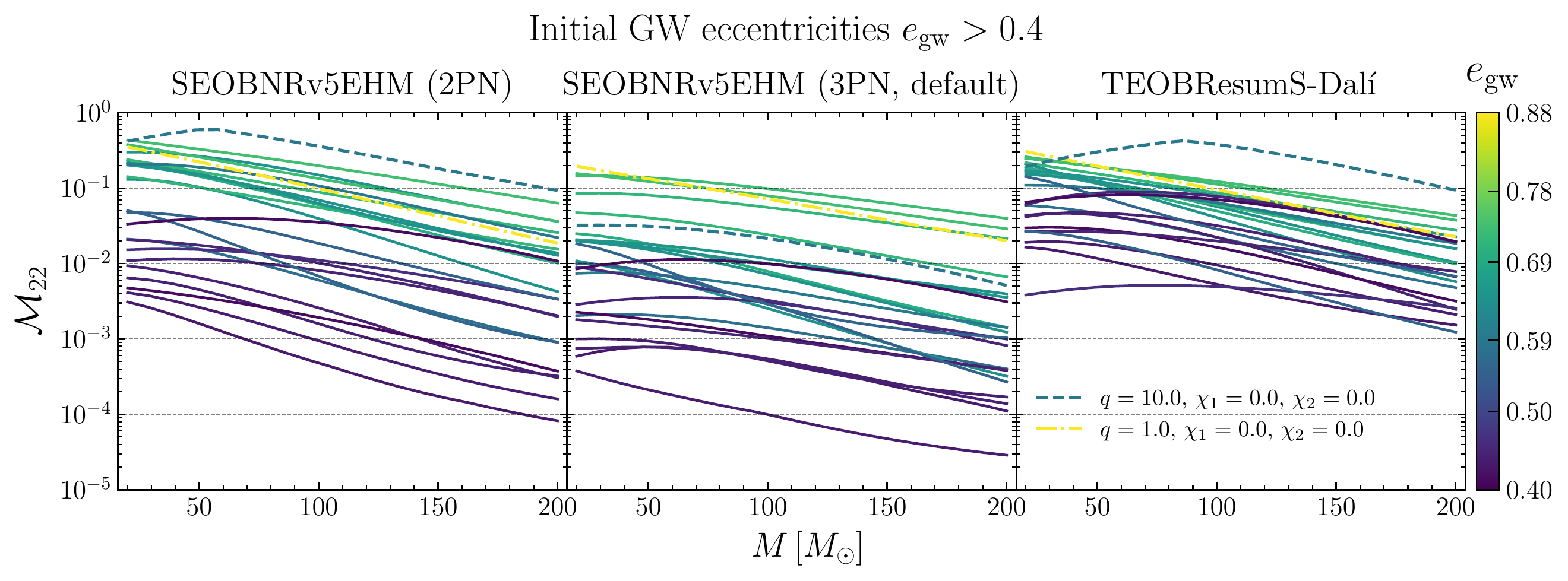}
\vspace{-25pt}
\caption{
\emph{Top panel:}
$(2,2)$-mode mismatches between a subset of 75 eccentric NR waveforms with initial GW eccentricities $ e _{ \text{gw}} < 0.4 $ and different eccentric, aligned-spin approximants, calculated over a range of total masses $ M \in [20, 200] \,\solarmass$.
The first column corresponds to a version of the \texttt{SEOBNRv5EHM} model with only 2PN eccentricity corrections in the EOB RR force and gravitational waveform modes, the second column corresponds to the \emph{default} \texttt{SEOBNRv5EHM} model which has 3PN eccentricity corrections, and the third column corresponds to the \texttt{TEOBResumS-Dal\'i} model which has 2PN eccentricity corrections.
The color of each curve indicates the initial value of the GW eccentricity $ e _{ \text{gw}} $ for each NR waveform.
The different line styles highlight the cases with the worst maximum mismatch.
\emph{Botttom panel:}
The same as in the top panel, but for 24 eccentric NR waveforms with initial GW eccentricities $ e _{ \text{gw}} > 0.4 $.
}
\label{fig:mismatch_spaghetti_all_ecc_2PN}
\end{figure*}

In this appendix, we show that the 3PN eccentricity corrections to the EOB RR force and waveform modes are important for achieving good accuracy for higher eccentricities.

Specifically, in Fig.~\ref{fig:mismatch_spaghetti_all_ecc_2PN}, we show the results for the $ (2,2) $-mode mismatch $ \mathcal M_{22} $ between the 99 eccentric NR waveforms employed in this work and a version of \texttt{SEOBNRv5EHM} which only has 2PN eccentricity corrections (first column), the \emph{default} version of the \texttt{SEOBNRv5EHM} model which has 3PN eccentricity corrections (second column), and the \texttt{TEOBResumS-Dal\'i} model~\cite{Nagar:2024dzj} (third column; see footnote \ref{fn:dali} for the details about the specific version employed) which employs generic-orbit Newtonian prefactors for the RR force and waveform modes, as well as 2PN eccentricity corrections for the radial component of the RR force.
The 99 waveforms are split into a subset of 75 waveforms with initial GW eccentricity below $ 0.4 $ (top panel) and a subset of 24 waveforms with initial GW eccentricity above $ 0.4 $ (bottom panel).
The mismatches are computed over a range of total masses between $20$ and $200\, \solarmass$ with the faithfulness function given in Eq.~\eqref{eq:faithfulness_ecc_22}.
We color each curve depending on the initial value of the GW eccentricity $ e _{ \text{gw}} $ as defined in Refs.~\cite{Ramos-Buades:2022lgf,Shaikh:2023ypz} (see footnote \ref{fn:e_gw} for the details about the specific version employed).
Note that the information in Fig.~\ref{fig:mismatch_spaghetti_all_ecc_2PN} regarding the models \texttt{SEOBNRv5EHM} (default) and \texttt{TEOBResumS-Dal\'i} is also presented in Fig.~\ref{fig:mismatch_spaghetti_all_ecc}, but in Fig.~\ref{fig:mismatch_spaghetti_all_ecc_2PN} we employ the initial value of GW eccentricity $ e _{ \text{gw}} = 0.4 $, instead of $ e _{ \text{gw}} = 0.5 $, to separate the set of NR waveforms.
Histograms of maximum mismatches across the considered total mass interval are presented in Fig.~\ref{fig:mismatch_distributions_2PN} of Sec.~\ref{sec:comparison_nr_eccentric}, where we also include information about the sky-and-polarization-averaged, SNR-weighted mismatches.

From Fig.~\ref{fig:mismatch_spaghetti_all_ecc_2PN}, we observe that the largest differences between the 2PN-version and the default \texttt{SEOBNRv5EHM} model appear for NR waveforms that have moderate-to-large values of initial GW eccentricities;
in particular, the 3PN eccentricity corrections become important for initial $ e _{ \text{gw}} \gtrsim 0.4$.
The accuracy improvement for large eccentricities is expected since a higher PN order would reduce the waveform dephasing at each (high-velocity) periastron passage.
In fact, without the 3PN eccentricity corrections, the accuracies of \texttt{SEOBNRv5EHM} and \texttt{TEOBResumS-Dal\'i} become comparable for systems with initial $ e _{ \text{gw}} \gtrsim 0.4$, which makes sense since the \texttt{TEOBResumS-Dal\'i} model employs 2PN eccentricity corrections to the EOB RR force.
(For low eccentricities, the accuracy of \texttt{SEOBNRv5EHM} is better due to the calibration to QC NR waveforms inherited from the \texttt{SEOBNRv5HM} model~\cite{Pompiliv5}, as discussed in Sec.~\ref{sec:comparison_nr_eccentric}.)

%%%%%%%%%%%%%%%%%%%%%%%%%%%
%%%%%%%%%%%%%%%%%%%%%%%%%%%
%\clearpage

\section{
Alternatives to the RR force and waveform modes
}
\label{sec:forces_modes}

\begin{figure*}%[H]
\includegraphics[width=\linewidth]
{spaghetti_22_forces}
\vspace{-25pt}
\caption{
$(2,2)$-mode mismatches between 99 eccentric NR waveforms and different versions of the \texttt{SEOBNRv5E} RR force and waveform modes, calculated over a range of total masses $ M \in [20, 200] \,\solarmass$.
The first panel corresponds to the default \texttt{SEOBNRv5EHM} model,
the second panel corresponds to the \texttt{SEOBNRv5HM} model with eccentric initial conditions,
and the third to fifth panels correspond to different model variations where the expressions for the EOB RR force and/or waveform modes either include or omit eccentricity corrections.
The color of each curve indicates the initial value of the GW eccentricity $ e _{ \text{gw}} $ for each NR waveform.
}
\label{fig:spaghetti_22_forces}
\end{figure*}

This appendix considers different prescriptions for the EOB RR force and waveform modes in the \texttt{SEOBNRv5EHM} model.

The \texttt{SEOBNRv5E} RR force and waveform modes are given by Eqs.~\eqref{eq:eccRRforce} and \eqref{eq:eob_modes}, which we repeat here for completeness
\begin{subequations}
\label{eq:eccRRforce_ap}
\begin{align}
\mF_\phi &= \mF_\phi^\text{modes}\, \mF_\phi^\text{ecc}(x,e,\zeta), \\
\mF_r &= \frac{p_r}{p_\phi} \ \mF_\phi^\text{modes} \,\mF_r^\text{ecc}(x,e,\zeta),\\
\mF_\phi^\text{modes} &= -\frac{M^2 \Omega}{8 \pi} \sum_{\ell=2}^8 \sum_{m=1}^{\ell} m^2\left|d_L \s h_{\ell m}^{\text{F}}\right|^2 \!,
\end{align}
\end{subequations}
\begin{equation}
\label{eq:eob_modes_ap}
h_{\ell m}^\text{F}=h_{\ell m}^\text{F,\,qc} (x) \,h_{\ell m}^\text{ecc}(x,e,\zeta).
\end{equation}
where $ h_{\ell m}^\text{F,\s\s qc} $ represents the factorization of QC modes employed in \texttt{SEOBNRv5HM} \cite{Pompiliv5}, and $ \mF_\phi^\text{ecc} $, $ \mF_r^\text{ecc} $, and $h_{\ell m}^\text{ecc} $  are the eccentricity corrections to the RR force and modes, respectively, given as functions of the Keplerian eccentricity $ e $, the relativistic anomaly $ \zeta $, and the dimensionless orbit-averaged orbital frequency $ x = \langle M \Omega \rangle^{2/3} $.
These eccentricity corrections contain nonspinning contributions up to the 3PN order, and are calculated in the companion paper~\cite{Gamboa:2024imd}.
We note that in \texttt{SEOBNRv5EHM} we employ the \textit{eccentric modes \eqref{eq:eob_modes_ap} in the RR force}, though this is not the only possible choice.
For example, one could employ the QC expressions for the modes in the factor $ \mF_\phi^\text{modes} $, as long as the eccentricity corrections $ \mF_\phi^\text{ecc} $ and $ \mF_r^\text{ecc} $ are modified accordingly.

In this appendix, we consider four alternatives to the default \texttt{SEOBNRv5E} RR force and waveform modes:
\begin{enumerate}
\item
\textit{\texttt{SEOBNRv5HM} RR force and modes.}
\\
We take the expressions employed in the QC model \texttt{SEOBNRv5HM} \cite{Pompiliv5} (modulo the change commented in footnote~\ref{fn:vphi}), and we supplement them with the prescription for eccentric initial conditions employed in \texttt{SEOBNRv5EHM}.
There is no dependence of the equations of motion on the Keplerian parameters $ (x, e, \zeta) $.
\item
\textit{QC RR force and QC modes.}
\\
Similarly, we take the \texttt{SEOBNRv5HM} expressions for the RR force and modes \cite{Pompiliv5}, but the dependence of the modes on the instantaneous angular velocity $ \Omega $ is replaced by a dependence on the orbit-averaged angular velocity $ \langle \Omega \rangle $.
This is achieved by setting $ \mF_\phi^\text{ecc} = \mF_r^\text{ecc} = h_{\ell m}^\text{ecc} = 1$ in Eqs.~\eqref{eq:eccRRforce_ap} and \eqref{eq:eob_modes_ap}.
The value of $ x $ entering in $ h_{\ell m}^\text{F,\,qc} $ is obtained from the coupled Keplerian equations~\eqref{eq:e}-\eqref{eq:xOmega}.
\item
\textit{Eccentric RR force and QC modes.}
\\
We use the complete expressions for $ \mF_\phi^\text{ecc} $, $ \mF_r^\text{ecc} $, and $h_{\ell m}^\text{ecc} $ in the RR force, but we set $h_{\ell m}^\text{ecc} = 1 $ for the modes given to the user.
\item
\textit{QC RR force and eccentric modes.}
\\
We set $ \mF_\phi^\text{ecc} = \mF_r^\text{ecc} = h_{\ell m}^\text{ecc} = 1$ in the RR force, but we use the complete expressions for $h_{\ell m}^\text{ecc}$ in the modes given to the user.
This choice is analogous to the \texttt{SEOBNRv4EHM} model~\cite{Ramos-Buades:2021adz}, in which eccentricity corrections are included only in the modes given to the user and not in the RR force.
\end{enumerate}

To quantify the accuracy of these alternative models, we show in Fig.~\ref{fig:spaghetti_22_forces} the corresponding $ (2,2) $-mode mismatches against the set of 99 eccentric NR simulations employed in this work.
One observation from this figure is that the eccentricity corrections in the RR force allow for a good performance for moderate eccentricities;
without eccentricity corrections in the RR force, the bulk of mismatches for simulations with moderate eccentricities lies around or above the $ 1\% $ level.
Another observation is that there is a similar accuracy for high-eccentricity simulations between the default \texttt{SEOBNRv5EHM} model and the model with eccentric RR force but QC modes.
This fact suggests that the current model is significantly limited by the RR force, and not by the waveform modes.
For example, the factorization of the eccentric RR force and/or the coupling to the Keplerian evolution equations are choices that could be limiting the model's accuracy.
%Another observation from Fig.~\ref{fig:spaghetti_22_forces} is that the model with eccentric RR force but QC modes is less accurate for low eccentricities than the default model.
%This means that the accuracy's bottleneck for low eccentricities is in the waveform modes.

%
\begin{figure*}%[h]
%\ContinuedFloat
%\centering
%\vspace{170pt}
\subfloat{\includegraphics[width=0.485\linewidth]{h22_sync}}
%\vspace{-10pt}
\hfill
\subfloat{\includegraphics[width=0.485\linewidth]{h22_desync}}

\subfloat{\includegraphics[width=0.48\linewidth]{traj_sync}}
\hspace{10pt}
%\hfill
\subfloat{\includegraphics[width=0.48\linewidth]{traj_desync}}

%\vspace{-10pt}
\caption{
Amplitude and frequency of the $ (2,2) $ mode (top panels), and the trajectory in the $ (r, \zeta) $ polar plane (bottom panels) are shown for a standard eccentric binary (left panels) and for a system affected by a desynchronization between the Keplerian and Hamiltonian evolution equations (right panels).
In the standard case, the envelopes of the amplitude and frequency exhibit a monotonic increase during the inspiral.
In contrast, the \textit{desynchronized} system displays unphysical modifications of such envelopes, which arise due to inaccuracies in the prediction of the periastron precession, characterized by the value of the relativistic anomaly $ \zeta $.
}
\label{fig:trajectories_waveforms}
\end{figure*}

Overall, accurate expressions for the RR force and waveform modes are required for a good performance.
A good accuracy in the QC limit is also an essential aspect of any eccentric waveform model, as the eccentric waveforms will have, at most, comparable accuracies to the underlying QC model.
Future eccentric models would benefit from methods that improve the accuracy for QC waveforms, as well as better treatments to the eccentric RR force and waveform modes.

%%%%%%%%%%%%%%%%%%%%%%%%%%%
%%%%%%%%%%%%%%%%%%%%%%%%%%%
%\clearpage

\section{
Desynchronization of the binary dynamics in the \texttt{SEOBNRv5EHM} model
}
\label{sec:desynchronization}

In this appendix, we discuss a problem appearing in the \texttt{SEOBNRv5EHM} model when it is applied in challenging regions of the parameter space of eccentric binaries.

The equations of motion employed in \texttt{SEOBNRv5EHM} are given by Eqs.~\eqref{eq:dot_r}-\eqref{eq:xOmega}.
This is a set of six ordinary differential equations plus an algebraic equation for the EOB variables $ (r, \phi, p_{r_*}, p_\phi) $ and the Keplerian parameters $ (x = \nolinebreak \langle M \Omega \rangle^{2/3}\!, e, \zeta)$;
thus, we employ the usual four EOB equations (Hamilton's equations), along with three equations in PN-expanded form.
This set of equations is formally consistent up to 3PN order, and it allows us to generate accurate eccentric waveforms, as demonstrated in Sec.~\ref{sec:comparison_nr_eccentric}.
However, this set is overdetermined since eccentric, aligned-spin binaries only have 4 degrees of freedom.
This poses a problem in certain regions of the parameter space (mostly large spins and eccentricity) in which the PN and EOB two-body dynamics are too different from each other. In these cases, coupling those equations introduces inaccuracies that ultimately \textit{desynchronize} the equations of motion.

In Fig.~\ref{fig:trajectories_waveforms}, we show examples of normal (synchronized) and desynchronized binaries.
The top panels show the amplitude and frequency of the $ (2,2) $ mode, while the bottom panels show the trajectory of the corresponding binary in the $ (r, \zeta) $ plane;
the two systems under consideration have the same binary parameters, except for the spin values.
In a synchronized system, the value of the relativistic anomaly $ \zeta $ tracks the periastron precession of the orbit, as required by Eq.~\eqref{eq:Kep_param};
hence, in the $ (r, \zeta) $ plane one should observe a nonprecessing ellipse which slowly shrinks due to the emission of GWs.\footnote{
The plots of the trajectories in Fig.~\ref{fig:trajectories_waveforms} have a nonsmooth behavior due to the discrete time steps taken by the numerical integrator.
}
Moreover, the waveform amplitude should exhibit a monotonically increasing envelope during the inspiral phase, since the orbit-averaged frequency increases as the binary components approach each other.
In contrast, in a problematic system, the evolution of $ \zeta $ is not consistent with the periastron precession predicted by Hamilton's equations.
This causes a desynchronization of the dynamics, as seen in the bottom right panel of Fig.~\ref{fig:trajectories_waveforms}, and an unphysical modification of the waveform, as represented in the top right panel.

Overall, the desynchronization of the equations of motion appears when there are significant differences between the two-body dynamics predicted by EOB and PN equations, and it causes an unphysical modification to the amplitude and frequency of the waveform.
Any challenging configuration is a potential trigger of this problem;
this is the case for very high eccentricities, high values of spins, large mass ratios, and/or very long evolutions.
All these configurations accumulate errors which could trigger a desynchronization, as seen in the right panels of Fig.~\ref{fig:trajectories_waveforms}.

By inspecting the monotonicity of the waveform's envelope, we determined a conservative region of the parameter space in which we detected no desynchronization issue.
This region is specified in Sec.~\ref{sec:robustness}, and we repeat it here for completeness:
mass ratios $ q \in [1, 20] $,
dimensionless spin components $ \chi_{1, \, 2} \in [-0.999, 0.999] $,
eccentricities $ e \in [0, 0.45] $,
and
relativistic anomalies $ \zeta \in [0, 2 \pi] $,
for total masses $ \geq 10 \, \solarmass $
at $ \langle f _{ \text{start}} \rangle = 20\,$Hz.
When using the model outside this region, we recommend visually inspecting some sample waveforms.

%%%%%%%%%%%%%%%%%%%%%%%%%%%
%%%%%%%%%%%%%%%%%%%%%%%%%%%
%\clearpage

\onecolumngrid

\section{
Summary of the eccentric NR waveforms employed in this work
}
\label{sec:tables}

\vspace{-20pt}

% Store the same table number for splitted table
\newcounter{tabletemp}
\setcounter{tabletemp}{\value{table}}

{
\renewcommand{\arraystretch}{0.7} % Row spacing
\begin{table}[H]%[ht]
\centering
\caption{
Summary of the 99 eccentric, aligned-spin SXS NR simulations employed in this work, along with the optimum parameters that produce the best-fitting waveforms for each of the eccentric, aligned-spin approximants considered in this work.
We include 28 publicly available waveforms (with SXS IDs: \texttt{SXS:BBH:0089}--\texttt{SXS:BBH:1374}) \cite{Hinder:2017sxy, Boyle:2019kee} plus 71 private waveforms \cite{Ramos-Buades:2022lgf}.
For each simulation, we list the mass ratio $ q = m_1 / m_2 \geq 1 $, the dimensionless spin components $ \chi_1 $ and $ \chi_2 $, the GW eccentricity $ e _{ \text{gw}} $ at a reference orbit-averaged $ (2,2) $-mode dimensionless frequency $ \langle M \omega_{22} \rangle $ (both corresponding to their starting values calculated with the \texttt{gw\_eccentricity} Python package; see footnote \ref{fn:e_gw}), and the number of periastron passages $ N_p $.
Additionally, for each waveform model, we list the initial (input) eccentricity $ e_0 $, dimensionless orbit-averaged orbital frequency $\langle M \Omega \rangle_0$, and the maximum $ (2,2) $-mode mismatch, $ \mathcal M _{ 22} ^{ \text{max}} \equiv \max_M \mathcal M _{ 22}$, calculated over the range of total masses between $ 20$ and $ 200 \, \solarmass $.
The procedure to find the optimum values for each model and for calculating the mismatches is detailed in Sec.~\ref{sec:faithfulness}.
The information in this table is also provided in an ancillary file (see footnote \ref{fn:ancillary}).
}
%\small
\vspace{8pt}
\label{tab:simulation_data}
\begin{tabular}{l | D{.}{.}{1.1} D{.}{.}{3.3} D{.}{.}{2.2} D{.}{.}{2.3} D{.}{.}{0.1} D{.}{.}{2.0} | D{.}{.}{1.3} D{.}{.}{0.2} D{.}{.}{0.2} | D{.}{.}{1.3} D{.}{.}{0.2} D{.}{.}{0.2} | D{.}{.}{1.3} D{.}{.}{0.2} D{.}{.}{0.2}} % The combination x.y increases or decreases the spacing to the left or right for each row, with each column aligned at the decimal point
 & \multicolumn{6}{c}{Physical properties} & \multicolumn{3}{|c}{\makecell[c]{\texttt{SEOBNRv4EHM}\\[-1pt]optimum values}} & \multicolumn{3}{|c}{\makecell[c]{\texttt{SEOBNRv5EHM}\\[-1pt]optimum values}} & \multicolumn{3}{|c}{\makecell[c]{\texttt{TEOBResumS-Dal\'i}\\[-1pt]optimum values}} \\
\hline
\hline
\multicolumn{1}{c|}{$\text{SXS ID}$} & \multicolumn{1}{c}{$q$} & \multicolumn{1}{c}{$\chi_{1}$} & \multicolumn{1}{c}{$\chi_{2}$} & \multicolumn{1}{c}{$e_\text{gw}$} & \multicolumn{1}{c}{$\langle M \omega_{22} \rangle$} & \multicolumn{1}{c|}{$N _{ \text{p}}$} & \multicolumn{1}{c}{$e_0$} & \multicolumn{1}{c}{$\langle M \Omega \rangle_0$} & \multicolumn{1}{c|}{\makecell[c]{$ \mathcal M_{22} ^{ \text{max}}$\\$[\%]$}} & \multicolumn{1}{c}{$e_0$} & \multicolumn{1}{c}{$\langle M \Omega \rangle_0$} & \multicolumn{1}{c|}{\makecell[c]{$ \mathcal M_{22} ^{ \text{max}}$\\$[\%]$}} & \multicolumn{1}{c}{$e_0$} & \multicolumn{1}{c}{$\langle M \Omega \rangle_0$} & \multicolumn{1}{c}{\makecell[c]{$ \mathcal M_{22} ^{ \text{max}}$\\$[\%]$}} \\
\hline
\hline
\texttt{SXS:BBH:0089} & 1.0 & -0.5 & 0.0 & 0.064 & 0.023 & 21 & 0.071 & 0.01 & 0.1 & 0.069 & 0.01 & 0.017 & 0.069 & 0.01 & 0.16 \\
\texttt{SXS:BBH:0321} & 1.22 & 0.33 & -0.44 & 0.064 & 0.037 & 8 & 0.101 & 0.014 & 0.117 & 0.069 & 0.017 & 0.007 & 0.072 & 0.014 & 0.134 \\
\texttt{SXS:BBH:0322} & 1.22 & 0.33 & -0.44 & 0.081 & 0.038 & 7 & 0.113 & 0.015 & 0.119 & 0.093 & 0.017 & 0.007 & 0.084 & 0.018 & 0.128 \\
\texttt{SXS:BBH:0323} & 1.22 & 0.33 & -0.44 & 0.135 & 0.036 & 8 & 0.18 & 0.015 & 0.201 & 0.147 & 0.017 & 0.011 & 0.15 & 0.015 & 0.097 \\
\texttt{SXS:BBH:0324} & 1.22 & 0.33 & -0.44 & 0.285 & 0.036 & 7 & 0.392 & 0.011 & 0.968 & 0.332 & 0.013 & 0.017 & 0.291 & 0.015 & 0.275 \\
\texttt{SXS:BBH:1136} & 1.0 & -0.75 & -0.75 & 0.099 & 0.041 & 4 & 0.144 & 0.017 & 0.203 & 0.106 & 0.018 & 0.028 & 0.103 & 0.017 & 0.162 \\
\texttt{SXS:BBH:1149} & 3.0 & 0.7 & 0.6 & 0.05 & 0.036 & 15 & 0.1 & 0.014 & 0.194 & 0.06 & 0.015 & 0.044 & 0.053 & 0.018 & 0.413 \\
\texttt{SXS:BBH:1169} & 3.0 & -0.7 & -0.6 & 0.048 & 0.029 & 13 & 0.062 & 0.013 & 0.173 & 0.05 & 0.014 & 0.018 & 0.057 & 0.012 & 0.109 \\
\texttt{SXS:BBH:1355} & 1.0 & 0.0 & 0.0 & 0.077 & 0.038 & 8 & 0.097 & 0.016 & 0.088 & 0.081 & 0.018 & 0.003 & 0.069 & 0.018 & 0.106 \\
\texttt{SXS:BBH:1356} & 1.0 & 0.0 & 0.0 & 0.133 & 0.028 & 14 & 0.173 & 0.012 & 0.153 & 0.16 & 0.012 & 0.007 & 0.15 & 0.012 & 0.127 \\
\texttt{SXS:BBH:1357} & 1.0 & 0.0 & 0.0 & 0.149 & 0.036 & 8 & 0.181 & 0.016 & 0.176 & 0.181 & 0.014 & 0.004 & 0.163 & 0.016 & 0.096 \\
%\hline
%\hline
\end{tabular}
\end{table}
}

\clearpage

% Assign the same table number
\setcounter{table}{\value{tabletemp}}
{
\renewcommand{\arraystretch}{0.7} % Row spacing
\begin{table}%[b]
\centering
\caption{ \emph{Continued}.}
%\small
\vspace{8pt}
\begin{tabular}{l | D{.}{.}{2.1} D{.}{.}{3.3} D{.}{.}{2.2} D{.}{.}{2.3} D{.}{.}{0.1} D{.}{.}{2.0} | D{.}{.}{1.3} D{.}{.}{0.2} D{.}{.}{0.2} | D{.}{.}{1.3} D{.}{.}{0.2} D{.}{.}{0.2} | D{.}{.}{1.3} D{.}{.}{0.2} D{.}{.}{0.2}} % The combination x.y increases or decreases the spacing to the left or right for each row, with each column aligned at the decimal point
 & \multicolumn{6}{c}{Physical properties} & \multicolumn{3}{|c}{\makecell[c]{\texttt{SEOBNRv4EHM}\\[-1pt]optimum values}} & \multicolumn{3}{|c}{\makecell[c]{\texttt{SEOBNRv5EHM}\\[-1pt]optimum values}} & \multicolumn{3}{|c}{\makecell[c]{\texttt{TEOBResumS-Dal\'i}\\[-1pt]optimum values}} \\
\hline
\hline
\multicolumn{1}{c|}{$\text{SXS ID}$} & \multicolumn{1}{c}{$q$} & \multicolumn{1}{c}{$\chi_{1}$} & \multicolumn{1}{c}{$\chi_{2}$} & \multicolumn{1}{c}{$e_\text{gw}$} & \multicolumn{1}{c}{$\langle M \omega_{22} \rangle$} & \multicolumn{1}{c|}{$N _{ \text{p}}$} & \multicolumn{1}{c}{$e_0$} & \multicolumn{1}{c}{$\langle M \Omega \rangle_0$} & \multicolumn{1}{c|}{\makecell[c]{$ \mathcal M_{22} ^{ \text{max}}$\\$[\%]$}} & \multicolumn{1}{c}{$e_0$} & \multicolumn{1}{c}{$\langle M \Omega \rangle_0$} & \multicolumn{1}{c|}{\makecell[c]{$ \mathcal M_{22} ^{ \text{max}}$\\$[\%]$}} & \multicolumn{1}{c}{$e_0$} & \multicolumn{1}{c}{$\langle M \Omega \rangle_0$} & \multicolumn{1}{c}{\makecell[c]{$ \mathcal M_{22} ^{ \text{max}}$\\$[\%]$}} \\
\hline
\hline
\texttt{SXS:BBH:1358} & 1.0 & 0.0 & 0.0 & 0.147 & 0.037 & 8 & 0.221 & 0.014 & 0.049 & 0.166 & 0.016 & 0.008 & 0.166 & 0.015 & 0.157 \\
\texttt{SXS:BBH:1359} & 1.0 & 0.0 & 0.0 & 0.145 & 0.037 & 8 & 0.18 & 0.016 & 0.081 & 0.159 & 0.017 & 0.016 & 0.162 & 0.016 & 0.145 \\
\texttt{SXS:BBH:1360} & 1.0 & 0.0 & 0.0 & 0.21 & 0.037 & 7 & 0.296 & 0.015 & 0.111 & 0.236 & 0.016 & 0.012 & 0.227 & 0.016 & 0.126 \\
\texttt{SXS:BBH:1361} & 1.0 & 0.0 & 0.0 & 0.216 & 0.039 & 7 & 0.281 & 0.016 & 0.127 & 0.241 & 0.016 & 0.013 & 0.249 & 0.014 & 0.189 \\
\texttt{SXS:BBH:1362} & 1.0 & 0.0 & 0.0 & 0.295 & 0.037 & 7 & 0.393 & 0.011 & 0.334 & 0.3 & 0.016 & 0.03 & 0.296 & 0.016 & 0.146 \\
\texttt{SXS:BBH:1363} & 1.0 & 0.0 & 0.0 & 0.317 & 0.037 & 7 & 0.396 & 0.011 & 0.343 & 0.319 & 0.015 & 0.018 & 0.311 & 0.015 & 0.124 \\
\texttt{SXS:BBH:1364} & 2.0 & 0.0 & 0.0 & 0.064 & 0.038 & 8 & 0.1 & 0.015 & 0.193 & 0.074 & 0.017 & 0.002 & 0.071 & 0.017 & 0.077 \\
\texttt{SXS:BBH:1365} & 2.0 & 0.0 & 0.0 & 0.089 & 0.036 & 9 & 0.109 & 0.016 & 0.178 & 0.112 & 0.015 & 0.002 & 0.106 & 0.014 & 0.076 \\
\texttt{SXS:BBH:1366} & 2.0 & 0.0 & 0.0 & 0.143 & 0.036 & 9 & 0.188 & 0.015 & 0.157 & 0.161 & 0.016 & 0.002 & 0.175 & 0.014 & 0.151 \\
\texttt{SXS:BBH:1367} & 2.0 & 0.0 & 0.0 & 0.136 & 0.039 & 8 & 0.154 & 0.018 & 0.058 & 0.161 & 0.016 & 0.007 & 0.144 & 0.018 & 0.13 \\
\texttt{SXS:BBH:1368} & 2.0 & 0.0 & 0.0 & 0.138 & 0.037 & 8 & 0.177 & 0.016 & 0.203 & 0.153 & 0.017 & 0.005 & 0.177 & 0.014 & 0.081 \\
\texttt{SXS:BBH:1369} & 2.0 & 0.0 & 0.0 & 0.276 & 0.037 & 8 & 0.355 & 0.019 & 0.219 & 0.282 & 0.017 & 0.015 & 0.282 & 0.017 & 0.161 \\
\texttt{SXS:BBH:1370} & 2.0 & 0.0 & 0.0 & 0.294 & 0.039 & 7 & 0.355 & 0.02 & 0.267 & 0.323 & 0.015 & 0.014 & 0.271 & 0.017 & 0.27 \\
\texttt{SXS:BBH:1371} & 3.0 & 0.0 & 0.0 & 0.084 & 0.036 & 10 & 0.113 & 0.015 & 0.076 & 0.11 & 0.014 & 0.005 & 0.088 & 0.017 & 0.083 \\
\texttt{SXS:BBH:1372} & 3.0 & 0.0 & 0.0 & 0.137 & 0.037 & 10 & 0.151 & 0.018 & 0.093 & 0.172 & 0.015 & 0.006 & 0.158 & 0.015 & 0.059 \\
\texttt{SXS:BBH:1373} & 3.0 & 0.0 & 0.0 & 0.137 & 0.038 & 10 & 0.168 & 0.017 & 0.149 & 0.154 & 0.017 & 0.004 & 0.144 & 0.016 & 0.103 \\
\texttt{SXS:BBH:1374} & 3.0 & 0.0 & 0.0 & 0.279 & 0.039 & 9 & 0.352 & 0.018 & 0.73 & 0.3 & 0.016 & 0.015 & 0.296 & 0.015 & 0.144 \\
\texttt{SXS:BBH:2517} & 1.0 & 0.0 & 0.0 & 0.031 & 0.033 & 11 & 0.049 & 0.013 & 0.166 & 0.037 & 0.013 & 0.005 & 0.035 & 0.013 & 0.123 \\
\texttt{SXS:BBH:2518} & 1.0 & 0.0 & 0.0 & 0.057 & 0.02 & 28 & 0.054 & 0.009 & 0.316 & 0.07 & 0.008 & 0.004 & 0.064 & 0.009 & 0.124 \\
\texttt{SXS:BBH:2519} & 1.0 & 0.0 & 0.0 & 0.051 & 0.023 & 21 & 0.058 & 0.01 & 0.303 & 0.053 & 0.011 & 0.003 & 0.05 & 0.011 & 0.121 \\
\texttt{SXS:BBH:2520} & 1.0 & 0.0 & 0.0 & 0.141 & 0.031 & 12 & 0.18 & 0.013 & 0.158 & 0.166 & 0.013 & 0.007 & 0.156 & 0.013 & 0.132 \\
\texttt{SXS:BBH:2521} & 1.0 & 0.0 & 0.0 & 0.302 & 0.023 & 18 & 0.421 & 0.008 & 2.873 & 0.372 & 0.009 & 0.024 & 0.332 & 0.01 & 0.588 \\
\texttt{SXS:BBH:2522} & 1.0 & 0.0 & 0.0 & 0.347 & 0.022 & 17 & 0.429 & 0.008 & 4.773 & 0.429 & 0.008 & 0.077 & 0.38 & 0.009 & 0.53 \\
\texttt{SXS:BBH:2523} & 1.0 & 0.0 & 0.0 & 0.433 & 0.019 & 19 & 0.361 & 0.01 & 10.742 & 0.45 & 0.009 & 0.078 & 0.47 & 0.008 & 1.655 \\
\texttt{SXS:BBH:2524} & 1.0 & 0.0 & 0.0 & 0.645 & 0.01 & 22 & 0.765 & 0.003 & 31.822 & 0.696 & 0.004 & 1.083 & 0.716 & 0.003 & 16.254 \\
\texttt{SXS:BBH:2525} & 1.0 & 0.0 & 0.0 & 0.574 & 0.016 & 16 & 0.696 & 0.005 & 24.156 & 0.676 & 0.005 & 0.212 & 0.68 & 0.004 & 5.901 \\
\texttt{SXS:BBH:2526} & 1.0 & 0.0 & 0.0 & 0.701 & 0.009 & 19 & 0.834 & 0.002 & 32.584 & 0.745 & 0.004 & 2.494 & 0.781 & 0.002 & 19.004 \\
\texttt{SXS:BBH:2527} & 1.0 & 0.0 & 0.0 & 0.877 & 0.007 & 14 & 0.95 & 0.0 & 37.939 & 0.917 & 0.001 & 19.65 & 0.865 & 0.001 & 30.593 \\
\texttt{SXS:BBH:2528} & 1.0 & 0.0 & 0.0 & 0.716 & 0.012 & 10 & 0.861 & 0.002 & 26.173 & 0.801 & 0.004 & 4.728 & 0.772 & 0.004 & 17.691 \\
\texttt{SXS:BBH:2529} & 2.0 & 0.0 & 0.0 & 0.038 & 0.026 & 18 & 0.041 & 0.011 & 0.223 & 0.043 & 0.012 & 0.002 & 0.038 & 0.012 & 0.101 \\
\texttt{SXS:BBH:2530} & 2.0 & 0.0 & 0.0 & 0.151 & 0.027 & 18 & 0.175 & 0.013 & 0.127 & 0.168 & 0.012 & 0.007 & 0.171 & 0.011 & 0.196 \\
\texttt{SXS:BBH:2531} & 2.0 & 0.0 & 0.0 & 0.304 & 0.023 & 19 & 0.406 & 0.008 & 4.014 & 0.329 & 0.01 & 0.012 & 0.366 & 0.008 & 0.723 \\
\texttt{SXS:BBH:2532} & 2.0 & 0.0 & 0.0 & 0.353 & 0.022 & 20 & 0.446 & 0.008 & 7.315 & 0.361 & 0.01 & 0.082 & 0.369 & 0.01 & 0.603 \\
\texttt{SXS:BBH:2533} & 2.0 & 0.0 & 0.0 & 0.439 & 0.018 & 22 & 0.35 & 0.01 & 12.329 & 0.507 & 0.007 & 0.038 & 0.459 & 0.008 & 1.972 \\
\texttt{SXS:BBH:2534} & 2.0 & 0.0 & 0.0 & 0.645 & 0.01 & 25 & 0.794 & 0.003 & 32.788 & 0.665 & 0.005 & 2.075 & 0.726 & 0.003 & 20.994 \\
\texttt{SXS:BBH:2535} & 2.0 & 0.0 & 0.0 & 0.717 & 0.012 & 11 & 0.879 & 0.002 & 29.46 & 0.762 & 0.005 & 8.587 & 0.823 & 0.002 & 21.742 \\
\texttt{SXS:BBH:2536} & 3.0 & 0.0 & 0.0 & 0.05 & 0.02 & 39 & 0.053 & 0.009 & 0.242 & 0.055 & 0.009 & 0.004 & 0.055 & 0.009 & 0.091 \\
\texttt{SXS:BBH:2537} & 3.0 & 0.0 & 0.0 & 0.154 & 0.023 & 26 & 0.211 & 0.01 & 0.175 & 0.159 & 0.011 & 0.003 & 0.176 & 0.01 & 0.141 \\
\texttt{SXS:BBH:2538} & 3.0 & 0.0 & 0.0 & 0.262 & 0.019 & 36 & 0.335 & 0.009 & 5.92 & 0.311 & 0.008 & 0.014 & 0.283 & 0.008 & 0.28 \\
\texttt{SXS:BBH:2539} & 3.0 & 0.0 & 0.0 & 0.202 & 0.026 & 21 & 0.264 & 0.011 & 0.519 & 0.216 & 0.012 & 0.019 & 0.248 & 0.01 & 0.107 \\
\texttt{SXS:BBH:2540} & 3.0 & 0.0 & 0.0 & 0.159 & 0.034 & 11 & 0.195 & 0.015 & 0.156 & 0.166 & 0.016 & 0.007 & 0.173 & 0.015 & 0.058 \\
\texttt{SXS:BBH:2541} & 3.0 & 0.0 & 0.0 & 0.31 & 0.02 & 29 & 0.386 & 0.008 & 10.277 & 0.325 & 0.009 & 0.026 & 0.382 & 0.007 & 0.747 \\
\texttt{SXS:BBH:2542} & 3.0 & 0.0 & 0.0 & 0.309 & 0.023 & 23 & 0.405 & 0.008 & 6.296 & 0.38 & 0.008 & 0.024 & 0.364 & 0.008 & 0.54 \\
\texttt{SXS:BBH:2543} & 3.0 & 0.0 & 0.0 & 0.349 & 0.022 & 22 & 0.382 & 0.009 & 11.32 & 0.362 & 0.01 & 0.017 & 0.427 & 0.008 & 1.025 \\
\end{tabular}
\end{table}
}

\clearpage

% Assign the same table number
\setcounter{table}{\value{tabletemp}}
{
\renewcommand{\arraystretch}{0.7} % Row spacing
\begin{table}%[!]
\centering
\caption{ \emph{Continued}.}
%\small
\vspace{8pt}
\begin{tabular}{l | D{.}{.}{2.1} D{.}{.}{3.3} D{.}{.}{2.2} D{.}{.}{2.3} D{.}{.}{0.1} D{.}{.}{2.0} | D{.}{.}{1.3} D{.}{.}{0.2} D{.}{.}{0.2} | D{.}{.}{1.3} D{.}{.}{0.2} D{.}{.}{0.2} | D{.}{.}{1.3} D{.}{.}{0.2} D{.}{.}{0.2}} % The combination x.y increases or decreases the spacing to the left or right for each row, with each column aligned at the decimal point
 & \multicolumn{6}{c}{Physical properties} & \multicolumn{3}{|c}{\makecell[c]{\texttt{SEOBNRv4EHM}\\[-1pt]optimum values}} & \multicolumn{3}{|c}{\makecell[c]{\texttt{SEOBNRv5EHM}\\[-1pt]optimum values}} & \multicolumn{3}{|c}{\makecell[c]{\texttt{TEOBResumS-Dal\'i}\\[-1pt]optimum values}} \\
\hline
\hline
\multicolumn{1}{c|}{$\text{SXS ID}$} & \multicolumn{1}{c}{$q$} & \multicolumn{1}{c}{$\chi_{1}$} & \multicolumn{1}{c}{$\chi_{2}$} & \multicolumn{1}{c}{$e_\text{gw}$} & \multicolumn{1}{c}{$\langle M \omega_{22} \rangle$} & \multicolumn{1}{c|}{$N _{ \text{p}}$} & \multicolumn{1}{c}{$e_0$} & \multicolumn{1}{c}{$\langle M \Omega \rangle_0$} & \multicolumn{1}{c|}{\makecell[c]{$ \mathcal M_{22} ^{ \text{max}}$\\$[\%]$}} & \multicolumn{1}{c}{$e_0$} & \multicolumn{1}{c}{$\langle M \Omega \rangle_0$} & \multicolumn{1}{c|}{\makecell[c]{$ \mathcal M_{22} ^{ \text{max}}$\\$[\%]$}} & \multicolumn{1}{c}{$e_0$} & \multicolumn{1}{c}{$\langle M \Omega \rangle_0$} & \multicolumn{1}{c}{\makecell[c]{$ \mathcal M_{22} ^{ \text{max}}$\\$[\%]$}} \\
\hline
\hline
\texttt{SXS:BBH:2544} & 3.0 & 0.0 & 0.0 & 0.647 & 0.012 & 21 & 0.775 & 0.003 & 32.722 & 0.731 & 0.004 & 1.991 & 0.707 & 0.004 & 15.352 \\
\texttt{SXS:BBH:2545} & 4.0 & 0.0 & 0.0 & 0.035 & 0.026 & 25 & 0.033 & 0.013 & 0.088 & 0.04 & 0.013 & 0.007 & 0.043 & 0.01 & 0.063 \\
\texttt{SXS:BBH:2546} & 4.0 & 0.0 & 0.0 & 0.151 & 0.026 & 24 & 0.206 & 0.011 & 0.187 & 0.178 & 0.011 & 0.009 & 0.164 & 0.012 & 0.064 \\
\texttt{SXS:BBH:2547} & 4.0 & 0.0 & 0.0 & 0.299 & 0.024 & 27 & 0.389 & 0.008 & 8.671 & 0.373 & 0.009 & 0.027 & 0.32 & 0.01 & 0.568 \\
\texttt{SXS:BBH:2548} & 4.0 & 0.0 & 0.0 & 0.354 & 0.022 & 27 & 0.391 & 0.009 & 14.398 & 0.407 & 0.009 & 0.046 & 0.435 & 0.007 & 0.987 \\
\texttt{SXS:BBH:2549} & 4.0 & 0.0 & 0.0 & 0.442 & 0.018 & 29 & 0.363 & 0.01 & 19.904 & 0.508 & 0.007 & 0.101 & 0.456 & 0.008 & 4.64 \\
\texttt{SXS:BBH:2550} & 4.0 & 0.0 & 0.0 & 0.651 & 0.012 & 24 & 0.845 & 0.002 & 35.601 & 0.736 & 0.004 & 2.036 & 0.748 & 0.003 & 17.126 \\
\texttt{SXS:BBH:2551} & 4.0 & 0.0 & 0.0 & 0.73 & 0.012 & 14 & 0.847 & 0.003 & 35.33 & 0.79 & 0.004 & 14.613 & 0.806 & 0.003 & 26.012 \\
\texttt{SXS:BBH:2552} & 6.0 & 0.0 & 0.0 & 0.036 & 0.026 & 32 & 0.038 & 0.012 & 0.042 & 0.043 & 0.012 & 0.006 & 0.045 & 0.011 & 0.081 \\
\texttt{SXS:BBH:2553} & 6.0 & 0.0 & 0.0 & 0.151 & 0.027 & 30 & 0.184 & 0.012 & 0.507 & 0.177 & 0.011 & 0.013 & 0.163 & 0.012 & 0.169 \\
\texttt{SXS:BBH:2554} & 6.0 & 0.0 & 0.0 & 0.311 & 0.022 & 33 & 0.372 & 0.009 & 13.991 & 0.315 & 0.011 & 0.25 & 0.376 & 0.008 & 1.049 \\
\texttt{SXS:BBH:2555} & 6.0 & 0.0 & 0.0 & 0.353 & 0.022 & 33 & 0.441 & 0.008 & 16.906 & 0.413 & 0.009 & 0.022 & 0.405 & 0.008 & 0.905 \\
\texttt{SXS:BBH:2556} & 6.0 & 0.0 & 0.0 & 0.443 & 0.018 & 36 & 0.358 & 0.01 & 25.419 & 0.487 & 0.008 & 0.078 & 0.54 & 0.006 & 4.747 \\
\texttt{SXS:BBH:2557} & 6.0 & 0.0 & 0.0 & 0.593 & 0.015 & 29 & 0.714 & 0.005 & 36.014 & 0.674 & 0.005 & 0.76 & 0.65 & 0.005 & 10.93 \\
\texttt{SXS:BBH:2558} & 6.0 & 0.0 & 0.0 & 0.735 & 0.011 & 17 & 0.903 & 0.001 & 40.169 & 0.827 & 0.003 & 15.636 & 0.802 & 0.003 & 24.848 \\
\texttt{SXS:BBH:2559} & 8.0 & 0.0 & 0.0 & 0.012 & 0.035 & 18 & 0.015 & 0.014 & 0.017 & 0.015 & 0.015 & 0.004 & 0.011 & 0.016 & 0.065 \\
\texttt{SXS:BBH:2560} & 8.0 & 0.0 & 0.0 & 0.154 & 0.026 & 38 & 0.21 & 0.011 & 0.823 & 0.186 & 0.011 & 0.009 & 0.193 & 0.01 & 0.145 \\
\texttt{SXS:BBH:2561} & 8.0 & 0.0 & 0.0 & 0.309 & 0.022 & 42 & 0.431 & 0.008 & 15.176 & 0.358 & 0.009 & 0.134 & 0.362 & 0.009 & 0.755 \\
\texttt{SXS:BBH:2562} & 8.0 & 0.0 & 0.0 & 0.361 & 0.021 & 41 & 0.372 & 0.01 & 21.433 & 0.429 & 0.008 & 0.032 & 0.425 & 0.008 & 1.416 \\
\texttt{SXS:BBH:2563} & 8.0 & 0.0 & 0.0 & 0.414 & 0.023 & 30 & 0.366 & 0.011 & 21.646 & 0.475 & 0.009 & 0.226 & 0.483 & 0.008 & 3.023 \\
\texttt{SXS:BBH:2564} & 10.0 & 0.0 & 0.0 & 0.015 & 0.036 & 23 & 0.019 & 0.015 & 0.036 & 0.017 & 0.016 & 0.005 & 0.017 & 0.014 & 0.063 \\
\texttt{SXS:BBH:2565} & 10.0 & 0.0 & 0.0 & 0.019 & 0.034 & 26 & 0.022 & 0.015 & 0.037 & 0.022 & 0.015 & 0.005 & 0.022 & 0.014 & 0.062 \\
\texttt{SXS:BBH:2566} & 10.0 & 0.0 & 0.0 & 0.449 & 0.022 & 33 & 0.367 & 0.012 & 29.35 & 0.488 & 0.01 & 0.356 & 0.456 & 0.01 & 8.952 \\
\texttt{SXS:BBH:2567} & 10.0 & 0.0 & 0.0 & 0.448 & 0.025 & 26 & 0.487 & 0.011 & 25.82 & 0.483 & 0.011 & 0.18 & 0.548 & 0.007 & 2.682 \\
\texttt{SXS:BBH:2568} & 10.0 & 0.0 & 0.0 & 0.593 & 0.014 & 40 & 0.772 & 0.003 & 55.771 & 0.643 & 0.006 & 3.235 & 0.703 & 0.004 & 42.414 \\
\texttt{SXS:BBH:3822} & 15.05 & 0.0 & 0.01 & 0.057 & 0.035 & 34 & 0.079 & 0.015 & 0.057 & 0.064 & 0.016 & 0.008 & 0.061 & 0.017 & 0.045 \\
\texttt{SXS:BBH:3823} & 14.9 & 0.0 & 0.0 & 0.142 & 0.03 & 46 & 0.168 & 0.014 & 1.023 & 0.151 & 0.014 & 0.057 & 0.175 & 0.012 & 0.133 \\
\texttt{SXS:BBH:3824} & 14.84 & 0.0 & 0.0 & 0.262 & 0.031 & 36 & 0.357 & 0.016 & 6.813 & 0.321 & 0.012 & 0.059 & 0.296 & 0.013 & 0.328 \\
\texttt{SXS:BBH:3825} & 15.27 & 0.0 & 0.0 & 0.353 & 0.031 & 30 & 0.476 & 0.01 & 19.816 & 0.367 & 0.015 & 0.237 & 0.423 & 0.011 & 1.346 \\
\texttt{SXS:BBH:3826} & 15.78 & 0.0 & 0.0 & 0.414 & 0.026 & 38 & 0.567 & 0.008 & 53.722 & 0.43 & 0.012 & 1.131 & 0.434 & 0.011 & 8.134 \\
\texttt{SXS:BBH:3827} & 18.07 & 0.0 & 0.0 & 0.153 & 0.031 & 51 & 0.184 & 0.014 & 2.376 & 0.163 & 0.014 & 0.286 & 0.177 & 0.013 & 0.11 \\
\texttt{SXS:BBH:3828} & 17.71 & 0.0 & 0.0 & 0.27 & 0.031 & 40 & 0.32 & 0.016 & 21.499 & 0.32 & 0.013 & 0.454 & 0.275 & 0.015 & 1.065 \\
\texttt{SXS:BBH:3829} & 17.98 & 0.0 & 0.0 & 0.347 & 0.03 & 36 & 0.43 & 0.011 & 20.275 & 0.374 & 0.014 & 0.16 & 0.386 & 0.012 & 0.739 \\
\texttt{SXS:BBH:3961} & 1.1 & -0.4 & -0.7 & 0.17 & 0.025 & 15 & 0.242 & 0.01 & 0.699 & 0.187 & 0.011 & 0.064 & 0.18 & 0.012 & 0.146 \\
\texttt{SXS:BBH:3962} & 1.1 & -0.4 & -0.7 & 0.117 & 0.023 & 19 & 0.184 & 0.009 & 0.678 & 0.143 & 0.009 & 0.027 & 0.133 & 0.01 & 0.172 \\
\texttt{SXS:BBH:3963} & 1.1 & -0.4 & -0.7 & 0.227 & 0.019 & 26 & 0.294 & 0.009 & 2.449 & 0.282 & 0.007 & 0.134 & 0.237 & 0.009 & 0.192 \\
\texttt{SXS:BBH:3964} & 1.1 & -0.4 & -0.7 & 0.366 & 0.022 & 15 & 0.454 & 0.007 & 4.586 & 0.398 & 0.008 & 0.212 & 0.406 & 0.007 & 0.378 \\
\texttt{SXS:BBH:3965} & 1.1 & -0.4 & -0.7 & 0.397 & 0.028 & 7 & 0.442 & 0.01 & 6.418 & 0.442 & 0.012 & 0.763 & 0.506 & 0.008 & 1.378 \\
\texttt{SXS:BBH:3966} & 10.0 & -0.75 & 0.0 & 0.465 & 0.032 & 9 & 0.493 & 0.01 & 15.211 & 0.643 & 0.007 & 0.901 & 0.505 & 0.011 & 0.516 \\
\texttt{SXS:BBH:3967} & 1.0 & -0.75 & -0.75 & 0.55 & 0.016 & 14 & 0.683 & 0.004 & 25.644 & 0.63 & 0.006 & 1.0 & 0.659 & 0.004 & 2.697 \\
\texttt{SXS:BBH:3968} & 1.0 & 0.75 & 0.75 & 0.555 & 0.013 & 31 & 0.693 & 0.004 & 28.025 & 0.568 & 0.006 & 1.832 & 0.564 & 0.006 & 14.239 \\
\texttt{SXS:BBH:3969} & 1.0 & 0.9 & 0.9 & 0.335 & 0.021 & 26 & 0.361 & 0.011 & 10.794 & 0.377 & 0.009 & 0.407 & 0.34 & 0.01 & 1.212 \\
\texttt{SXS:BBH:3970} & 2.0 & 0.9 & 0.9 & 0.314 & 0.022 & 29 & 0.457 & 0.008 & 14.605 & 0.344 & 0.01 & 0.572 & 0.393 & 0.008 & 1.983 \\
\texttt{SXS:BBH:4385} & 18.05 & 0.0 & 0.0 & 0.096 & 0.037 & 34 & 0.111 & 0.018 & 0.271 & 0.105 & 0.017 & 0.049 & 0.102 & 0.017 & 0.109 \\
\hline
\hline
\end{tabular}
\end{table}
}

\clearpage

\twocolumngrid

%%%%%%%%%%%%%%%%%%%%%%%%%%%
%%%%%%%%%%%%%%%%%%%%%%%%%%%
%%%%%%%%%%%%%%%%%%%%%%%%%%%

\clearpage

\bibliography{../references}
%To compile the references, one must compile the BibTex on the TexShop menu, or use the command cmd+shift+b, everytime there is a change in the references.
%To autocomplete a reference, start writing \cite{
%and then the letter of the autor and press F5

%To autogenerate a citation key in bibdesk, use Command+K

\end{document}